\documentclass[fleqn,usenatbib]{mnras}

\usepackage{mathptmx}

\usepackage[T1]{fontenc}

\DeclareRobustCommand{\VAN}[3]{#2}
\let\VANthebibliography\thebibliography
\def\thebibliography{\DeclareRobustCommand{\VAN}[3]{##3}\VANthebibliography}

\usepackage{graphicx}	
\usepackage{amsmath, tabu}	
\usepackage{amssymb}	
\usepackage{adjustbox}
\usepackage{makecell, multirow}
\usepackage{array}
\usepackage[dvipsnames]{xcolor}
\usepackage{placeins}
\usepackage{tabularx}
\usepackage{tablefootnote}
\usepackage{threeparttable}

\newcommand{\reff}[1]{{#1}}

\newcommand{\Msun}{\ensuremath{\,{\rm M}_\odot}}                  
\newcommand{\Lsun}{\ensuremath{\,{\rm L}_\odot}}                  
\newcommand{\Teff}{\ensuremath{T_{\rm eff}}}
\newcommand{\dsct}{$\delta$ Scuti}
\newcommand{\gdor}{$\gamma$ Doradus}
\newcommand{\kms}{\,km\,s$^{-1}$}                                 
\newcommand{\vsini}{\ensuremath{v\sin i}}                         
\newcommand{\Msunnom}{\hbox{$\mathcal{M}^{\rm N}_\odot$}}
\newcommand{\Rsunnom}{\hbox{$\mathcal{R}^{\rm N}_\odot$}}
\newcommand{\Lsunnom}{\hbox{$\mathcal{L}^{\rm N}_\odot$}}

\title[Pulsating Eclipsing Binary Stars]{Characterization of the \dsct\ eclipsing binary KIC 4851217 and its tertiary companion as well as detection of tidally tilted pulsations}

\author[Z.\ Jennings et al.]{
Z.\ Jennings$^{1}$\thanks{E-mail: z.jennings@keele.ac.uk},
J.\ Southworth$^{1}$, S.\ A.\ Rappaport$^{2}$, T.\ Borkovits$^{3,4,5,6,7}$, G.\ Handler$^{8}$, D.\ W.\ Kurtz$^{9,10}$
\\
$^{1}$\,Astrophysics Group, Keele University, Staffordshire, ST5 5BG, UK \\
$^{2}$\,Department of Physics, Kavli Institute for Astrophysics and Space Research,
MIT, Cambridge, MA 02139, USA\\
$^{3}$\,Baja Astronomical Observatory of University of Szeged, H-6500 Baja, Szegedi út, Kt. 766, Hungary\\
$^{4}$\,HUN-REN-SZTE Stellar Astrophysics Research Group, H-6500 Baja, Szegedi út, Kt. 766, Hungary\\
$^5$\,Konkoly Observatory, HUN-REN Research Centre for Astronomy and Earth Sciences, H-1121 Budapest, Konkoly Thege Miklós út 15-17, Hungary \\
$^6$\,ELTE Gothard Astrophysical Observatory, H-9700 Szombathely, Szent Imre h. u. 112, Hungary \\
$^7$\,HUN-REN-ELTE Exoplanet Research Group, H-9700 Szombathely, Szent Imre h. u. 112, Hungary \\
$^{8}$\,Nicolaus Copernicus Astronomical Center, Polish Academy of Sciences, ul. Bartycka 18, PL-00-716 Warszawa, Poland\\
$^{9}$\,Centre for Space Research, Physics Department, North West University, Mahikeng 2745, South Africa\\
$^{10}$\,Jeremiah Horrocks Institute, University of Central Lancashire, Preston PR1 2HE, UK
}

\date{Accepted XXX. Received YYY; in original form ZZZ}

\pubyear{2022}

\begin{document}
\label{firstpage}
\pagerange{\pageref{firstpage}--\pageref{lastpage}}
\maketitle

\begin{abstract}
Stellar theory enables us to understand the properties of stars at different stages of their evolution, and contributes to other fields of astrophysics such as galactic and exoplanet studies. Assessing the accuracy of stellar theories necessitates high precision, model-independent measurements of the properties of real stars, such as those obtainable for the components of double lined eclipsing binaries (DLEBs), while asteroseismology offers probing power of the stellar interior if one or both components pulsate. KIC\,4851217 is a DLEB containing two late A-type stars and exhibits pulsations of the \dsct\ type. By analysing high resolution HERMES and moderate resolution ISIS spectra, jointly with \emph{Kepler} and TESS light curves, we measured the masses, radii and effective temperatures of the components to precisions of $\sim$0.5, $\sim$1.1 and $\sim$1 per cent, respectively. We additionally report the discovery and characterisation of a tertiary M-dwarf companion. Models of the system's spectral energy distribution agree with an age of 0.82 Gyr, with the more massive and larger secondary component near the end of the main sequence lifetime. An examination of the pulsating component's pulsation frequencies reveals 39 pulsation multiplets that are split by the orbital frequency. For most of these, it is evident that the pulsation axes have been tilted into the orbital plane. This makes KIC\,4851217 a tidally tilted pulsator (TTP). This precisely characterized \dsct\ DLEB is an ideal candidate for advancing intermediate-mass stellar theory, contributing to our understanding of hierarchichal systems as well as to the topic of TTPs. 
\end{abstract}





\section{Introduction}
\label{sec:introduction}
Stars serve as one of the universe's foundational components, creating elements and giving rise to galaxies; accurate understanding of stellar structure and evolution is essential for understanding the universe's history as well as galaxies and exoplanets \citep{SilvaAguirre_2018}. In a broader sense, stars are natural laboratories that allow for the advancement of physics by studying processes which cannot be replicated on Earth.

Discriminating among different stellar theories and improving them requires assessing their accuracy, and this necessitates measuring the properties of real stars.  High precision and model independence of the measurements is essential for their effectiveness as constraints \citep{Torres_2010}. Measurements satisfying these criteria can be made for the components of double lined eclipsing binaries (DLEBs). The characterisation of a DLEB relies on the combination of modelling the eclipses and the radial velocity (RV) curves of both components, where each analysis contributes a subset of the information required to obtain model-independent, high-precision \citep[e.g., better than 1\%;][]{Southworth_2015_DEBCat} measurements of the components' masses and radii. For this reason, DLEBs are routinely used to critically assess stellar evolution theory \citep[e.g.,][]{Stancliffe_2015,delBurgo_2018}. Further information is accessible for the components of DLEBs by, e.g., modelling the spectral energy distribution (SED) of the system, or the components' atmospheres (see Sections \ref{sec: SED fitting} and \ref{sec: atmospheric parameters}).

Intricate processes (e.g., mixing, magnetism and convection) occurring in the stellar interior are difficult to calibrate using constraints from DLEBs alone, and additional constraints are \reff{needed} in order to overcome their simplified descriptions in stellar models. Suitable constraints can be obtained for pulsating stars using asteroseismology, and combined with conventional constraints if they exist in a \reff{DLEB}. The conventional properties act to constrain the pulsation properties as well as allowing for the appropriate stellar models to be used when comparing theoretical pulsation frequencies to the observed ones \citep{Liakos_2020,Feng_2021, Sekaran_2021}. These synergies make DLEBs with pulsating components invaluable for advancing stellar theory and the number of such systems reported in the literature is increasing with more detections of pulsating stars in eclipsing binaries (EBs) \citep{Gaulme_Guzik_2019b,Chen_2023}, thanks to space missions such as CoRoT \citep[e.g.,][]{Maceroni_2013}, \emph{Kepler} \citep[e.g.,][]{Southworth_2011, Guo_2019}, and TESS \citep[e.g.,][]{Lee_2018,MeBowman22mn,MeVanreeth22mn}. 

Pulsating EBs also offer the opportunity to study the effects of binarity on pulsations. Tidally excited modes occur when harmonics of the forcing frequency associated with the \emph{dynamical} tide of an eccentric binary come close to an eigen-frequency of a free oscillation; this mostly leads to gravity modes at integer multiples of the orbital frequency \citep{Welsh_2011, Hambleton_2013, Fuller_2017}. On the other hand, the \emph{equilibrium} tide of circular binaries may cause deformation of pulsation mode cavities resulting in perturbed self-excited modes \citep{Polfliet_Smeyers_1990, Reyniers_Smeyers_2003b, Reyniers_Smeyers_2003a, ParkJang_2020}.  \citet{Liakos_Niarchos_2017} find a threshold orbital period for \dsct\ stars below which the dominant pulsation period is correlated with the orbital period, i.e., influenced by binarity. \citet{Kahraman_2017} almost doubled this threshold \citep{Liakos_2020_KIC4851217} considering eclipsing systems only. 

Almost 300 EBs containing \dsct\ components have been announced \citep[e.g.,][]{Zhoe_2010, Soydugan_2011, Liakos_2012, Liakos_Niarchos_2017, Gaulme_Guzik_2019, Chen_2022, Kahraman_2022}.
The \dsct\ stars are early A to F variables and their luminosity class ranges from dwarf to giant \citep{Kahraman_2022, Aerts_book}. They pulsate in nonradial and radial pressure modes (p~modes) with periods ranging between 15 min and 8 h \citep{Aerts_book, Uttyerhoeven_2011}, driven by the $\kappa$ mechanism acting in the partial ionization zone of He\,II \citep{Pamyatnykh_1999, Antoci_2014, Murphy_2020}. The mass range of \dsct\ stars, between 1.5 and 2.5\Msun\ \citep{Aerts_book}, places them within the transition region of lower-mass stars with convective envelopes to higher-mass stars with predominantly radiative envelopes and thin convection zones \citep{Bowman_2017, Yang_2021}. The pulsations of \dsct\ stars are therefore excellent for probing stellar processes in a mass range over which major changes to the interior structure take place. 

$\delta$ Scuti stars can exist with more than one companion \citep[e.g.,][]{Hareter_2008}. Variations in the primary and secondary eclipse times (ETVs) can indicate that the EB is gravitationally bound to a third component \citep{Rappaport_2013}. This is because the ETVs are, in this case, the result of the barycentric motion of the EB's centre of mass (the \emph{outer} orbit) about the third body, i.e., the light travel time effect (LTTE). However, there are alternative mechanisms that cause ETVs, such as mass transfer between the components \citep{Conroy_2014}, so a considerable time span of the observations, comparable to one outer orbital cycle, is typically required to determine confidently that the signal is due to a tertiary component. The study of triple star systems gives new insights into the physics of EBs. The orbital architecture and masses of the constituents can contribute to our understanding of processes that form multiple systems \citep{Rappaport_2013}; the general interpretation for the formation of close binaries is that they become hardened over time through interactions with a third body \citep{Conroy_2014}. See \citet{Borkovits_2022} for a review of EBs in dynamically interacting close, multiple systems.

In this work, we present a comprehensive analysis of KIC 4851217. This object is a detached DLEB in a close orbit with a period of $2.470$ d and shows \dsct\ pulsations, some of which are tidally tilted pulsations (TTPs) meaning that the pulsation axis has been tilted into the orbital plane, in many cases along the tidal axis itself \citep[e.g.,][]{Handler_2020, Kurtz_2020, Rappaport_2021}. ETVs are detected in the O-C diagram (see Section \ref{sec: preliminary light curve solutions}) which we successfully modelled as a combination of the LTTE due to a third body and apsidal motion of the EB orbit. Thus, the object is an ideal candidate for deriving constraints on stellar structure from its pulsations and dynamically derived fundamental parameters, studying the effects of tides on pulsations from its TTPs, as well as contributing to our understanding of hierarchical systems. 

KIC 4851217 was previously studied by \citet{Liakos_2020_KIC4851217}, who presented a detailed light curve, spectroscopic, and seismic analysis using RVs derived by \citet{Helminiak_2019} in their high resolution ($R\sim50\,000$) spectroscopic monitoring of 22 bright objects in the \emph{Kepler} eclipsing binary catalogue (KEBC) \citep{Prsa_2011, Kirk_2016}. A frequency analysis was also performed by \citet{fedurco_parimucha_gajdos_2019} on the \emph{Kepler} data; these authors concluded that the detected oscillations are due to tidally focussed pulsation modes. In addition to other previous studies mentioned by \citet{Liakos_2020_KIC4851217}, KIC 4851217 was detected by \citet{Gaulme_Guzik_2019} in their systematic search for pulsators in the KEBC; similarly, \citet{Chen_2022} detected the object in their search for \dsct\ pulsators in the catalogues of TESS EBs by \citet{Prsa_2022} and \citet{Shi_2022}. None of the these studies report the detection of a tertiary companion. Only a long-term parabolic trend in the primary and secondary ETVs was noted in the studies by \citet{Gies_2012, Gies_2015} and \citet{Conroy_2014}. Our work is complementary to the previous studies. We present and analyse additional, higher resolution ($R=85\,000$) spectroscopic observations, while the inclusion of TESS photometry allows us to report the discovery and characterization of the tertiary component for the first time. In this analysis, we denote the hotter primary star in the inner EB as star Aa and the secondary, star Ab, is the one eclipsed during secondary eclipse; the tertiary body is star B. 

Section \ref{sec:observations} describes the observations and in Section \ref{sec: preliminary analysis} we perform a preliminary analysis of the photometric light curves and spectral energy distribution (SED) of KIC\,4851217. We present a detailed spectroscopic analysis in Section \ref{sec:Spectroscopic-Analysis} and analyse the light curves in Section \ref{sec:lc:wd}. In Section \ref{sec: combined anlysis}, we perform a simultaneous analysis of the RVs, light curves, ETVs and SED (jointly) from which estimations of the components' physical properties follow. We also determine the physical properties of the components based on the modelling of the individual subsets of these data in Section \ref{tab:absdim} to demonstrate the extractable information from each, and then compare the two sets of results in Section \ref{sec: comparison of individual and combined results}. We introduce the pulsations for this object in Section \ref{sec: pulsation analysis}, but will present their full analyses in a follow-up study (in prep). Section \ref{sec:discussion} discusses the results, and concluding remarks are given in Section \ref{sec:conclusion}.

\section{Observations}
\label{sec:observations}

\subsection{\textit{Kepler} photometry}

\begin{figure*}
    \centering
    \includegraphics[width = 0.9\textwidth]{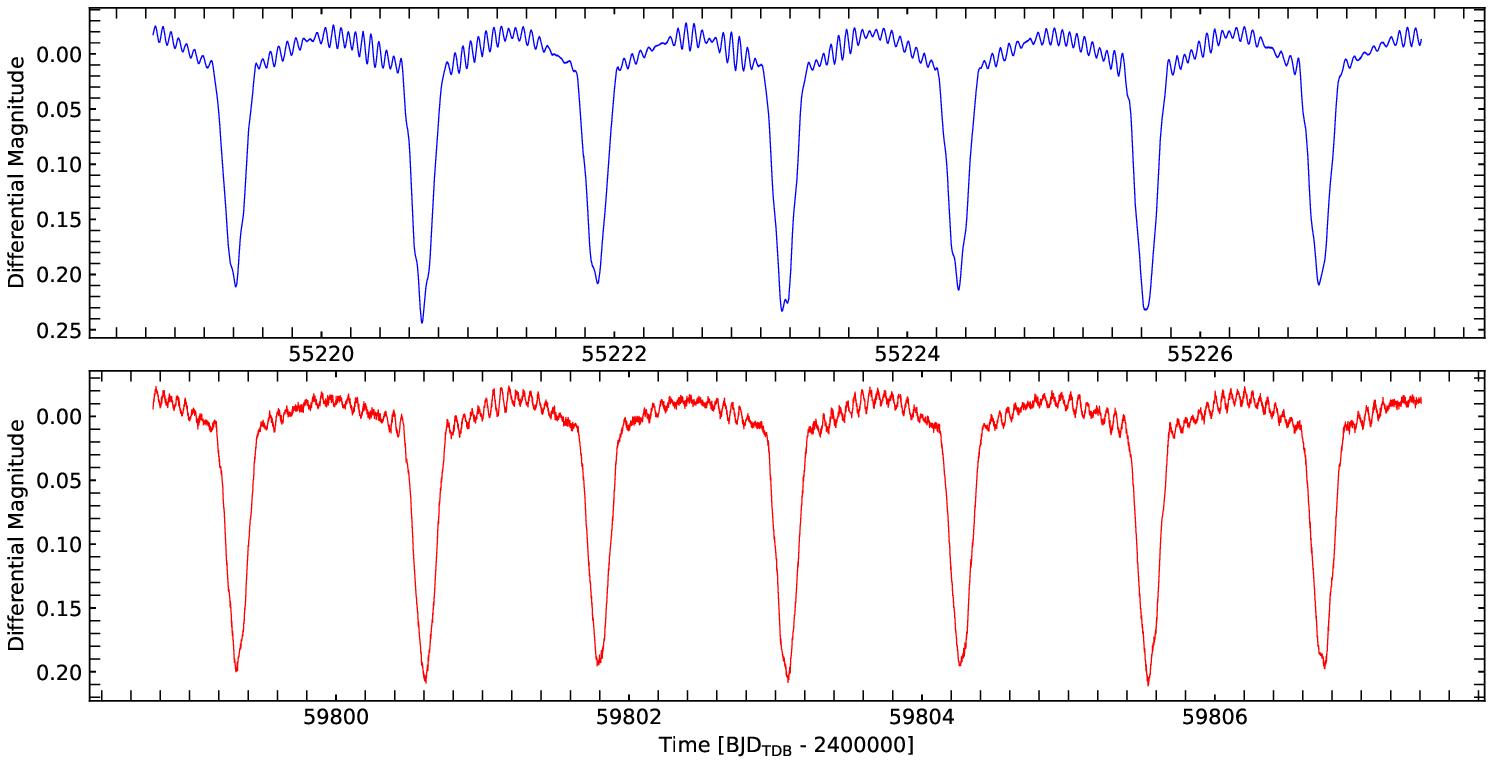}
    \caption{Top panel; a representative part of the \emph{Kepler} simple aperture photometry light curve from Quarter 4. Bottom panel; same as the top panel but for the TESS sector 55 SAP light curve.}
    \label{fig: Kepler and TESS raw light curves}
\end{figure*}

The \emph{Kepler} mission \citep{Koch_2010}, which was launched in 2009 March, continuously monitored approximately 150\,000 main-sequence stars in the direction of the constellations Cygnus and Lyra. There are a multitude of advantages associated with pointing to a single sky region, including concentrating on the best available star field, optimising the spacecraft design, and simplifying operations \citep{Koch_2010}. The most important advantage in the context of the current work is being able to monitor stars for multiple years with a high duty cycle, allowing for highly detailed asteroseismic investigations. The photometric precision achieved by the {\it Kepler} observations was designed to be sufficient to detect a single 6.5-hr transit from an Earth-sized planet passing in front of a 12th-magnitude G2 star at the 4$\sigma$ level \citep{Borucki_2010, Koch_2010}, in order to achieve the primary science objective of the satellite. Thus, \emph{Kepler} collected a large amount of high-quality data for many stellar systems. \emph{Kepler} collected the data for its targets by summing individual images into either 29.424-min long cadence (LC) bins or 58.85-s short cadence (SC) bins. The data were further grouped into quarters defined by successive $90^{\circ}$ rotations of the spacecraft every three months to keep the solar arrays pointed toward the Sun during its Earth-trailing heliocentric orbit. KIC 4851217 was observed in seven quarters (2, 4, 9, 13, 15--17) in SC mode between 2009 June and 2013 May and in 15 quarters (0--5, 7--9, 11--13, 15--17) in LC mode. Most quarters consist of three months of observations by \emph{Kepler} except quarters 0 (ten days), 1 (one month) and 17 (32 days), and SC data are only available for one out of three months during quarters 2 and 4.


\subsection{TESS photometry}


The Transiting Exoplanet Survey Satellite (TESS; \citealt{Ricker_2015}) searches for transiting planets by observing the nearest and brightest stars via an all-sky survey. During its 2\,yr primary mission (2018--2020), TESS collected data for 200\,000 main-sequence dwarfs with spectral types from F5 to M5, pre-selected according to transit detectability, using a 2\,min (SC) sampling cadence. Further data were collected for all stars within the field of view ($24^{\circ} \times 96^{\circ}$) with a 30\,min (LC) cadence -- these are the full frame images (FFIs) \citep{Ricker_2015}. For the extended missions (2020--2022 and 2022--2025), TESS implemented 20\,s cadence monitoring of selected targets in addition to the existing 2\,min cadence, alongside 10\,min FFIs in the first extension and 200\,s FFIs in the second extension. One patch of sky is observed in each sector and 13 sectors together cover most of one hemisphere of the sky; TESS will have gathered data for 97\% of the sky by the end of the second extended mission. Due to data downlink to Earth, there is a gap in the observations during each sector. 

KIC 4851217 has been observed in SC mode by TESS in five sectors as of 2023. These are sectors 14 and 15 (2019 July 18 to August 15), 41 (2021 July 23 to August 20), and 54 and 55 (2022 July 9 to September 1).

\subsection{WHT spectroscopy}

Spectroscopic observations were carried out using the ISIS spectrograph on the 4.2-m William Herschel Telescope (WHT) at La Palma. ISIS has two arms split by a dichroic so it can observe two wavelength intervals simultaneously. We used ISIS to acquire 17 observations in 2011 June (over 4 nights) and 14 observations in 2012 July (over 7 nights). 

A 0.5$^{\prime\prime}$ slit was used to obtain the highest possible spectral resolution. In the 2011 run the slit change mechanism was not working properly so the slit width was set manually to somewhere close to the intended 0.5$^{\prime\prime}$.

In the blue arm we used the H2400B grating to obtain spectra covering the 4200--4550\,\AA\ wavelength interval. The reciprocal dispersion was 0.11\,\AA\,px$^{-1}$ and the resolution was approximately 0.22\,\AA. The standard 5300\,\AA\ dichroic was used to split the blue and red arms.

In the red arm we used the R1200R grating to obtain spectra covering the 6100--6730\,\AA\ wavelength interval. The reciprocal dispersion was 0.26\,\AA\,px$^{-1}$ and the resolution was approximately 0.52\,\AA.

\subsection{HERMES spectroscopy}

A total of 41 spectroscopic observations of KIC 4851217 were obtained using the cross-dispersed fibre-fed \'echelle spectrograph HERMES \citep[High Efficiency and Resolution Mercator Échelle Spectrograph;][]{Raskin_2011} on the 1.2-m Mercator telescope at La Palma. The high-efficiency mode was used, giving spectra with a resolving power of $R = 85\,000$. These observations were obtained between 2011 April and 2012 July.

\section{Preliminary Analysis}\label{sec: preliminary analysis}
\subsection{Ephemeris} \label{sec: preliminary light curve solutions}

In all, there are 438 primary and 442 secondary eclipse times extracted from the {\it Kepler} light curves, as well as 54 primary and 50 secondary eclipse times derived from the TESS light curves. These data span 13.3 yr. The method we used for determining the eclipse mid-times has been discussed in several previous papers \citep[see][]{Borkovits_2015, Borkovits_2016}. Pulsations can affect the measurement of mid-eclipse times \citep[e.g.,][]{Borkovits_2014} but in the present case the pulsation amplitudes are negligible compared to the eclipse amplitudes, and the only potential influence of the pulsating behaviour might be the slightly larger scatter in the primary eclipse times compared to the secondary's (suggesting the pulsations might belong to, or are stronger in, the secondary). However, it is evident that both the ETV curves in Fig.\,\ref{fig: ETV model} are well-defined. 

The best-fit linear ephemeris for these eclipse times is given by:
\begin{equation}
T_{\rm TDB}(E) = 2456016.12186(13) + 2.47028992(34)\,E,
\end{equation}
where TDB stands for the barycentric dynamical time scale and $E$ for the epoch number. The ETV curve that results from subtracting out this linear ephemeris is shown in Fig.~\ref{fig: ETV model}.  We subtracted 0.042 days from the primary ETV curve so as to bring it visually closer to the secondary ETV curve, but this was done only after we analyzed the curves for an outer orbit.  A first look at these ETV curves shows three interesting features: (i) there is clearly non-linear behaviour that likely indicates the presence of a third body; (ii) the two curves drift upward, indicating that our trial linear fit to the eclipse times has some residual term to be fitted for; and (iii) the two ETV curves are slowly, but clearly, converging (by $\sim$0.003 d over 13 yr), thereby indicating a possible apsidal motion.

\subsection{Preliminary ETV analysis}\label{sec: preliminary ETV analysis}
\label{Sect:ETV_preliminary}

\begin{figure}
    \centering
    \includegraphics[width = \columnwidth]{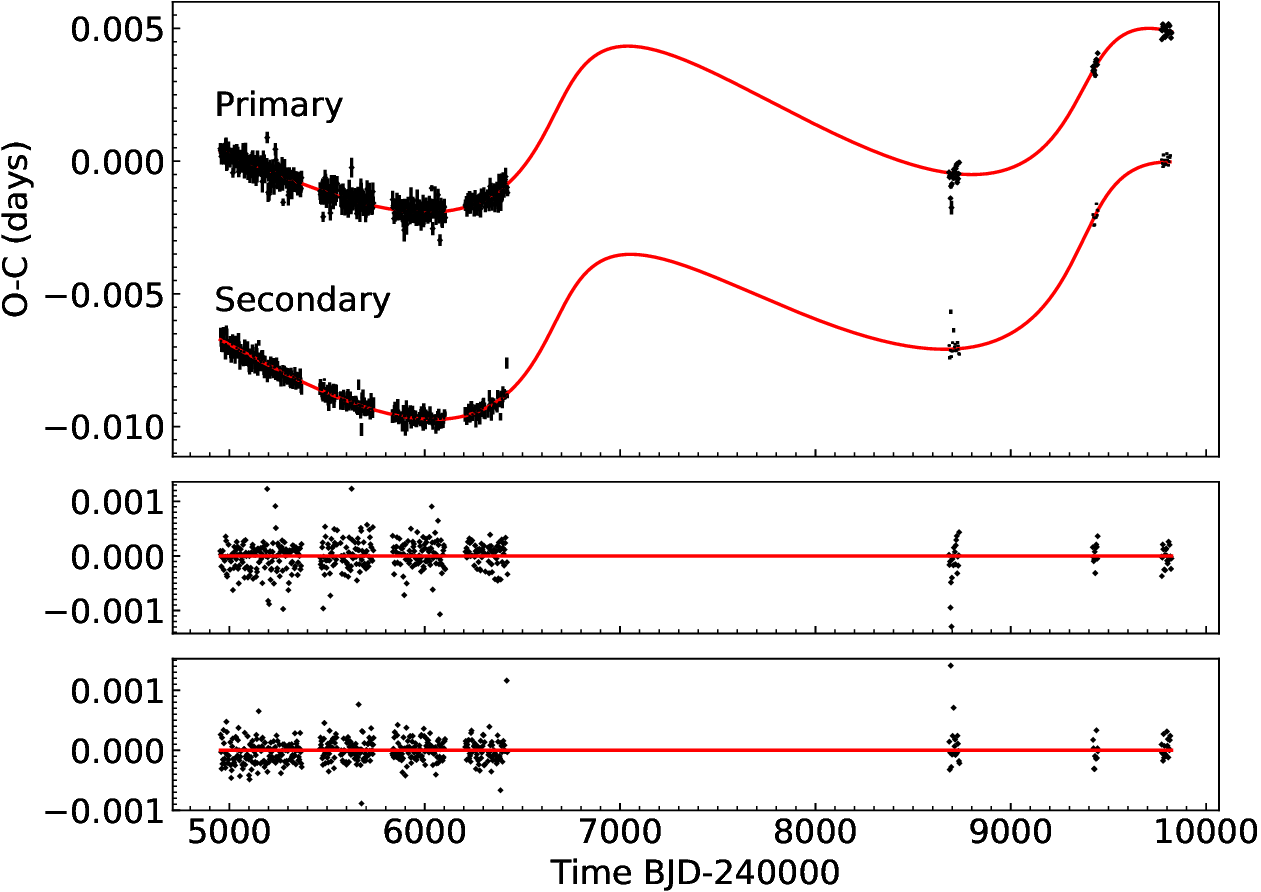}
    \caption{Model fits (red lines) to the ETV curves constructed from the measured times of primary and secondary mid-eclipses, where that of the former is shifted by $-$0.042 d. The points before day 7000 are from {\it Kepler}, while the later points are from TESS data.}
    \label{fig: ETV model}
\end{figure}

We first tried to fit an outer orbit to the ETV curves shown in Fig.~\ref{fig: ETV model}.  As noted, in addition to the obvious non-linear behaviour in the ETV curves that likely indicates an outer orbit, the two curves are slightly converging toward each other.  If the non-linear behaviour is due to the classic light travel time effect (LTTE)\footnote{In Sect.\,\ref{sec: combined anlysis} we show that the dynamical delays are negligible in this system.}, the ETV curves of both the primary and secondary eclipses should run parallel to each other.  Since they do not, we take this to tentatively suggest that there is apsidal motion in the EB.  As we show later in Sect.~\ref{sec: combined anlysis}, this is too large an effect to be driven by the third body.  Therefore, for now we assume that any apsidal motion in the EB is due to the classical effect from mutually induced tides, and treat it as such in our preliminary fit of the ETV curves.  

	The expression we fit is as follows:
\begin{eqnarray}
{\rm ETV}(t) & = & t_0 + dP_{\rm in}(t-t_i)/P_{\rm in} + {\rm LTTE}(t) \\
& \pm & \frac{e_{\rm in} P_{\rm in}}{\pi} \cos[\omega_{\rm in}(t_i) +2 \pi (t-t_i)/P_{\rm aps}]
\end{eqnarray}
where $t_i$ is simply defined as the start of the observations on BJD 2454953.90098, and is not a free parameter, and the plus and minus symbol refers to the primary and secondary eclipse times, respectively.  In all, there are four terms comprising ten free parameters: (i) an arbitrary offset time for the ETVs, $t_0$; (ii) a linear term in time that corrects the EB period, $dP_{\rm in}$; (iii) the LTTE effect that accounts for the outer orbit with five free parameters, $P_{\rm out}$, $a_{\rm out,eb} \, \sin i$,\footnote{The is the projected semimajor axis of the EB around the centre of mass of the triple system.} $e_{\rm out}$, $\omega_{\rm out}$, and $\tau_{\rm out}$, with their usual meanings; and (iv) the apsidal motion term which has three free parameters: $e_{\rm in}$, $\omega_{\rm in}$, and $P_{\rm aps}$, where the ``in'' subscript refers to the `inner orbit", i.e., that of the EB, and $P_{\rm aps}$ is the period of the apsidal motion.

The red curves superposed on each of the ETV curves in Fig.~\ref{fig: ETV model} are the result of a Levenberg-Marquardt minimization of $\chi^2$.  The best-fit parameters are summarized in Table \ref{tab: ETV results}, where the cited uncertainties were derived from an MCMC \citep{Ford_2005} evaluation of parameter space.  The outer orbital period is fairly well determined at $2700 \pm 40$ d (note that there are nearly two full outer periods in the span of the data train). \reff{The orbital cycle that the \emph{Kepler} data project to in the TESS epoch is precisely determined to within about $0.05\%$ of the inner orbital cycle given the $\sim 1$\,min accuracy of the $\sim440$ \emph{Kepler} eclipse times; the shape of the curves between these epochs is a consequence of the fairly high outer-orbital eccentricity of $e_{\rm out} \simeq 0.55 \pm 0.01$.}  The inferred mass function is $f(M) = 0.0033$\,M$_\odot$, which in turn provides a rough-estimate that the mass of the third body is about 0.4\,M$_\odot$ for an assumed outer orbital inclination angle near 90$^\circ$\footnote{\reff{The assumption of $i=90^{\circ}$ yields a minimum value because $M_{\rm ter}^3 = f(M)*M_{\rm tot}^2/{\rm sin^3}i$ for $M_{\rm ter}$ and $M_{\rm tot}$ the tertiary body mass and total mass of the system, respectively.}} and a total mass of the EB near 4\,M$_\odot$. 
	
The fit to the apsidal motion yields a well-defined apsidal period for the EB of $160 \pm 5$ yr. 
An additional bonus from fitting the precise ETV times for apsidal motion is that we also find remarkably precise values of $e_{\rm in}$ and $\omega_{\rm out}$ of $0.03173 \pm 0.00008$ and $170.2^\circ \pm 1.7^\circ$, respectively. 

\begin{table} \caption{\label{tab: ETV results} Results from the ETV model.}
\centering
\begin{threeparttable}
\begin{tabular}{l r@{\,$\pm$\,}l}
\hline
    Parameter & \multicolumn{2}{c}{Result}\\
     \hline
    $\tau_{\rm out}$ [days] \tnote{a} &    6747&   85 \\
    $P_{\mathrm{out}}$ [days]  &    2676 &  43 \\
    $A_{\mathrm{ltte}}$ [days]  \tnote{b}&    0.00317    &0.00015 \\
    $e_{\mathrm{out}}$         &    0.55    &0.03 \\
    $\omega_{\mathrm{out}}$\tnote{c} [$^{\circ}$] & 21   & 10 \\
    $P_{\mathrm{aps}}$ [years] &    163  & 13 \\
    $\omega_{\rm in}$ [$^{\circ}$] \tnote{c}&    170.2  &  1.7 \\
    $e_{\rm in}$               &    0.03174    &0.00008 \\
    $\mathrm{d}P_{\rm in}$ [days]             &    $1.89\times 10^{-6}$ & $0.10\times 10^{-6}$ \\
    \hline
\end{tabular}
\begin{tablenotes}
            \item[a] \reff{Time of periastron passage.}
            \item[b] \reff{Amplitude of the LTTE.}
            \item[c] \reff{Argument of periastron.}
        \end{tablenotes}
    \end{threeparttable}
\end{table}


\subsection{SED fitting}\label{sec: SED fitting}

\begin{figure*}
\begin{center}
\includegraphics[width=0.495 \textwidth]{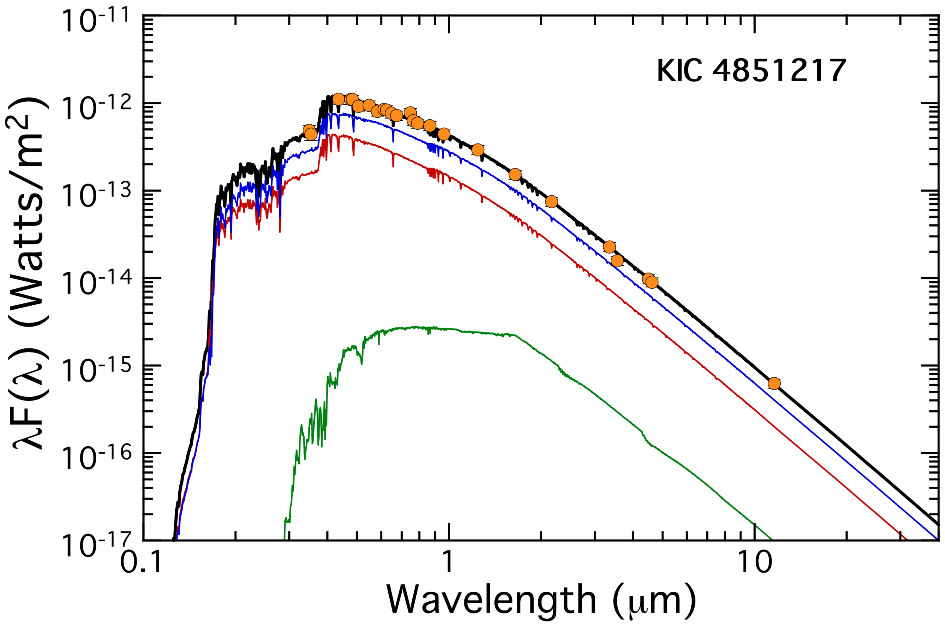} 
\includegraphics[width=0.475 \textwidth]{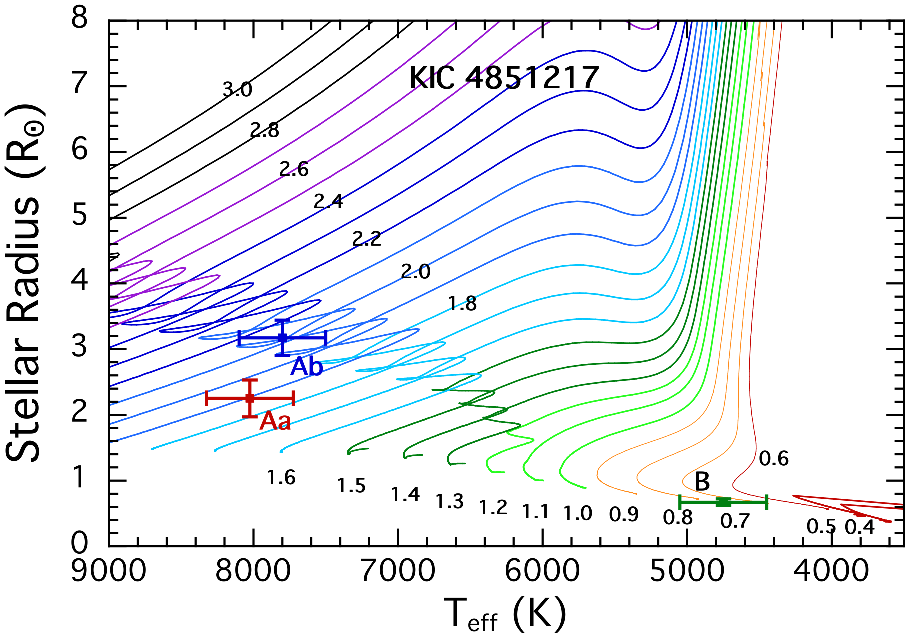}
\caption{{\it Left panel:} An illustrative SED fit to the KIC 4851217 system.  The orange points are the observed SED values take from VizieR (see text), while the blue, red, and green curves are the model SED curves for the secondary, primary, and tertiary stars, respectively.  Black is the sum of the individual stellar contributions. {\it Right panel:} The corresponding locations of the three stars on the MIST stellar evolution tracks (see text).  The numbers labelling the tracks are the stellar mass in M$_\odot$.  Note that the primary star, Aa, is the hotter, but less massive of the binary pair. }
\label{fig:sed} 
\end{center}
\end{figure*}

In this section we attempt to see what can be learned about the system parameters using only information from the spectral energy distribution (SED) of the triple system.  We find 25 SED points on the VizieR\footnote{\url{http://vizier.unistra.fr/vizier/sed/}}(\citealt{ochsenbein00}) SED database between 0.35 $\mu$m and 11.6 $\mu$m. These are shown in Fig.~\ref{fig:sed}.  We assign fixed uncertainties of 10 per cent on all fluxes to take into account the fact that there are frequent eclipses of this depth occurring. The purpose of the SED fitting at this stage of the analysis is to provide some initial insights into the system parameters.  

In order to fit three model atmospheres to a single SED curve, it is important to have at least a few other constraints in order to produce anything like a unique solution.  Here we adopt the following set of conditions and assumptions: (1) there are three stars in the system which are co-evolutionary and have experienced no prior mass transfer events; (2) star B contributes $\lesssim 10$ per cent of the system light, otherwise it would have been detected in the RV data (see Sect.~\ref{sec:RV-analysis}); and (3) the hotter primary star in the eclipsing binary, Aa, has a temperature ratio with the secondary star, Ab, of $T_{\rm eff, Ab}/T_{\rm eff, Aa} = 0.975 \pm 0.007$ based on the ratio of eclipse depths. Finally, we note that the large amplitudes of the ellipsoidal variations in the light curve (of $\sim$4 per cent full amplitude) imply that one or more of the stars \reff{must have evolved to a significantly larger radius than the zero-age main sequence value for its mass.} 
In that case, in order to nudge the solutions in the right direction, we assume that Ab is the slightly more massive and evolved star of the pair based on previous results \citep{Liakos_2020}, with (4) $M_{\rm Aa}/M_{\rm Ab} \lesssim 0.95$ and (5) $R_{\rm Aa}/R_{\rm Ab} \lesssim 0.95$.  The details of these latter two constraints are unimportant as long as the best-fit answers for the masses and radii are well away from these constraint boundaries.

The other constraints are (1) we take the Gaia distance of $1127 \pm 20$\,pc, and use it as a Gaussian prior; (2) we use the MESA\footnote{Modules for Experiment in Stellar Astrophysics.} Isochrones and Stellar Tracks (MIST) stellar evolution tracks for an assumed solar composition (\citealt{paxton11}; \citealt{paxton15}; \citealt{paxton19}; \citealt{dotter16}; \citealt{choi16}) to compute the stellar radii and $T_{\rm eff}$ values given the mass and the age of the system; and (3) we utilize stellar atmosphere models from \citet{castelli03}.  

The fitting is done via an MCMC code specifically constructed for this problem as described in \citet{Rappaport_2022}.  There are 5.5 fitted parameters which are: $M_{\rm Aa}$, $M_{\rm Ab}$, $M_{\rm B}$, the system age, interstellar extinction ($A_V$), and a consistency check on the distance.

\begin{table}
\centering
\caption{KIC 4851217 parameters determined from the SED fit only.}
\begin{tabular}{lcc}
\hline
Parameter & Value & Uncertainty  \\
\hline
$M_{\rm Aa}$  [M$_\odot$]  & 1.89 &  0.13  \\
$R_{\rm Aa}$ [R$_\odot$] & 2.25 & 0.28  \\
$T_{\rm eff, Aa}$  [K] & 8025 &  300  \\
$M_{\rm Ab}$ [M$_\odot$] & 2.12 &  0.08  \\
$R_{\rm Ab}$ [R$_\odot$]  & 3.17 &  0.27  \\
$T_{\rm eff, Ab}$ [K]  & 7800 &  300  \\
$M_{\rm B}$  [M$_\odot$] & 0.69 &  0.06  \\
$R_{\rm B}$ [R$_\odot$] & 0.67 &  0.05  \\
$T_{\rm eff, B}$  [K] & 4750 &  300  \\
system age [Myr] & 865 &  120  \\
$A_V$    & 0.20 &  0.10  \\
distance [pc] & 1128 &  19  \\
\hline
\end{tabular}

\label{tbl:parms_sed} 
\end{table}

The results of the SED fit are shown in Fig.~\ref{fig:sed} and in Table \ref{tbl:parms_sed}.  The values in Table \ref{tbl:parms_sed} are the median values of the posterior distributions, while the error bars are the rms scatter of the posterior distributions around the mean.  The fit to the SED points in Fig.~\ref{fig:sed} shows the 25 measured flux values at wavelengths between 0.35 $\mu$m and 11.6 $\mu$m, as well as the modelled flux for each of the three stars individually (blue, red, and green curves) and the total flux (black curve). In the right panel of Fig.~\ref{fig:sed} we show where the stars with the inferred properties would lie in the $R-T_{\rm eff}$ plane.  As can clearly be seen, the secondary star (Ab) is the more massive and evolved, and is in the evolutionary `loop' corresponding to contraction of the hydrogen-depleted core after leaving the main sequence.  While the primary star (Aa) has definitely evolved off the ZAMS, it has not yet arrived at the evolutionary `loop' in the $R-T_{\rm eff}$ plane. It is difficult to say much about the tertiary star except that our results are consistent with it contributing $\lesssim 1$ per cent of the system light, and having a mass $\lesssim 1$\,M$_\odot$.  

Consulting Table \ref{tbl:parms_sed}, we see that the masses are determined to $\sim$6 per cent accuracy, about 10 per cent in the radii, and $\sim$300 K for $T_{\rm eff}$.  The distance is nicely consistent with the \textit{Gaia} result\footnote{Modelling the SED entails deriving the intrinsic properties and scaling by the distance to match the observed fluxes.}.  The system age of $\sim$800 Myr is, not surprisingly, what is expected for 2 M$_\odot$ stars that are just leaving the MS.  It is gratifying to see that our final, much more accurate stellar parameter set for the inner EB, found from all the available data, agree to within the 1$\sigma$ error bars in Table \ref{tbl:parms_sed} (see Table \ref{tab: appendix- full comparison}).

Overall, the SED fit, with just a few reasonable assumptions and constraints, yields remarkably useful first estimates of the stellar parameters of the system.  

\section{Spectroscopic Analysis }\label{sec:Spectroscopic-Analysis}

\subsection{Radial velocities}\label{sec:RV-analysis}
The 41 HERMES spectra were reduced and \'echelle orders were merged with the standard HERMES pipeline. The 31 ISIS spectra were reduced using \textsc{pamela} and \textsc{molly} \citep{Marsh_2014, Marsh_2019}. Normalization was carried out using the method of \cite{Xu_2019}. Template spectra were synthesised using \textsc{ispec} \citep{ispec} for the components of the inner EB; the atmospheric parameters of these templates were determined from a preliminary analysis of the ISIS spectra and were in agreement with those derived from the SED fitting in Section \ref{sec: SED fitting}. Each set of templates was synthesised according to the resolution of either instrument, which in velocity space satisfies 1.56 km~s$^{-1}$  for the HERMES spectra and 7.46 km~s$^{-1}$  for the blue arm of the ISIS spectra. 

The projected rotational velocity \vsini\ of each component was estimated by cross-correlating the observations against both the primary and secondary templates for a range of \vsini\ values between 10 and 150\kms\ in steps of 10\kms, and then interpolating between the peaks of the cross correlation functions (CCFs). Observations less than 0.125 times the orbital phase away from an eclipse were omitted from the calculation to avoid issues associated with blending between the spectral lines of the components near phases of conjunction. 
For the HERMES observations, this approach yielded $\vsini_{\rm Aa} = 43.9 \pm 0.5$\kms\ and $\vsini_{\rm Ab} = 61.6 \pm 0.3$\kms, which are in excellent agreement with our adopted values derived from the atmospheric analysis of the disentangled HERMES spectra (see Section \ref{sec: atmospheric parameters}). For the ISIS observations, this approach yielded $v\sin i_{\rm Aa} = 31.7 \pm 0.5 $\kms\ and $v\sin i_{\rm Ab} = 55.7 \pm 0.7$\kms, where the discrepancies are likely due to the lower velocity resolution.  In any case, these are the values that maximise the peaks of the CCFs for each set of templates so we broadened them to these values \citep{gray_2005, pyastronomy}.

RVs were measured using our implementation of \textsc{todcor} \citep{Zucker&Mazeh_1994} using the region between 4400\,--\,4800\,\AA\ on the HERMES spectra, and between 4380\,--\,4580\,\AA\ on the blue arm of the ISIS observations. These regions were chosen because of the presence of many well-resolved lines, which makes them reliable indicators of RV, and the absence of broad lines, i.e., the Balmer series, compared to other regions. We excluded RVs derived from observations taken near phases of conjunction because these RVs contain little or no information about the velocity amplitudes of the components and are prone to yielding anomalous RVs due to severe blending of the spectral lines.

Blending between the main correlation peaks and sidelobes introduces systematic shifts in RVs derived from double-lined spectra at any phase, and the dependence on phase is expected to be complex \citep{Latham_1996}. To mitigate this effect, we performed an initial fit to the extracted RVs using the SciPy package \textsc{curvefit} \citep{2020SciPy-NMeth} and then synthesized the observed orbit by adding synthetic spectra weighted by the relative light contributions of each component, as derived from the \textsc{todcor} light ratio, after applying Doppler shifts according to the initial fit. We used the exact same procedure to extract the known RVs from the simulated orbit and calculated their discrepancies which were applied to our actual RVs as corrections. This method has been utilized in, e.g., \cite{Latham_1996, Torres_1997, Torres_2000, Torres&Ribas_2002, Southworth_Clausen_2007}. The process was carried out separately for the HERMES and ISIS observations. 

\begin{figure}
    \centering
    \includegraphics[width = \columnwidth]{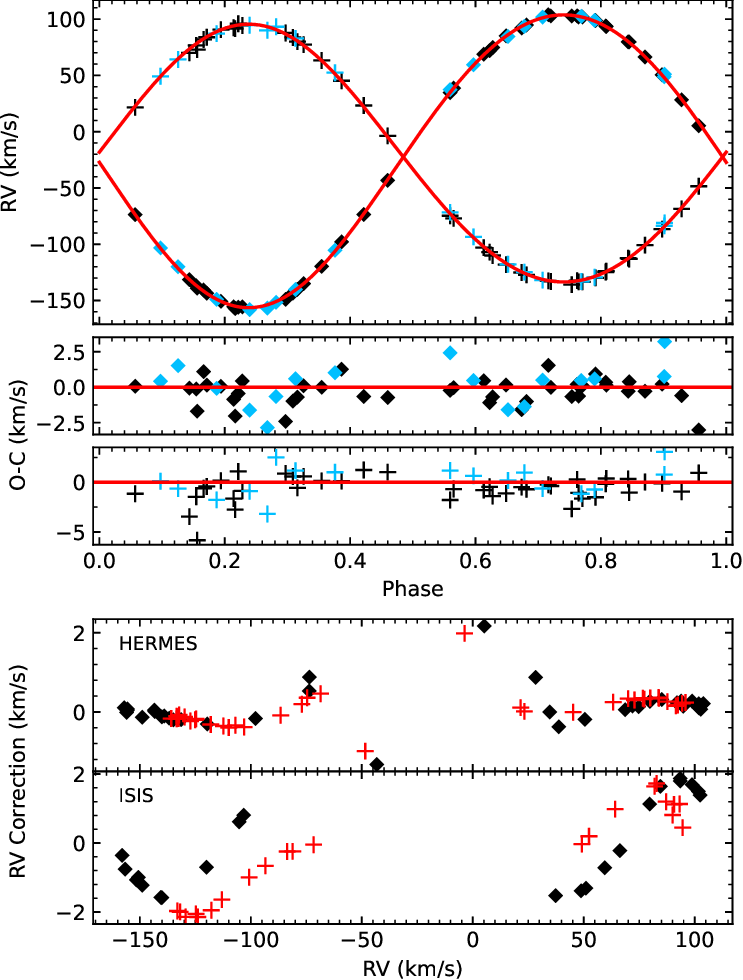}
    \caption{\label{fig: RV_fit} Orbital fit to the corrected RVs. Diamonds/crosses indicate the primary/secondary RVs and those corresponding to the HERMES/ISIS instrument are black/blue. Corrections that were applied to the primary (black diamonds) and secondary (red crosses) RVs are shown in the lower panels for both instruments.}
\end{figure}

We modelled the corrected RVs from both instruments jointly. The result is shown in Fig.\,\ref{fig: RV_fit} in the top panel and the corresponding orbital parameters for KIC\,4851217 are given in Table \ref{tab:orbital_params}. We attempted to fit for the centre-of-mass (CM) acceleration due to the third body but the results were not significant. This suggests that the third body's influence is negligible over the time-span of the spectroscopic observations. This is expected; the $\sim0.003$-d amplitude of the LTTE estimated in Section \ref{sec: preliminary ETV analysis} translates to a $\sim2.5$ \kms\ velocity amplitude of the EB CM, while our RVs only span $\sim15$ per cent of the outer orbital period. 

Our final values for the light ratio were obtained by repeating the RV extraction using templates corresponding to our adopted atmospheric parameters derived in Section \ref{sec: atmospheric parameters}. These values correspond to $\ell_{\rm Ab}/\ell_{\rm Aa} = 1.83 \pm 0.02$ and $\ell_{\rm Ab}/\ell_{\rm Aa} = 1.95 \pm 0.12$ for the HERMES and ISIS spectra, respectively.  Using the updated templates had a negligible impact on the resulting orbital parameters, as expected since RVs depend on the relative locations of spectral lines while $\ell_{\rm Ab}/\ell_{\rm Aa}$ is more sensitive to their shapes and depths. 

\begin{table}\caption{\label{tab:orbital_params}Orbital parameters}
   \centering
    \begin{tabular}{l r@{\,$\pm$\,}l r@{\,$\pm$\,}l}
    \hline 
     & \multicolumn{2}{c}{Primary} & \multicolumn{2}{c}{Secondary}\\
     \hline
     $K$ (\kms)     & \multicolumn{2}{c}{$130.11 \pm 0.13$}  & \multicolumn{2}{c}{$114.59 \pm 0.23$}   \\
     $\gamma$ (\kms)& \multicolumn{4}{c}{$-22.51 \pm 0.11$}   \\
     $e$            & \multicolumn{4}{c}{$0.032 \pm 0.001$} \\
     $\omega$ ($^\circ$)& \multicolumn{4}{c}{$170.8 \pm 2.0$} \\
     $T_{\rm per}$ (BJD$_{\rm TDB}$)      & \multicolumn{4}{c}{$2456016.649 \pm 0.013$}\\
     rms (\kms)      & \multicolumn{2}{c}{1.11}&\multicolumn{2}{c}{1.56}\\       
     \hline
\end{tabular} \end{table}

Fig.~\ref{fig: RV_fit} shows the corrections that were applied to the RVs as a function of RV in the bottom two panels. Applying the corrections to the HERMES RVs led to a 0.08 and 0.2 per cent increase in the velocity amplitude of the primary and secondary, respectively, where the latter translates to a 0.6 per cent increase in the mass, which is significant considering that we aim to achieve precisions of $\sim$0.5 per cent. The corresponding values for the ISIS spectra are a 0.5 and 1.2 percent increase in the velocity amplitudes, translating to a 1.5 and 3.5 per cent increase in the mass of the primary and secondary stars, respectively, which is very significant. This highlights the importance of the RV corrections for the reliable determination of the masses of the stars.

We did not include RVs published by other authors because firstly, we aim to contribute an independent analyses and secondly, the quality of previously published orbits do not suggest their addition would aid in achieving the desired precision here. 
\subsection{Spectral disentangling}\label{sec: spectral disentangling}
The spectral disentangling technique allows for the spectra of the individual components to be separated out from the composite binary spectra whilst simultaneously optimizing the orbital parameters of the system. We use the implementation \textsc{fd3} by \citet{fd32}, which works in the Fourier domain, to disentangle the HERMES observations in three spectral regions: (1) 4700\,--\,5000\,\AA, which contains the H$\beta$ line, (2) 5050\,--\,5300\AA, which contains the Mg b triplet associated with transitions in neutral magnesium, (3) 6480\,--\,6640\AA, which contains the H$\alpha$ line. An initial run was performed with values for the input parameters taken from Table \ref{tab:orbital_params} and allowed to vary to within three times their error bar to explore the possibility that \textsc{fd3} might predict different orbital parameters. In all three cases, 100 optimization runs each consisting of 1000 iterations did not converge to a solution with a smaller $\chi^2$ than at the starting point. We therefore separated the spectra with the orbital parameters fixed to the values in Table \ref{tab:orbital_params} for subsequent runs. We ignored the presence of the third body since it is not detected spectroscopically as demonstrated in Section \ref{sec:RV-analysis}. 

While we know that the secondary star is almost twice as bright as the primary star from the \textsc{todcor} analysis, the absence of observations taken during eclipse means that it is favourable to assume equal light contributions using \textsc{fd3} and then rescale the results according to the actual light contributions of the stars, as explained in \citet{FD3}. This renormalization of the resulting disentangled spectra heavily relies on an accurate value for the light ratio of the system. Due to the sensitivity of the \textsc{todcor} light ratio on the choice of stellar parameters of the templates \citep[as discussed in Section \ref{sec:RV-analysis} and in ][]{Jennings_KIC985}, we utilize a method which we find to be largely insensitive to relatively small differences in the stellar parameters of the templates to derive an independent value for $\ell_{\rm Ab}/\ell_{\rm Aa}$. 

Here, we estimate $\ell_{\rm Ab}/\ell_{\rm Aa}$ by minimizing the sum of the square residuals between the observed binary spectra and synthetic composite spectra, where the latter were calculated by adding Doppler-shifted synthetic spectra generated by \textsc{ispec} weighted by light fractions corresponding to trial values for $\ell_{\rm Ab}/\ell_{\rm Aa}$ \citep[e.g.,][]{Jennings_KIC985}. For the synthetic spectra, we used the \Teff\ values given in Table \ref{tbl:parms_sed} derived from the analysis of the SED, Doppler shifts corresponding to the RVs derived in Section \ref{tab:orbital_params}, and searched in a grid of 12 values for $\ell_{\rm Ab}/\ell_{\rm Aa}$ between 1.1 and 2.5. To ensure optimal normalization of the raw observations, we decided to normalize them at each iteration of the fit by dividing by a second-order polynomial whose coefficients were set as free parameters. The best estimate for the light ratio was then taken as the minimum of a polynomial fit to the sum of the squared residuals against $\ell_{\rm Ab}/\ell_{\rm Aa}$.

\begin{figure}
    \centering
    \includegraphics[width = \columnwidth]{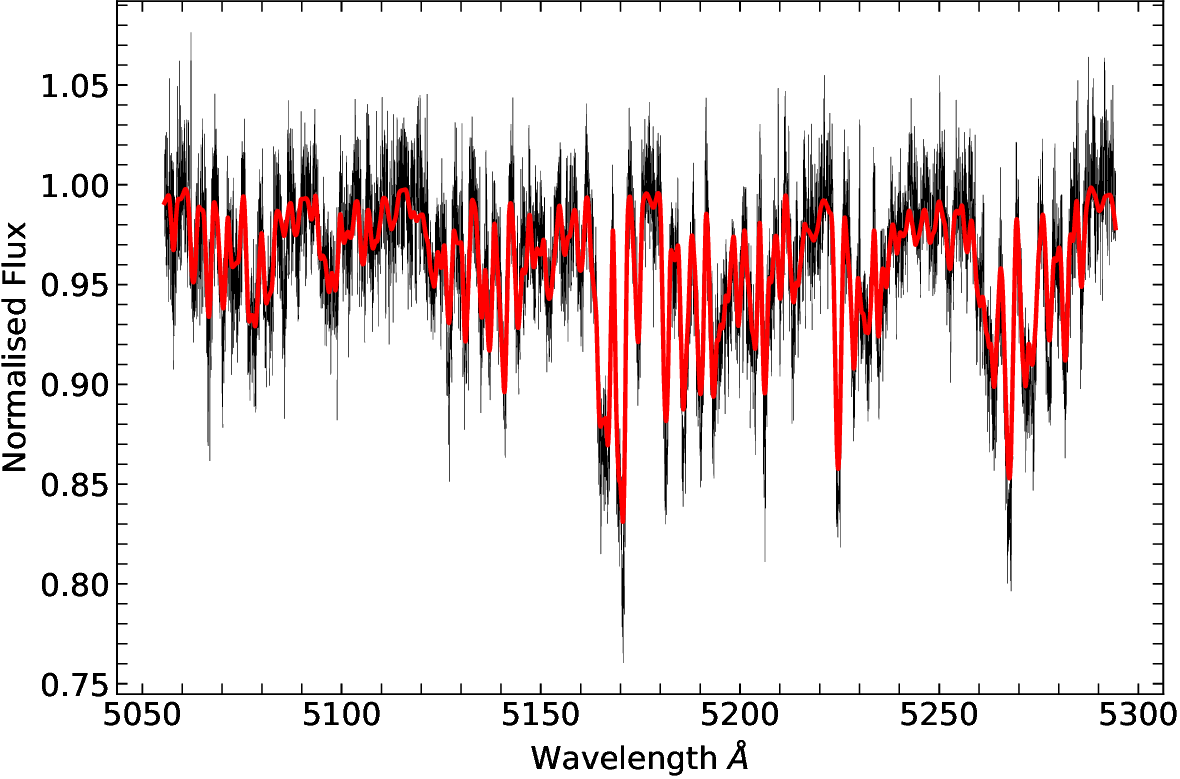}
    \caption{The observation at phase 0.762 optimally normalized as described in the text (black) for the spectral region between $5050-5300$\,\AA. The best fitting composite synthetic spectrum is overplotted in red, and was calculated using the best estimate for the light ratio from this region.}
    \label{fig: LR det MGIII}
\end{figure}

This process was carried out on a spectral segment within the region used to extract the RVs (4400-4600\,\AA), the Mg~b triplet (5050--5300\,\AA) region, as well as regions between 5300-5500\,\AA\ and 5500-5700\,\AA\ because these spectral regions showed a relatively large number of well-resolved lines compared to other spectral regions. We then used the five observations closest to positions of quadrature for each spectral segment and the minimization was carried out using the SciPy Python package \textsc{minimize} \citep{2020SciPy-NMeth}. The results were averaged over the observations for each spectral region and are given in Table \ref{tab: light ratios}. The optimally normalized observation at phase 0.762 is plotted in Fig.\,\ref{fig: LR det MGIII} with the best-fitting composite synthetic spectrum overplotted for the region containing the Mg~b triplet.
\begin{table}
\caption{\label{tab: light ratios} Results for the light ratio determination.}
   \centering
   \begin{tabular}{l r@{\,$\pm$\,}l r@{\,$\pm$\,}l} 
   \hline
   Wavelength range [\AA] & \multicolumn{2}{c}{$\ell_{\rm Ab}/\ell_{\rm Aa}$} \\
   \hline
   4400--4800 & 2.06 & 0.07\\
   5050--5300 & 1.96 & 0.03\\
   5300--5500 & 1.88 & 0.06\\
   5500--5700 & 1.99 & 0.05\\
   Adopted & 1.96 & 0.11 \\
   \hline
   \end{tabular}
\end{table}

The average of the light ratios estimated from each spectral segment satisfied $\ell_{\rm Ab}/\ell_{\rm Aa} = 1.96 \pm 0.02$. This value is consistent with the \textsc{todcor} light ratio derived from the ISIS spectra but inconsistent with that derived from the HERMES spectra. Thus, we inflate the error bar to be consistent with the weighted average of those two values and present this as our adopted value for $\ell_{\rm Ab}/\ell_{\rm Aa}$ in Table \ref{tab: light ratios}. We then use this value to normalize all observations in the Mg~b triplet, H$\alpha$, and H$\beta$ regions by optimizing the coefficients of a second-order polynomial against synthetic templates, as described above. 

\begin{figure}
    \centering
    \includegraphics[width = \columnwidth]{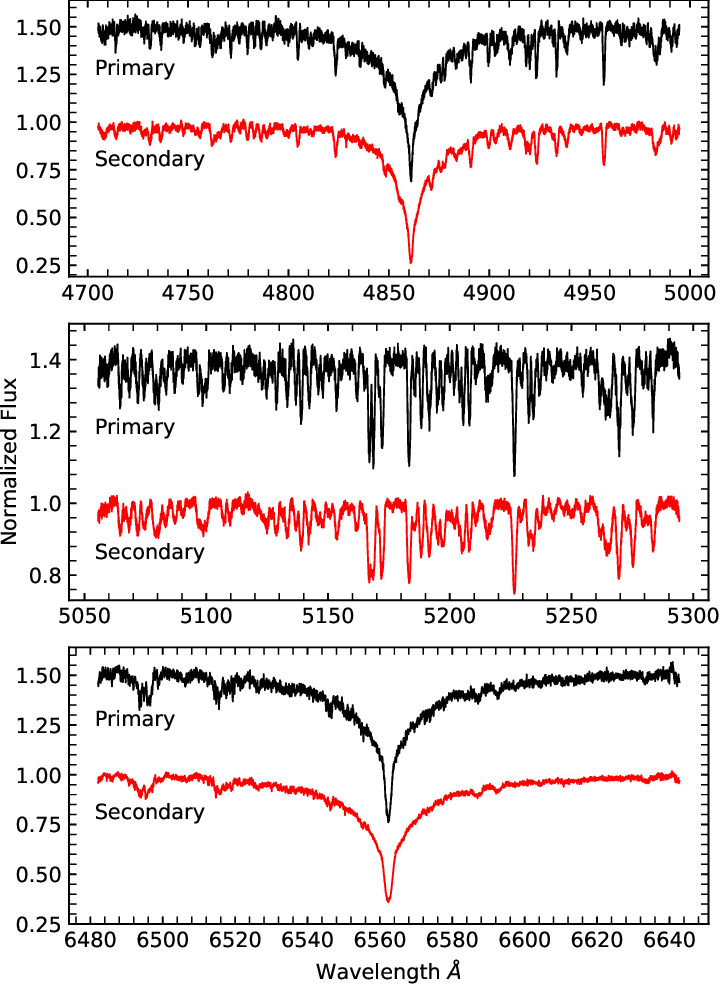}
    \caption{Disentangled component spectra for the H$\beta$ (top), Mg b triplet (middle), and H$\alpha$ (bottom) regions.}
    \label{fig:disentangled spectra}
\end{figure}

We performed disentangling as described at the start of this section on the normalized spectra and rescaled the results, as described in \citet{FD3}, using our adopted value for $\ell_{\rm Ab}/\ell_{\rm Aa}$. Our disentangled component spectra for KIC\,4851217 are shown in Fig.\ref{fig:disentangled spectra} for the Mg~b triplet, H$\beta$ and H$\alpha$ regions.
One of the benefits of spectral disentangling is the increased signal-to-noise (S/N) and this is obvious when comparing the middle panel of Fig.\,\ref{fig:disentangled spectra} to Fig.\,\ref{fig: LR det MGIII}.

The normalization of the observed spectra and the light ratio used to re-scale the disentangled spectra are possible sources of uncertainty that may propagate into the atmospheric analysis. Thus, we computed two more sets of disentangled spectra. For the first set, we normalized the observed spectra by optimizing the coefficients of a polynomial against synthetic templates with differing atmospheric parameters, i.e., $\Delta \Teff = 150$\,K and $\Delta \rm[M/H] = 0.1 dex$. We did not adjust $\log(g)$ because this is reliably determined dynamically and we do not attempt to derive its value from the atmospheric analysis. For the second set of additional disentangled spectra, we varied the value of $\ell_{\rm Ab}/\ell_{\rm Aa}$ used to rescale the spectra within the error bar reported in Table \ref{tab: light ratios}. 

Thus, in this section we have derived an independent value for the light ratio of the EB which we find to be more reliable than the values derived using \textsc{todcor}. We then used this light ratio to normalize our observed binary spectra against synthetic spectra, as well as calculate our primary set of disentangled spectra for each component. We also carried out the normalization and disentangling with adjusted values for the atmospheric parameters of the templates and light ratio, yielding two extra sets of disentangled spectra. These extra sets of disentangled spectra are used to estimate systematic uncertainties in the atmospheric parameters which we derive in the next section.

\subsection{Atmospheric parameters}\label{sec: atmospheric parameters}

\begin{table*}
\caption{\label{tab: atmospheric parameters} Atmospheric parameters for the components of KIC 4851217. See text for descriptions related to calculations of the adopted values in column four.}
   \centering
    \begin{tabular}{l r@{\,$\pm$\,}l r@{\,$\pm$\,}l r@{\,$\pm$\,}l r@{\,$\pm$\,}l} 
    \hline 
    Parameter & \multicolumn{2}{c}{H$\beta$} & \multicolumn{2}{c}{H$\alpha$} & \multicolumn{2}{c}{Mg b triplet} &\multicolumn{2}{c}{Adopted} \\
    Wavelength range (\AA) & \multicolumn{2}{c}{4800--5000} & \multicolumn{2}{c}{6480--6640} & \multicolumn{2}{c}{5050--5300} & \multicolumn{2}{c}{(see text) }   \\
    \hline
    Primary \\
     \Teff [K] & 7810 & 100 & 7880 & 140 & 7890 & 330 & 7830 & 80    \\
     $[\rm M/H]$[dex] & \multicolumn{2}{c}{0.0}  & \multicolumn{2}{c}{0.0} & 0.06 & 0.22 & 0.02 & 0.11  \\
     $v_{\rm mic}$[\kms] & 3.2 & 0.4 & 5.0 & 3.2 & 2.9 & 0.5 & 3.1 & 0.3\\
    $v \sin i$ [\kms] & \multicolumn{2}{c}{43.6}  & \multicolumn{2}{c}{43.6}  &  43.6 & 4.6 & 43.6 & 4.6  \\
    \hline
    Secondary     \\
    \Teff [K]   &  7720 & 90 & 7680 & 120 & 7860 & 390 & 7700 & 70 \\
     $[\rm M/H]$[dex]  &  \multicolumn{2}{c}{0.0}  & \multicolumn{2}{c}{0.0}& -0.03 & 0.27  & -0.10 & 0.15  \\
    $v_{\rm mic}$[\kms] & 3.3 & 0.3 & 4.0 & 1.2 & 3.1 & 0.5  &3.3&0.3  \\
    $v \sin i$ [\kms] & \multicolumn{2}{c}{61.6}    & \multicolumn{2}{c}{61.6}   & 61.6 & 7.0 & 61.6 & 7.0  \\
    
    \hline
\end{tabular}
\end{table*}
\begin{figure*}
    \centering
    \includegraphics[width = 0.9\textwidth]{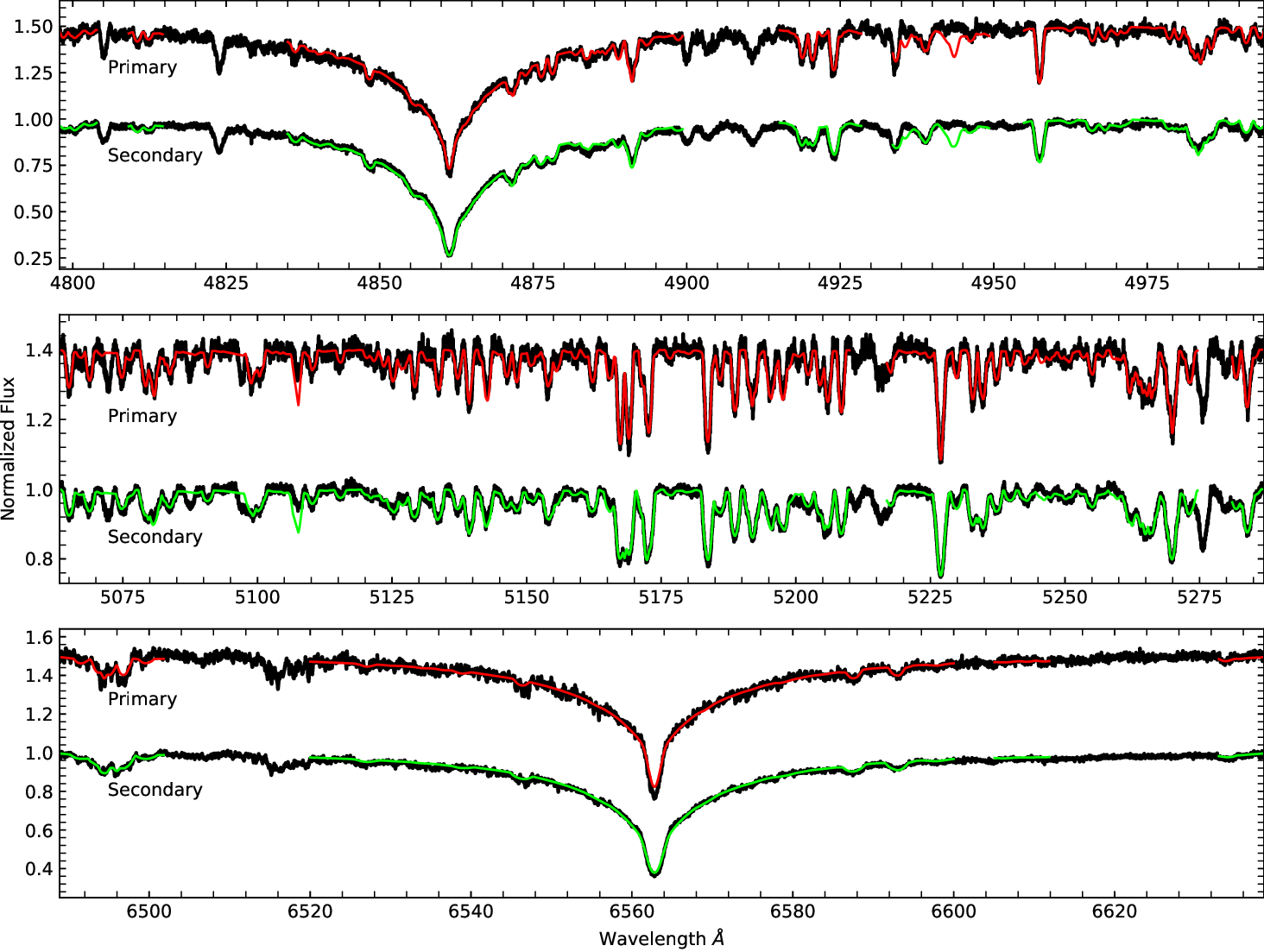}
    \caption{Synthetic spectral fits to the H$\beta$ (top), Mg b triplet (middle), and H$\alpha$ (bottom) regions. Observed data are shown in black and the best-fitting synthetic spectra are shown in red for the primary and green for the secondary. Gaps in the synthetic spectra result from the fact that the synthesis was only carried out for spectral regions containing the pre-selected line-masks (see text). The primary spectrum is offset by +0.5 for presentation purposes.}
    \label{fig: spectral fits}
\end{figure*}
We used the tools included in \textsc{ispec} to estimate the S/N of the disentangled component spectra. This led to an average S/N of $\sim 82$ and $\sim156$ for the primary and secondary, respectively. Estimates for the errors on the disentangled fluxes then follow by dividing them by the S/N. 

Atmospheric parameters were determined via synthetic spectral fits using the \textsc{ispec} framework. By default, we opted to use the MARCS models \citep{Gustaffson_2008} because these are adequate for dwarf stars \citep{ispec} but we also considered the ATLAS9 models \citep{Kurucz_2005, Kirby_2011, Meszaros_2012} to explore wider ranges in \Teff\ and estimate systematic uncertainties. We combined the MARCS models with the solar abundances from \citet{Grevesse_2007} to conform with the choice by \citet{ispec}, where they report better precisions in the resulting parameters, and used the \emph{Gaia} ESO Survey atomic line list. Our synthesis is only performed on the pre-selected line-masks provided by \textsc{ispec}, which are based on the \emph{Gaia} ESO survey atomic linelist, and these only extend to 4800\,\AA\ in the blue so part of our H$\beta$ region was not included in the fits. 

In all cases, $\log (g)$ is fixed to the dynamical values derived from the combined analysis of the light and RV curves because it is more precise than the spectroscopic value. The macroturbulent velocity was fixed to zero for two reasons: (1) the convective envelope is relatively deep in early-F and late-A stars, so we expect granulation signatures to be relatively weak, (2) for surface velocity fields to be directly detectable requires projected rotational velocities of $\lesssim13$ \kms\ for stars with $\Teff \sim7500$ K \citep{Landstreet_2009}; our estimates for \vsini\ are $\sim$3 and $\sim$5 times this threshold for the primary and secondary, respectively (see Table \ref{tab: atmospheric parameters}). 

First, we fitted for the spectral region containing the Mg b triplet (5050\,--\,5300\,\AA\,) mainly to determine \vsini\ because this region is free of strong lines, i.e, the Balmer series where the line profiles are heavily influenced by Stark broadening mechanisms. We then constrained \Teff\ by fitting for the Balmer regions with \vsini\ fixed. We expect \Teff\ to be better determined from Balmer lines because their profiles are highly temperature sensitive and are insensitive to $\log(g)$ for stars with $\Teff \lesssim 8000$\,K \citep{Smalley_2005, Bowman_2021}; we expect the dynamical $\log(g)$ to be accurate but any uncertainties in fixing its value are thus minimized. We also fixed [M/H] to zero in the Balmer regions because our solution from the Mg b triplet region was consistent with solar (see Table \ref{tab: atmospheric parameters}) and, in any case, Balmer lines are less sensitive to the influence of metallicity.   

We repeated the fits at each spectral region using the Kurucz, Castelli, and APOGEE ATLAS9 models to investigate the systematic uncertainty associated with our preferred choice of atmospheric model. For \Teff, the MARCS, Castelli and APOGEE models gave consistent results but the Kurucz models predicted larger \Teff\ values by around $\sim 150$K for both components and in both the H$\alpha$ and H$\beta$ regions. We also carried out the full process on the two extra sets of disentangled spectra that were calculated in the previous section to estimate the uncertainty associated with our estimation for $\ell_{\rm Ab}/\ell_{\rm Aa}$ as well as the normalization of the raw observations (see Section \ref{sec: spectral disentangling}). In each investigation, we took the standard deviations of the results as the estimates for the associated systematic uncertainties. Final error bars were then calculated by adding these values in quadrature to the formal error bars from the least squares fits. Results for the fitted parameters from the fits to each spectral region are given in the first three columns of Table \ref{tab: atmospheric parameters} (fixed parameters are given without an error bar) and Fig.\,\ref{fig: spectral fits} displays the best-fitting synthetic spectra against the observations. 

Values for \Teff\ are poorly constrained in the Mg b triplet region. This could be explained by correlations between \Teff\ and [M/H] being complicated due to, e.g., line blanketing effects, and this is compounded by the fact both those parameters are correlated with the microturbulent velocity $v_{\rm mic}$. However, these effects are less pronounced for Balmer lines so our adopted values for \Teff\ (fourth column of Table \ref{tab: atmospheric parameters}) are the weighted averages of the results from the H$\alpha$ and H$\beta$ regions only. This decision is corroborated by the insensitivity of Balmer lines to $\log(g)$ for stars with $\Teff\lesssim8000$\,K. We adopted the weighted average of the results from all three regions for the final value of $v_{\rm mic}$ and finally note that our values for \vsini\ are consistent with synchronous rotation.

The correlations between \Teff, [M/H] and $v_{\rm mic}$ may be the cause of the large uncertainties in the values for [M/H] derived from the Mg b triplet region. In an attempt to better constrain the values for [M/H], we repeated the fits in the Mg b triplet region except we additionally fixed \Teff\ and $v_{\rm mic}$ to our adopted values. Here we obtain [M/H] $=0.02\pm0.06$ for the primary and [M/H] $=-0.10\pm0.05$ for the secondary. As expected, these efforts have reduced the uncertainties on [M/H] significantly but the values do not satisfy the assumption of coevality. Additionally, we also noticed our adopted values for $v_{\rm mic}$ are larger than the empirical values calculated using \textsc{ispec}'s built-in relation constructed based on \emph{Gaia} FGK benchmark stars \citep{Jofre_2014}. These empirical values for $v_{\rm mic}$ correspond to $2.5$\,km\,s$^{-1}$ and $2.4$\,km\,s$^{-1}$, and result in an increase in [M/H] by 0.10\,dex and 0.14\,dex for the primary and secondary, respectively. We added these differences in quadrature to the uncertainties on the updated values for [M/H] reported above and present them as our adopted values in the fourth column of Table \ref{tab: atmospheric parameters}. These efforts have reduced the uncertainties on [M/H] by about a factor of two compared to the previously derived values reported in the third column of Table \ref{tab: atmospheric parameters}, but yield the same conclusions that both components are of solar abundance to within the uncertainties.

In summary, we have derived atmospheric parameters for the components of KIC\,4851217 by performing synthetic spectral fits in three spectral regions. Our uncertainties on the parameters take into account those associated with the normalization of the observations, choice of atmospheric models, and light ratio used to rescale the disentangled spectra. The uncertainties on our adopted values for [M/H] take into account the observed strong anti-correlation with $v_{\rm mic}$. Adopted values are presented in the fourth column of Table\,\ref{tab: atmospheric parameters}.

\section{Light curve analysis with the Wilson-Devinney Code}\label{sec:lc:wd}

To obtain the light curve solution, we considered only the \textit{Kepler} SC observations, as these have a much better time resolution than the \textit{Kepler} LC observations and a lower scatter than the TESS data. We first used version 43 of the {\sc jktebop}\footnote{\texttt{http://www.astro.keele.ac.uk/jkt/codes/jktebop.html}} code \citep{Me13aa}, chosen because it is fast, to model each month of data separately. Star Ab is too deformed for this code to give reliable results, so this analysis was only used to determine the orbital phase of each datapoint. The data were then phase-binned them into a total of 352 data points. Orbital phases around the eclipses were sampled every 0.001 in phase, whilst those away from the eclipses had a sampling of 0.01 phases. This process removed the shifts in eclipse times due to the third body (neglecting the extremely small changes over the course of one month), averaged out the pulsation signature, and reduced the number of observations by three orders of magnitude.

We then analysed the phase-binned light curve using the Wilson-Devinney \citep{WilsonDevinney71apj,Wilson79apj} code. This code uses modified Roche geometry to model the shapes of stars, so is applicable to stars that are significantly deformed. We used the 2004 version of the code ({\sc wd2004}) driven using the {\sc jktwd} wrapper \citep{Southworth_2011}. The user guide which accompanies the Wilson-Devinney code \citep{WilsonVanhamme04} includes a description of all input and output quantities.

We quickly arrived at a good solution to the light curve through a process of trying a large number of different modelling options available in {\sc wd2004}. Our default solution was obtained in Mode $=$ 0 with a numerical precision of $N=60$, the mass ratio and \Teff\ values of the stars fixed at the spectroscopic values in Table \ref{tab: atmospheric parameters}, synchronous rotation, gravity darkening exponents of 1.0 for both stars, the simple reflection model, limb darkening implemented according to the logarithmic law with the non-linear coefficients fixed, and using the Cousins $R$ filter as a proxy for the \textit{Kepler} response function. The fitted parameters comprised the potentials, albedos, light contributions and linear limb darkening coefficients of the two stars, plus the orbital inclination, eccentricity, argument of periastron, and third light. 

The best fit to the light curve corresponds to a light ratio between the components of approximately 1.5, which is in significant disagreement with the spectroscopic value. We therefore forced the solution to agree with the spectroscopic light ratio, finding that the solution is almost as good (as expected given the additional imposed constraint). We adopt the latter results corresponding to the fixed, spectroscopic light ratio given that a purely photometric light ratio is less reliable for partially eclipsing systems \citep{Jennings_KIC985}, as well as to ensure internal consistency between analyses. 

\begin{figure}
\includegraphics[width=\columnwidth]{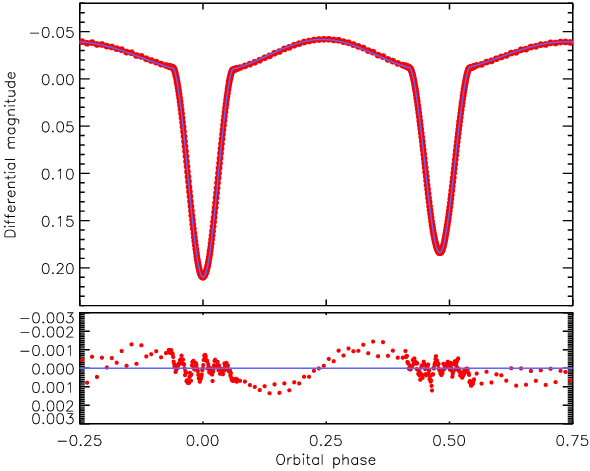} \\
\caption{\label{fig:wd} The best-fitting WD model (blue line) to the \textit{Kepler} SC
phase-binned light curve of KIC 4851217 (red filled circles). The residuals of the fit
are plotted in the lower panel using a greatly enlarged y-axis to bring out the detail.}
\end{figure}

The uncertainties in the fitted parameters are dominated by the uncertainty in the spectroscopic light ratio, model choices and the numerical integration limit, because the Poisson noise in the binned light curve is negligible. We evaluated the uncertainties individually for all relevant sources and added them in quadrature for each fitted parameter. The sources include the spectroscopic light ratio, chosen numerical precision, mass ratio, mode of operation of {\sc wd2004} (0 or 2), rotation rates (varied by 10 per cent), gravity darkening, whether or not to include third light, choice of limb darkening law (logarithmic versus square-root) and choice of filter (Cousins $R$ versus $I$).

\begin{table}
\caption{Summary of the parameters for the {\sc wd2004} solution of the phase-binned light curve of KIC 4851217. Detailed descriptions of the control parameters can be found in the WD code user guide \citep{WilsonVanhamme04}. Uncertainties are only quoted when they have been robustly assessed by comparison of a full set of alternative solutions.}
\begin{tabular}{llcccc} \hline
Parameter                            & {\sc wd2004} name    & Value               \\
\hline
{\it Control and fixed parameters:} \\
{\sc wd2004} operation mode          & {\sc mode}           & 0                   \\
Treatment of reflection              & {\sc mref}           & 1                   \\
Number of reflections                & {\sc nref}           & 1                   \\
LD law                               & {\sc ld}             & 2 (logarithmic)     \\
Numerical grid size (normal)         & {\sc n1, n2}         & 60                  \\
Numerical grid size (coarse)         & {\sc n1l, n2l}       & 60                  \\[3pt]
{\it Fixed parameters:} \\
Mass ratio                           & {\sc rm}             & 1.135               \\
Phase shift                          & {\sc pshift}         & 0.0                 \\
$T_{\rm eff}$ star\,Aa (K)           & {\sc tavh}           & 7834                \\
$T_{\rm eff}$ star\,Ab (K)           & {\sc tavh}           & 7701                \\
Gravity darkening exponents          & {\sc gr1, gr2}       & 1.0                 \\
Rotation rates                       & {\sc f1, f2}         & 1.0, 1.0            \\
Logarithmic LD coefficients          & {\sc y1a, y2a}       & 0.618, 0.628        \\ [3pt]
{\it Fitted parameters:} \\
Star\,Aa potential                   & {\sc phsv}           & $6.78   \pm 0.12  $ \\
Star\,Ab potential                   & {\sc phsv}           & $5.537  \pm 0.061 $ \\
Orbital inclination (\degr)          & {\sc xincl}          & $76.86  \pm 0.12  $ \\
Orbital eccentricity                 & {\sc e}              & $0.0324 \pm 0.0049$ \\
Argument of periastron (\degr)       & {\sc perr0}          & $161    \pm 19    $ \\
Bolometric albedo of star\,Aa        & {\sc alb1}           & $1.4    \pm 0.5   $ \\
Bolometric albedo of star\,Ab        & {\sc alb2}           & $1.1    \pm 0.3   $ \\
Star\,Aa light contribution          & {\sc hlum}           & $4.34   \pm 0.17  $ \\
Star\,Ab light contribution          & {\sc clum}           & $8.52   \pm 0.17  $ \\
Star\,Aa linear LD coefficien        & {\sc x1a}            & $0.640  \pm 0.046 $ \\
Star\,Ab linear LD coefficien        & {\sc x2a}            & $0.734  \pm 0.032 $ \\
Fractional radius of star\,Aa        &                      & $0.1790 \pm 0.0024$ \\
Fractional radius of star\,Ab        &                      & $0.2509 \pm 0.0032$ \\[3pt]
\hline
\label{tab:WD}
\end{tabular} \end{table}

The best-fitting parameters and uncertainties are given in Table\,\ref{tab:WD}. Third light is negligible, which places an upper limit on the brightness of the third component. Of greatest importance is that we have managed to measure the volume-equivalent fractional radii to precisions of approximately 1.5 per cent. Our results differ significantly from those of \citet{Helminiak_2019}, who relied on the {\sc jktebop} code in their work. The eccentricity and argument of periastron also agree well with the spectroscopic values in Section \ref{sec:RV-analysis}.

A plot of the solution is shown in Fig.\,\ref{fig:wd} where significant structure can be seen in the residuals. The short-period wiggles in the residuals during eclipse are likely due to spatial resolution of the pulsations plus possible commensurabilities between the orbital period and pulsation periods. The cause of the slower variation seen outside eclipse is unclear but may be related to imperfect treatment of the mutual irradiations of the stars, residual pulsation effects, or Doppler beaming \citep{Zucker++07apj}. However, we calculated an estimate for the amplitude of the Doppler beaming effect for this system of 0.38 ppt, which is well below the residuals in Fig.\,\ref{fig:wd}. It is interesting that a similar variation was seen in the TESS light curve of $\zeta$ Phe \citep{Me20obs} but with the opposite sign versus orbital phase. 

We found that the albedos of the stars must be fitted to obtain the best solution, although their values are sensitive in particular to the passband used. As mentioned above, the specified numerical precision contributed to the uncertainty in the fitted parameters; the significance of this uncertainty is unexpected and merits further exploration, but a detailed analysis is beyond the scope of the current work. Finally, we reran the analysis with a light curve from which the main pulsations had been removed, finding that this had a negligible effect on the results.



\section{Physical properties}

In this section, we undertake a comprehensive and combined analysis of all the various data sets available for this source. \reff{This is carried out using the software {\sc Lightcurvefactory} \citep{Borkovits_2013,Borkovits_2019}.} The result is a unified set of all the system parameters, both stellar and orbital. \reff{All results obtained from this comprehensive analysis are given in Table \ref{tab: syntheticfit_KIC4851217}. We then compare those results to the parameters that we extracted from the individual analysis of each of the datasets in the previous sections (i.e., the ETVs, SED, RVs and light curves), as well as the physical properties of the system that can be derived from those individual subsets of parameters which we present below in Section \ref{sec: physical properties individual}.} This approach to extracting information from various parts of the data, vs.~what can be done by a single global modelling is instructive for cases where the data sets are not so rich. 



\subsection{Physical properties of the EB from the individual analyses}\label{sec: physical properties individual}
\begin{table} 
\caption{\label{tab:absdim} Physical properties of KIC 4851217 derived from the independent analysis of the photometric and spectroscopic data. The units labelled with a `N' are given in
terms of the nominal solar quantities defined in IAU 2015 Resolution B3 \citep{Prsa+16aj}. The synchronous rotational velocity $v_{\rm sync}$ is reported for the period of the system and corresponding radii measurements.}
\setlength{\tabcolsep}{4pt}
\begin{tabular}{l r@{\,$\pm$\,}l r@{\,$\pm$\,}l} \hline
Parameter                     & \multicolumn{2}{c}{Star Aa} & \multicolumn{2}{c}{Star Ab} \\
\hline
Mass ratio                                      & \multicolumn{4}{c}{$1.1354 \pm 0.0025$} \\
Semimajor axis (\Rsunnom)                       & \multicolumn{4}{c}{$12.263 \pm 0.015$}  \\
Mass (\Msunnom)                                 &    1.899 & 0.008   &    2.156 & 0.007   \\
Radius (\Rsunnom)                               &    2.195 & 0.030   &    3.077 & 0.039   \\
Surface gravity ($\log$[cgs])                   &    4.034 & 0.011   &    3.796 & 0.011   \\
$v_{\rm sync}$ (\kms)                           &   45.0   & 0.6     &   63.0   & 0.8     \\
\Teff\ (K)                                      &     7830 & 80      &     7700 & 70      \\
Luminosity $\log(L/\Lsunnom)$                   &    1.214 & 0.018   &    1.477 & 0.018   \\
Absolute bolometric magnitude                   &    1.706 & 0.046   &    1.047 & 0.044   \\
Interstellar extinction $E(B-V)$ (mag)          & \multicolumn{4}{c}{$0.04  \pm 0.02$}    \\
Distance (pc)                                   & \multicolumn{4}{c}{$1115 \pm 17$}       \\
\hline
\end{tabular} \end{table}

Before undertaking the combined analysis of all the various data subsets, we first derive the physical properties of the inner EB of KIC\,4851217 from the spectroscopic and photometric results derived from the individual analyses of those data, which are presented in Tables \ref{tab:orbital_params}, \ref{tab: atmospheric parameters} and \ref{tab:WD}. We used the $K_{\rm Aa}$ and $K_{\rm Ab}$ values from Table \ref{tab:orbital_params}, the orbital period from Section\,\ref{sec: preliminary light curve solutions}, and the fractional radii, orbital inclination and eccentricity from Section\,\ref{sec:lc:wd}. These were fed into the {\sc jktabsdim} code \citep{Me++05aa}, modified to use the IAU system of nominal solar values \citep{Prsa+16aj} plus the NIST 2018 values for the Newtonian gravitational constant and the Stefan-Boltzmann constant. Error bars were propagated via a perturbation analysis. The results are given in Table\,\ref{tab:absdim}.

We determined the distance to the system using optical $BV$ magnitudes from APASS \citep{Henden+12javso}, near-IR $JHK_s$ magnitudes from 2MASS \citep{Cutri+03book} converted to the Johnson system using the transformations from \citet{Carpenter01aj}, and surface brightness relations from \citet{Kervella+04aa}. The interstellar reddening was determined by requiring the optical and near-IR distances to match,
and is consistent with zero: $E(B-V) = 0.02 \pm 0.02$ mag. We found a final distance of $1115 \pm 17$\,pc, which is in good agreement with the distance of $1127 \pm 20$\,pc from the \textit{Gaia} DR3 parallax \citep{Gaia16aa,Gaia21aa}, as well as the value from the SED fit in section \ref{sec: SED fitting}. This distance only agrees with that given in Table \ref{tab: syntheticfit_KIC4851217} to within 2.3~$\sigma$ but note that the colour excess used to obtain that value is larger by 4.2~$\sigma$.  

\subsection{An independent, joint light curve, radial velocity curve and ETV analysis with \textsc{Lightcurvefactory}}\label{sec: combined anlysis}

\reff{For the independent and combined analysis of the RVs derived in Sect.\,\ref{sec:RV-analysis}, \emph{Kepler} and TESS lightcurves outlined in Sect.\ref{sec:observations} and ETVs measured in Sect.\ref{Sect:ETV_preliminary}, we used the software package {\sc Lightcurvefactory} \citep{Borkovits_2013,Borkovits_2019}.} This code is able to {\it simultaneously} handle multi-passband light curves, RVs and ETVs of different orbital configurations of hierarchical few-body systems, from simple binary stars up to sextuple star systems.

Thus, with the use of this software package we analysed KIC~4851217 directly as a hierarchical triple star system. In practice, this means that for each time of the observations, the software calculated the 3D Cartesian coordinates and velocities of all three constituent stars and then synthesized the observable stellar fluxes and RVs of each star accordingly.  Moreover, the mid-eclipse times for the ETV curves were also calculated directly from the relative, sky-projected distances of the stellar disks, without the use of any analytic formulae which are often used for fitting RV and/or ETV curves. {\sc Lightcurvefactory} has a built-in numerical integrator to calculate the stars' positions and velocities directly from the perturbed equations of motion. However, in the current situation, due to the large distance of the low-mass tertiary component, we found that the only detectable departure from pure Keplerian motions of both the inner and outer subsystems may come from the constant-rate apsidal motion of the inner pair, which is dominated by the tidal distortions of the inner binary stars. Therefore, instead of numerically integrating the stellar motions, we calculated the stellar positions only with the use of the usual analytic formulae describing the two (inner and outer) Keplerian motions, and with the assumption that the argument of pericentre of the inner orbit varied linearly in time. 

This additional analysis, using {\sc Lightcurvefactory}, was also independent in the sense that we used (partly) different sets of the analysed data. In the first rounds we used folded, binned and averaged \textit{Kepler} SC data, but in the present situation the entire dataset was binned into 1000 phase-cells (equal in length) and, hence, there was no difference in the sampling between the in-eclipse and out-of-eclipse sections of the light curve.
However, we used the very same ETV and RV data which were analysed earlier in Sects. \ref{sec: preliminary ETV analysis} and ~\ref{sec:RV-analysis}.

For the parameter optimization, and to explore parameter phase space, we used the built-in MCMC solver contained in the software package. We tried different sets of the stellar and orbital parameters to be adjusted. In our final solutions we adjusted the following parameters:
\begin{itemize}
    \item[(i)] Eight plus one parameters related to orbital elements describing the two Keplerian orbits, as follows: $e_1\cos\omega_1$, $e_1\sin\omega_1$, and $i_1$ giving the eccentricity, argument of periastron and the inclination of the inner orbit; furthermore, the parameters of the wide, outer orbit: $P_2$, $e_2\cos\omega_2$, $e_2\sin\omega_2$, $i_2$, and its periastron passage time, $\tau_2$. Moreover, we also adjusted the constant apsidal advance rate of the inner orbit $\Delta\omega_1$.
    \item[(ii)] Three parameters connected to the stellar masses: primary star's mass, $M_\mathrm{Aa}$, the mass ratio of the inner pair, $q_\mathrm{1}$, and, finally the mass function of the outer orbit $f_2(M_\mathrm{B})$.
    \item[(iii)] Four mainly light-curve connected parameters: the duration of the primary eclipse ($\Delta t_\mathrm{pri}$) is an observable which is strongly connected to the sum of the fractional radii of the EB stars; the ratio of the radii and the effective temperatures of the two EB stars ($R_\mathrm{Ab}/R_\mathrm{Aa}$; $T_\mathrm{Ab}/T_\mathrm{Aa}$), and, finally, the passband-dependent extra (contaminated) light: $\ell_{\mathrm{Kepler}}$ \footnote{\reff{We fit for the passband-dependent extra (contaminated) light $\ell_{\rm Kepler}$, i.e., additional light captured within the \emph{Kepler} photometric aperture.}}.
\end{itemize}
Furthermore, nine additional parameters were internally constrained (or derived), as follows:
\begin{itemize}
    \item [(i)] The orbital period of the EB, $P_1$, and the time of an inferior conjunction $\mathcal{T}_1^{\mathrm{inf}}$ of the secondary star (i.e., the mid-time of a primary eclipse) were constrained via the ETV curves \citep[see appendix A of][]{Borkovits_2019}.
    \item[(ii)] Even though in the current system, the light contribution of the distant tertiary is negligible, the code needs the effective temperature, $T_\mathrm{B}$, and the radius, $R_\mathrm{B}$ of the third component. These parameters were calculated internally simply according to the main-sequence mass-luminosity and mass-radius relations of \citet{Tout_1996}.
    \item[(iii)] The systemic radial velocity ($\gamma$) was derived internally at the end of each trial step by minimizing the value of $\chi^2_\mathrm{RV}$.
    \item [(iv)] Finally, note that similar to our previous modelling efforts, we applied a logarithmic limb-darkening law of which the coefficients for each star were interpolated from passband-dependent tables downloaded from the {\sc Phoebe} 1.0 Legacy page\footnote{\url{http://phoebe-project.org/1.0/download}}. These tables are based on the \citet{castelli03} atmospheric models and were originally implemented in former versions of the {\sc Phoebe} software \citep{Prsa_2005}.
\end{itemize}
Finally, the following parameters were kept fixed: the effective temperature of the primary star was set to $T_\mathrm{Aa}=7834$\,K, i.e., to the same value which was used in the {\sc wd2004} model. Moreover, since both EB members are hot, radiative stars, their gravity darkening exponents and bolometric albedos were set to unity and, opposite to the {\sc wd2004} model, all these parameters were fixed.

We also carried out a second type of complex photodynamical modelling with {\sc Lightcurvefactory}, where, besides the above described datasets, we included in the analysis a simultaneous fit of the observed, net SED of the triple system to a model SED.  The model SED is constructed from precomputed \texttt{PARSEC} tables of stellar evolutionary tracks \citep{PARSEC} which are built into {\sc Lightcurvefactory}. In the case of this latter type of analysis, the code calculates the radii, effective temperatures and selected passband magnitudes of each component separately with iteration from the three dimensional grids of [mass; metallicity; age] triplets \citep[see][for a detailed description of the process]{Borkovits_2020}. In this astrophysical model-dependent analysis, naturally, the temperature and stellar radii-related parameters are no longer adjusted or kept fixed, but are interpolated from the \texttt{PARSEC} grids in each trial step. New adjusted parameters are the stellar metallicity [$M/H$] and (logarithmic) age\footnote{These parameters, technically, can be set separately for each star, but in practice, we generally assume coeval stellar evolution and, moreover, identical chemical compositions of all the stars in a given multi-stellar system and, hence, we adjust only one global age and metallicity parameter.}. Moreover, two additional parameters, the interstellar extinction, $E(B-V)$, and the distance to the system are fitted for as well. The $E(B-V)$ is also adjusted in each step, but the distance calculated at each step is done a posteriori by minimizing the value of $\chi^2_\mathrm{SED}$.

\begin{table*}
 \centering
 \caption{Orbital and astrophysical parameters of KIC\,4851217 from the joint photodynamical light curve, RV and ETV solution with and without the involvement of the stellar energy distribution and \texttt{PARSEC} isochrone fitting.}
 \label{tab: syntheticfit_KIC4851217}
\begin{tabular}{@{}lllllll}
\hline
 & \multicolumn{3}{c}{without SED+\texttt{PARSEC}} & \multicolumn{3}{c}{with SED+\texttt{PARSEC}} \\ 
\hline
\multicolumn{7}{c}{orbital elements} \\
\hline
   & \multicolumn{6}{c}{subsystem}  \\
   & \multicolumn{2}{c}{Aa--Ab} & A--B & \multicolumn{2}{c}{Aa--Ab} & A--B  \\
  $P_\mathrm{anom}$ [days] & \multicolumn{2}{c}{$2.4703999_{-0.0000027}^{+0.0000027}$} & $2716_{-16}^{+26}$ & \multicolumn{2}{c}{$2.4703997_{-0.0000029}^{+0.0000026}$} & $2725_{-15}^{+16}$   \\
  $a$ [R$_\odot$] & \multicolumn{2}{c}{$12.22_{-0.02}^{+0.02}$} & $1349_{-6}^{+14}$ & \multicolumn{2}{c}{$12.20_{-0.01}^{+0.02}$} & $1355_{-9}^{+9}$ \\
  $e$ & \multicolumn{2}{c}{$0.03102_{-0.00004}^{+0.00004}$} & $0.64_{-0.04}^{+0.05}$ & \multicolumn{2}{c}{$0.03101_{-0.00004}^{+0.00004}$} & $0.67_{-0.04}^{+0.03}$ \\
  $\omega$ [deg]& \multicolumn{2}{c}{$168.4_{-0.6}^{+0.6}$} & $15_{-3}^{+3}$ & \multicolumn{2}{c}{$168.4_{-0.5}^{+0.6}$} & $15_{-2}^{+3}$ \\ 
  $i$ [deg] & \multicolumn{2}{c}{$77.32_{-0.12}^{+0.11}$} & $70_{-14}^{+33}$ & \multicolumn{2}{c}{$77.24_{-0.06}^{+0.07}$} & $76_{-14}^{+20}$ \\
  $\tau$ [BJD - 2400000]& \multicolumn{2}{c}{$55742.4719_{-0.0031}^{+0.0031}$} & $56694_{-28}^{+32}$ & \multicolumn{2}{c}{$54951.9443_{-0.0037}^{+0.0044}$} & $56686_{-20}^{+27}$\\
  $\Delta\omega$ [deg/yr] & \multicolumn{2}{c}{$2.40_{-0.06}^{+0.06}$} & ... & \multicolumn{2}{c}{$2.39_{-0.06}^{+0.06}$} & ... \\
  \hline
  mass ratio $[q=M_\mathrm{sec}/M_\mathrm{pri}]$ & \multicolumn{2}{c}{$1.137_{-0.003}^{+0.003}$} & $0.122_{-0.014}^{+0.016}$ & \multicolumn{2}{c}{$1.140_{-0.003}^{+0.003}$} & $0.120_{-0.011}^{+0.021}$ \\
  $K_\mathrm{pri}$ [km\,s$^{-1}$] & \multicolumn{2}{c}{$129.99_{-0.12}^{+0.16}$} & $3.20_{-0.24}^{+0.45}$ & \multicolumn{2}{c}{$129.94_{-0.09}^{+0.09}$} & $3.43_{-0.32}^{+0.28}$ \\ 
  $K_\mathrm{sec}$ [km\,s$^{-1}$] & \multicolumn{2}{c}{$114.39_{-0.34}^{+0.35}$} & $27.40_{-3.41}^{+1.94}$ & \multicolumn{2}{c}{$114.02_{-0.24}^{+0.29}$} & $28.29_{-2.86}^{+1.18}$ \\ 
  $\gamma$ [km\,s$^{-1}$] & \multicolumn{3}{c}{$-22.183_{-0.034}^{+0.034}$} & \multicolumn{3}{c}{$-22.178_{-0.032}^{+0.028}$} \\
  \hline  
  \multicolumn{7}{c}{derived apsidal motion related parameters} \\
  \hline
  $P_\mathrm{apse}^\mathrm{fit}$ [yr] & \multicolumn{2}{c}{$150_{-4}^{+4}$} & ...  & \multicolumn{2}{c}{$151_{-4}^{+4}$} & ... \\
  $P_\mathrm{apse}^\mathrm{theo}$ [yr]& \multicolumn{2}{c}{$154_{-2}^{+2}$} & $207000_{-90700}^{+42000}$  & \multicolumn{2}{c}{$152_{-1}^{+1}$} & $171000_{-28900}^{+49900}$ \\
  $\Delta\omega_\mathrm{tide}$ [deg/yr] & \multicolumn{2}{c}{$2.21_{-0.02}^{+0.02}$} & $6\times10^{-8}$ & \multicolumn{2}{c}{$2.25_{-0.02}^{+0.02}$} & $7\times10^{-8}$ \\
  $\Delta\omega_\mathrm{GR}$ [deg/yr] & \multicolumn{2}{c}{$0.11_{-0.01}^{+0.01}$} & $1.7\times10^{-6}$ & \multicolumn{2}{c}{$0.11_{-0.01}^{+0.01}$} & $1.9\times10^{-6}$ \\
  $\Delta\omega_\mathrm{3b}$ [deg/yr] & \multicolumn{2}{c}{$0.018_{-0.003}^{+0.006}$} & $0.0017_{-0.0006}^{+0.0008}$ & \multicolumn{2}{c}{$0.020_{-0.005}^{+0.007}$} & $0.0021_{-0.0005}^{+0.0005}$ \\
 \hline 
\multicolumn{7}{c}{stellar parameters} \\
\hline
   & Aa & Ab &  B  & Aa & Ab &  B  \\
  \hline
 \multicolumn{7}{c}{Relative quantities} \\
  \hline
 fractional radius [$R/a$] & $0.1719_{-0.0024}^{+0.0025}$ & $0.2511_{-0.0010}^{+0.0010}$  & $0.00033_{-0.00004}^{+0.00004}$ & $0.1728_{-0.0010}^{+0.0011}$ & $0.2520_{-0.0006}^{+0.0006}$  & $0.00034_{-0.00003}^{+0.00006}$ \\
 fractional flux [in $Kepler$-band] & $0.3238_{-0.0077}^{+0.0092}$  & $0.6686_{-0.0066}^{+0.0084}$    & $0.0007_{-0.0002}^{+0.0003}$ & $0.3269_{-0.0032}^{+0.0032}$ & $0.6681_{-0.0039}^{+0.0030}$ & $0.0005_{-0.0001}^{+0.0004}$  \\
 \hline
 \multicolumn{7}{c}{Physical Quantities} \\
  \hline 
 $M$ [$M_\odot$] & $1.876_{-0.012}^{+0.012}$ & $2.132_{-0.009}^{+0.009}$ & $0.489_{-0.058}^{+0.064}$ & $1.865_{-0.008}^{+0.011}$ & $2.125_{-0.005}^{+0.008}$ & $0.477_{-0.044}^{+0.083}$ \\
 $R$ [$R_\odot$] & $2.101_{-0.031}^{+0.031}$ & $3.069_{-0.012}^{+0.013}$ & $0.448_{-0.050}^{+0.061}$ & $2.108_{-0.013}^{+0.016}$ & $3.075_{-0.008}^{+0.008}$ & $0.461_{-0.043}^{+0.083}$ \\
 $T_\mathrm{eff}$ [K]& $7834$ & $7741_{-9}^{+9}$ & $3749_{-58}^{+98}$ & $7997_{-45}^{+45}$ & $7882_{-31}^{+36}$ & $3451_{-96}^{+224}$ \\
 $L_\mathrm{bol}$ [L$_\odot$] & $14.92_{-0.44}^{+0.45}$ & $30.36_{-0.28}^{+0.28}$ & $0.036_{-0.009}^{+0.015}$ & $16.28_{-0.34}^{+0.51}$ & $32.76_{-0.58}^{+0.69}$ & $0.027_{-0.007}^{+0.021}$ \\
 $M_\mathrm{bol}$ & $1.81_{-0.03}^{+0.03}$ & $1.03_{-0.01}^{+0.01}$ & $8.36_{-0.39}^{+0.33}$ & $1.74_{-0.03}^{+0.02}$ & $0.98_{-0.02}^{+0.02}$ & $8.69_{-0.64}^{+0.33}$ \\
 $M_V           $ & $1.78_{-0.03}^{+0.03}$ & $1.01_{-0.01}^{+0.01}$ & $9.96_{-0.60}^{+0.47}$ & $1.69_{-0.03}^{+0.02}$ & $0.91_{-0.02}^{+0.02}$ & $10.51_{-0.97}^{+0.50}$ \\
 $\log g$ [dex]   & $4.068_{-0.012}^{+0.012}$ & $3.794_{-0.004}^{+0.004}$ & $4.825_{-0.011}^{+0.030}$ & $4.060_{-0.005}^{+0.004}$ & $3.789_{-0.002}^{+0.002}$ & $4.788_{-0.075}^{+0.043}$ \\
 \hline
age [Gyr] & \multicolumn{3}{c}{$-$} & \multicolumn{3}{c}{$0.824_{-0.009}^{+0.006}$} \\
$[\mathrm{M/H}]$ [dex] & \multicolumn{3}{c}{$-$} & \multicolumn{3}{c}{$0.076_{-0.011}^{+0.011}$} \\
$E(B-V)$ [mag]    & \multicolumn{3}{c}{$-$} & \multicolumn{3}{c}{$0.133_{-0.009}^{+0.008}$} \\  
extra light $\ell_4$ [in \textit{Kepler}-band] & \multicolumn{3}{c}{$0.005_{-0.003}^{+0.006}$} & \multicolumn{3}{c}{$0.005_{-0.002}^{+0.004}$} \\
$(M_V)_\mathrm{tot}$ & \multicolumn{3}{c}{$0.57_{-0.01}^{+0.01}$} & \multicolumn{3}{c}{$0.48_{-0.02}^{+0.02}$} \\
distance [pc]     &\multicolumn{3}{c}{$-$} & \multicolumn{3}{c}{$1074_{-6}^{+6}$} \\
\hline
\end{tabular}

\end{table*}


The resulting median values of the posteriors of the adjusted and several derived parameters, together with their $1\sigma$ uncertainties for both kinds of analyses are tabulated in Table~\ref{tab: syntheticfit_KIC4851217}.

Among the parameters in Table~\ref{tab: syntheticfit_KIC4851217} are the derived apsidal motion parameters (around the middle of the table), which need further explanations. Here $P_\mathrm{apse}^\mathrm{fit}$ is simply that apsidal motion period that can be calculated easily from the fitted apsidal motion parameter ($\Delta\omega$). The other tabulated parameters, however, come from theory. $\Delta\omega_\mathrm{tide,GR,3b}$ are the theoretical (equilibrium) tide, general relativistic and third-body perturbation contributions, while $P_\mathrm{apse}^\mathrm{theo}$ is the theoretical apsidal motion period calculated from the summing of the three components. The calculations of these components are summarized in Appendix C of \citet{Borkovits_2015} and, discussed in detail in Sect.~6.2 of \citet{Kostov_2021}. Note also that the tidal contribution depends on the apsidal motion constants ($k_2$) of the two components. For a good agreement with the fitted apsidal motion rate, we set $k_2^\mathrm{Aa}=k_2^\mathrm{Ab}=0.00113$, which is in marginal accord with theoretical apsidal motion constants for such hot, radiative stars \citep[see, e.~g.,][]{Claret_2021}. Comparing the theoretically calculated apsidal advance rates, one can readily see that the apsidal motion is clearly dominated by the tidal distortions of the binary members, and the dynamically forced apsidal motion is perfectly negligible, as it was assumed a priori (see Sect.~\ref{Sect:ETV_preliminary}).

\begin{figure}
\includegraphics[width=\columnwidth]{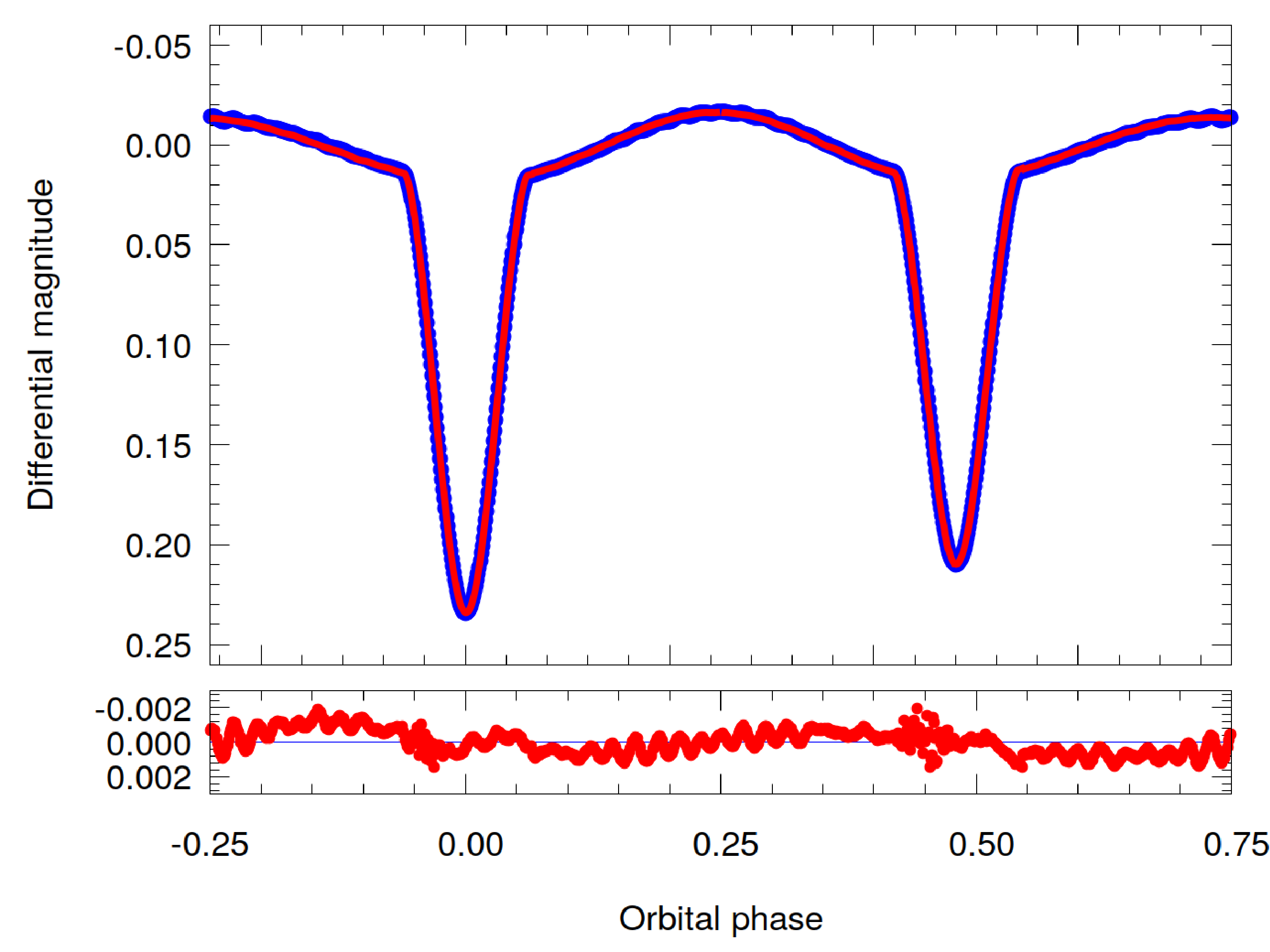} \\
\caption{\label{fig:lightcurvefactory} The best-fitting {\sc Lightcurvefactory} model (red line) to the \textit{Kepler} SC
phase-binned light curve of KIC 4851217 (blue filled circles). The residuals of the fit
are plotted in the lower panel using a greatly enlarged y-axis to bring out the detail. A very characteristic feature of the phase-folded curve, which can be seen better in the residual panel, that the phase-folding process did not averaged out fully the stellar pulsations, which indicate some connection between the oscillations and the orbital revolution.}
\end{figure}

\subsection{Comparison of physical properties from the individual and combined analyses}\label{sec: comparison of individual and combined results}

In this work we have done separate and independent analyses for subsets of the system parameters using subsets of the data, including RV data, ETV curves, SED fitting, and light curve analysis, in addition to a simultaneous joint analysis of all the data.  Here we compare how the results of the analyses of the various subsets of the data compare with those from the joint analysis.  Numerical comparisons are given in Table \ref{tab: appendix- full comparison} both as a percentage difference with respect to the values from the joint analysis and in terms of the mutual sigmas of the two approaches.  

In general, for the vast majority of the parameters we find agreement between the results using the data subsets versus the full joint solution at  the $\lesssim 1.5$\,$\sigma$ level.  In some cases the discrepancy for some of the non-essential parameters (e.g., the $\gamma$ velocity and colour excess) rises to the $3-4\,\sigma$ level.  This provides a caveat that we should not take these particular results too seriously at the quoted level of uncertainty. There is one particular parameter, namely the eccentricity of the EB, that is discrepant at the 8\,$\sigma$ level.  From our fit to the ETV data alone we found $e_{\rm in} = 0.03174 \pm 0.00008$ while from the joint analysis the result is $e_{\rm in} = 0.03102 \pm 0.00004$. The values are only discrepant by $0.00074 \pm 0.00009$, but nonetheless suggest there is a systematic error not accounted for here.  


Table \ref{tab: appendix- full comparison} also serves to summarize the parameters available from each of the subsets of data. Regarding the discrepancies larger than $\sim1.5\,\sigma$, we note the different levels of constraint that each type of analysis is subject to. For example, the results from the modelling of each individual subset of data are subject to the lowest level of constraint, relatively speaking, while those from case\,2 of the combined analysis are subject to the highest level of constraint. In the latter case, all the available observational constraints are imposed but note that the results are not entirely model-independent. Finally, we note the longer list of parameters reported in Table \ref{tab: syntheticfit_KIC4851217} for the tertiary component; notably, absolute estimates for its mass follow from the estimation for the outer orbital inclination. 

\begin{table}\caption{\label{tab: appendix- full comparison} Full table of comparisons $\Delta$ calculated for both sets of results from the combined analysis against those obtainable from the individual analyses, given as percentages. Also given are these discrepancies in units of the quadrature addition of the uncertainties $\sigma$.}
\centering
\setlength{\tabcolsep}{3pt}
\begin{tabular}{>{$}l<{$} r@{\,\%\,\,(}l@{\,$\sigma$)} c r@{\,\%\,\,(}l@{\,$\sigma$)}}

\hline
       &   \multicolumn{2}{c}{$\Delta$ Case\,1 } & &
           \multicolumn{2}{c}{$\Delta$ Case\,2} \\
\hline
\multicolumn{6}{l}{Orbital parameters from RV analysis}\\
K_{\rm Aa}    &  -0.092 &     -0.6 & &  -0.131 &     -1.8 \\
K_{\rm Ab}    &  -0.175 &     -0.5 & &  -0.497 &     -1.5 \\
\gamma        &  -1.45 &       2.8 & &   -1.48 &      2.9 \\
e             &  -3.06 &     -1.0 & &  -3.09 &     -1.0 \\
\omega        &  -1.41 &     -1.2 & &  -1.41 &     -1.2 \\\\
\multicolumn{6}{l}{Parameters from ETV curve analysis}\\
q             &   0.141 &      0.4 &  &  0.405 &      1.2 \\
\tau_3       &  -0.786 &     -0.6 &   &-0.904 &     -0.7 \\
P_3           &   1.50 &      0.8 &  &  1.83 &      1.1 \\
e_3           &  16.2 &      1.5 &  & 21.6 &      2.4 \\
\omega_3      & -28.6 &     -0.6 &  &-28.6 &     -0.6 \\
P_{\rm aps}   &  -7.98 &     -1.0 &  & -7.59 &     -0.9 \\
\omega        &  -1.06 &     -1.0 &  & -1.06 &     -1.0 \\
e             &  -2.27 &     -8.0 &  & -2.30 &     -8.2 \\\\
\multicolumn{6}{l}{Parameters from SED fitting}\\
M_{\rm Aa}    &  -2.80 &     -0.4 &  & -3.37 &     -0.5 \\
M_{\rm Ab}    &   0.566 &      0.1 &  &  0.236 &      0.1 \\
M_B           & -36.5 &     -3.2 &  &-38.1 &     -2.9 \\
R_{\rm Aa}    &  -9.83 &     -0.8 &  & -9.53 &     -0.8 \\
R_{\rm Ab}    &   0.294 &      0.03 &  &  0.490 &      0.1 \\
R_B           & -36.9 &     -3.3 &  &-35.1 &     -2.6 \\
T_{\rm eff, Aa}    &  -2.30 &     -0.6 &  & -0.262 &     -0.1 \\
T_{\rm eff, Ab}    &  -0.463 &     -0.1 &  &  1.35 &      0.3 \\
T_{\rm eff, B}           & -21.9   &   -3.3 & & -28.1 &     -3.6 \\
{\rm age}     & \multicolumn{2}{c}{N/A}& &   0.365 &      0.03 \\
{\rm distance}&\multicolumn{2}{c}{N/A}& &  -4.62 &     -2.9 \\
E(B-V)        &\multicolumn{2}{c}{N/A} & &  33.0 &      1.1 \\\\
\multicolumn{6}{l}{Parameters from the atmospheric analysis}\\
T_{\rm eff, Aa}&   0.000 &      0.0 & &   2.08 &      1.8 \\
T_{\rm eff, Ab}&   0.532 &      0.6 & &   2.36 &      2.2 \\
\multicolumn{6}{l}{Light curve analysis parameters} \\
r_{\rm Aa}           &  -3.97 &     -2.1 & &  -3.46 &     -2.4 \\
r_{\rm Ab}           &   0.080 &      0.1 &  &  0.438 &      0.3 \\\\
\multicolumn{6}{l}{Physical properties derived in Sect. \ref{sec: physical properties individual}}\\
a             &  -0.351 &     -1.7 &  & -0.514 &     -2.5 \\
M_{\rm Aa}    &  -1.21 &     -1.6 &  & -1.79 &     -2.5 \\
M_{\rm Ab}    &  -1.11 &     -2.1 &  & -1.44 &     -2.9 \\
R_{\rm Aa}    &  -4.28 &     -2.2 &  & -3.96 &     -2.6 \\
R_{\rm Ab}    &  -0.260 &     -0.2 &  & -0.065 &     -0.01 \\
\log(g)_{\rm Aa} &   0.843 &      2.1 &&    0.645 &      2.2 \\
\log(g)_{\rm Ab} &  -0.053 &     -0.2 & &  -0.184 &     -0.6 \\
\log(L/\Lsun)_{\rm Aa}&  -3.30 &     -1.8 &&   -0.165 &     -0.1 \\
\log(L/\Lsun)_{\rm Ab} &   0.339 &      0.3 &&    2.57 &      1.9 \\
E(B-V)   &   \multicolumn{2}{c}{N/A}&&  233 &      4.2 \\
{\rm distance}   &  \multicolumn{2}{c}{N/A}&&   -3.68 &     -2.3 \\
\hline
\end{tabular}
\end{table}


\section{Pulsation Analysis}\label{sec: pulsation analysis}

\citet{fedurco_parimucha_gajdos_2019} first reported pulsations in KIC 4851217. They detected a large number of pulsation frequencies of the $\delta$ Scuti type, many of which are spaced by the orbital frequency. These authors interpreted those pulsations as sequences of sectoral modes. \citet{Liakos_2020_KIC4851217} argued that the highest-amplitude pulsations originate in the secondary star on the basis of a comparison of the amplitudes during primary and secondary eclipse. In what follows, we present a preliminary analysis of the pulsations in this system as a precursor for a more detailed analysis (paper in preparation).

To this end, we used the {\sc Period04} software \citep{2005CoAst.146...53L}. This package produces amplitude spectra by Fourier analysis and can also perform multi-frequency least-squares sine-wave fitting. It also includes advanced options, such as the calculation of optimal light-curve fits for multiperiodic signals including harmonic, combination, and equally spaced frequencies which is essential for the analysis to be presented.

We have examined the {\it Kepler} LC and SC data and chose to analyse the LC data. The SC data do show some peaks at higher frequency than the LC Nyquist frequency. Those lie in the $35 - 40$\,d$^{-1}$ range,  but can be seen to be, at least primarily, harmonics and combinations of the pulsation modes at half that frequency range. The {\it Kepler} LC data, which span 1459.5\,d after removal of the Q0 and Q2 data that show large drifts, give higher frequency resolution. The higher-frequency harmonics and combinations do reflect about the Nyquist frequency down into the lower-frequency range, where they lie in the $10 - 15$\,d$^{-1}$ range, but at lower amplitude than we are analysing and hence can be neglected. The {\it Kepler} data are more precise and of longer time span than the TESS data ($\Delta T = 1140.9$\,d). Minor complications of using those data are that KIC 4851217 shows pulsational amplitude variations during the 4-yr time base of {\it Kepler} observations, as do a large fraction of $\delta$ Scuti pulsators \citep[e.g.,][]{2016MNRAS.460.1970B}, and that there are ETVs (Section~\ref{sec: preliminary light curve solutions}).

The first step in the analysis therefore is to determine the average value of the orbital frequency during the time of {\it Kepler} observations, and then to fit a harmonic series to remove that as a heuristic representation of the orbital light variations from the LC data. The average orbital frequency obtained was $\nu_{\rm orb}=0.40481179(2)$\,d$^{-1}$.

Owing to the ETVs and amplitude variations, we have subdivided the data set into four parts (with comparable time bases and numbers of data points): [Q1,Q3--5], Q7--Q9, Q11-Q13, Q15--Q17. We established the frequencies using the full data set for best accuracy, but then determined the amplitudes and phases of the signals from the four data subsets. For the detection of additional frequencies we then merged the residuals of those four data subsets into a single light curve and computed residual Fourier spectra, mostly free of artefacts from pulsational amplitude variations, from it. During this process it became clear that there is a multitude of pulsational signals, often spaced by multiples of the orbital frequency.

\begin{figure*}
    \centering
    \includegraphics[width = 0.88\textwidth]{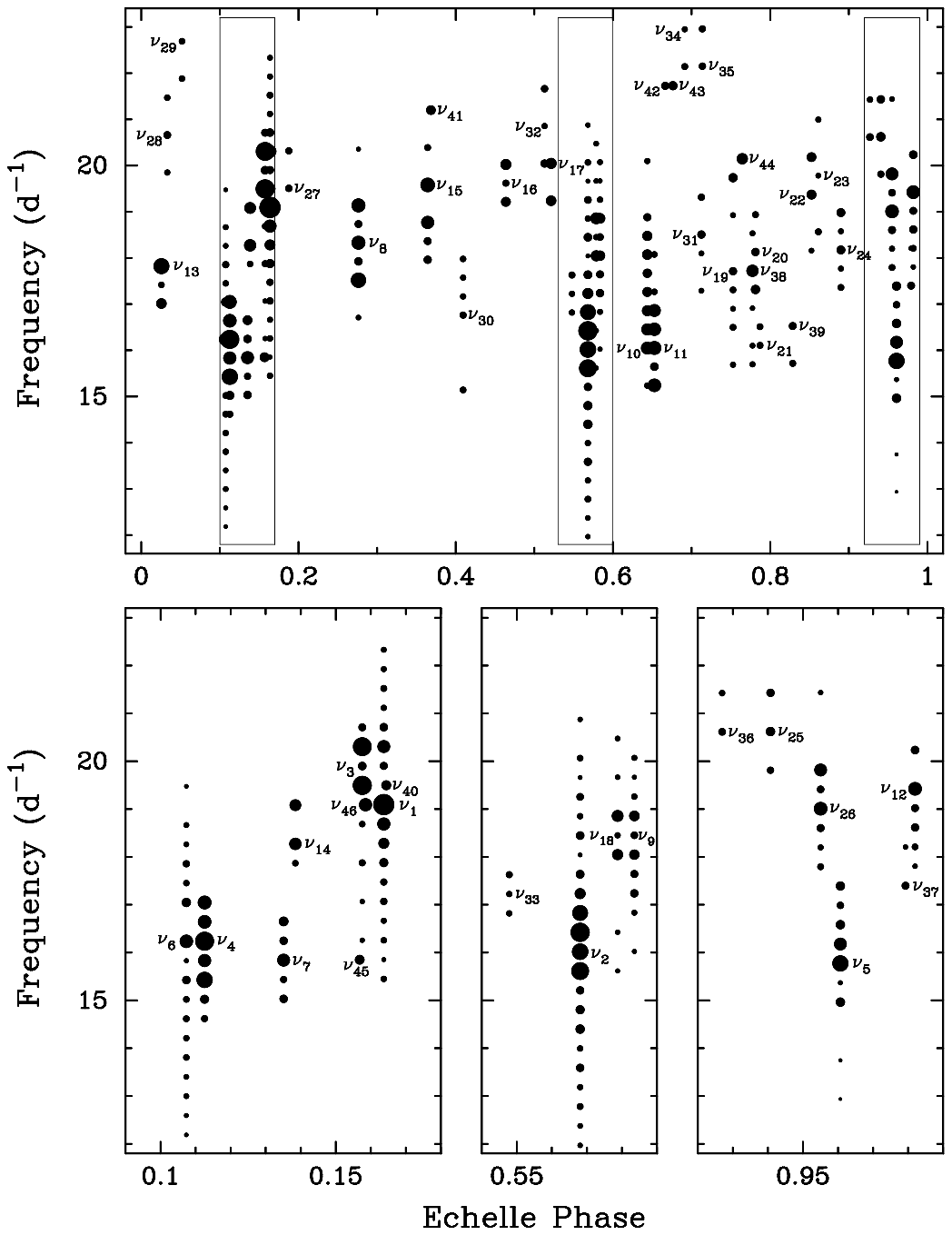}
    \caption{\label{fig:echelle}
    \reff{The \'echelle diagram of the pulsations. Here the frequency of a pulsation is plotted along the vertical axis while its `echelle phase’ is shown on the horizontal axis, In turn, the echelle phase is just the pulsation frequency mudulo the orbital frequency. Points that are vertically aligned are, by construction, frequencies that are spaced by an integer multiple of the orbital frequency. The echelle phase tells us where the pulsation lands with respect to actual or projected orbital harmonics. The lower panels are zooms into the most crowded regions. The size of the plot symbols is proportional to the amplitude of the signals.}}
\end{figure*}

In such a situation one needs to be careful about the application of S/N criteria regarding frequency detection, as this may lead to overly optimistic numbers of detections \citep{2014MNRAS.439.3453B}. In a first step, we therefore only accepted signals with amplitudes exceeding 0.05 mmag, corresponding to $S/N=25$ following \citet{1993A&A...271..482B}. We then computed an \'echelle diagram using those frequencies with respect to the orbital frequency (see \citet{Jayaraman_2022} for an explanation) and looked for additional possible components of the emerging multiplet structures. For multiplet components to be accepted, we demanded them to be exactly equally spaced in frequency by multiples of the orbital frequency within {\sc Period04}, and that their amplitude exceeds 0.012 mmag ($S/N=6$). The \'echelle diagram obtained in this way is shown in Fig. \ref{fig:echelle}, and the list of pulsation frequencies is given in Table \ref{tab:freq}. 

The \'echelle diagram shown in Fig.\ \ref{fig:echelle} is both very rich and complicated; \reff{the number of modes is typical for a \dsct\ star, but the large number of tidally split multiplets (some of which with many components) make this object particularly interesting.} The various multiplets are numbered according to the frequency scheme given in Table \ref{tab:freq}.  We have made an initial tally of the different types of multiplets (see Table \ref{tbl:multiplets}) that we see in the \'echelle diagram.  There are some 11 multiplets that we consider to be clear cases of dipole pulsation modes; these have either two elements separated by $2 \nu_{\rm orb}$ or the same but with an additional central element.  We count eight multiplets that are clear representations of quadrupole mode pulsations.  These have either two elements separated by $4 \nu_{\rm orb}$, the same but with an additional central element, or one with all five elements.  These 19 multiplets are a good indication of tidally tilted pulsations \citep[e.g.,][]{Handler_2020, Kurtz_2020, Fuller_2020, Rappaport_2021}.

In addition, there are 16 multiplets that resemble dipole or quadrupole pulsations but have either (i) clear asymmetries in the element amplitudes or (ii) extra elements beyond the $\pm 1 \nu_{\rm orb}$ or $\pm 2 \nu_{\rm orb}$ elements. These are more difficult to interpret. Interestingly, there are three cases of multiplets with long strings of elements (i.e., $\gtrsim 10$).  These may be caused by the eclipses which can obscure or enhance pulsations by removing some of the geometric cancellation \citep[e.g., eclipse mapping; see][and references therein]{Lampens_2019}.  And, since these latter events occur for only a short portion of the orbit, they can produce long strings of harmonics in the Fourier transform.  Finally, there are eight singlets.  These are possibly (i) radial modes, (ii) non-radial modes that are not tidally tilted, or (iii) part of a tidally tilted triplet, one component of which remains aligned with the orbital angular momentum axis, but is a standing rather than circulating wave \citep{Zhang_2023}.

The derivation of individual pulsational mode identifications from the runs of the pulsation amplitudes and phases over the orbit \citep[see][]{Jayaraman_2022} is out of the scope of the present paper (similarly is the question of which star pulsates for which a concrete determination would require asteroseismic modelling; insights can also be drawn via eclipse mapping). However, the complexity and multitude of the detected signals clearly argue against an interpretation in terms of tidally focused modes as put forward by \citep{fedurco_parimucha_gajdos_2019}.

\begin{table}
\centering
\caption{Multiplet counts.}
\begin{tabular}{lc}
\hline
\hline
Multiplet Type & Number \\
\hline
   Singlets & 8  \\
   Clear dipoles & 11  \\
   Clear quadrupoles & 8  \\
   Irregular & 16   \\
   Long strings & 3   \\
 \hline
\hline 
\label{tbl:multiplets} 
\end{tabular}
\end{table}

\section{Discussion}\label{sec:discussion}
Considering our model-independent methods of analysis of the photometric and spectroscopic data (i.e., using the results from the modelling of the individual data subsets and their joint analysis via method 1 of the combined analysis in Section \ref{sec: combined anlysis}), we measured the masses of the components of the inner EB to 0.5 and 0.4 per cent precision on average for star Aa and star Ab, respectively, with a mutual agreement at the $\sim1.5\,\sigma$ and $\sim2\,\sigma$ level; their radii were measured to 1.4 and 0.8 per cent precision on average with mutual agreement at the $\sim 2\,\sigma$ and $\sim 0.2\,\sigma$ level. We measured the components' \Teff s to $\sim1$ per cent precision via the atmospheric analysis of their disentangled spectra in Section \ref{sec: atmospheric parameters}, and used the result for star Aa to fix its \Teff\ in method 1 of the combined analysis; the results for the \Teff\ of star Ab obtained from both these methods were consistent to $\sim0.5\,\sigma$ (see Table \ref{tab: appendix- full comparison}).

Table \ref{tab: appendix- liakos comparisons} presents comparisons of our model-independent results against those reported by \citet{Liakos_2020_KIC4851217}. Masses are in agreement, as expected since \citet{Liakos_2020_KIC4851217} used velocity amplitudes reported by \citet{Helminiak_2019}, which are consistent with our values to within $1\,\sigma$ in all cases (see Tables \ref{tab:orbital_params} and \ref{tab: syntheticfit_KIC4851217}). However, our radii are discrepant by $\sim7\,\sigma$ and $\sim8\,\sigma$ compared to those by \citet{Liakos_2020_KIC4851217}, where they find near-equivalent radii between the components, i.e., $k\sim1$, compared to our values of around $k\sim1.4$ and $k\sim1.5$. Our values for \Teff\ are in modest agreement with those of \citet{Liakos_2020_KIC4851217} but the discrepancy in radii for both components has led to different conclusions as to which star contributes most of the system's light; they find $L_1/L_T \sim 0.585$ and $L_2/L_T \sim 0.415$, where $L_T$ is the total light of the system. These values were determined by \citet{Liakos_2020_KIC4851217} from fitting the light curves alone, where $k$ and $\ell_{\rm Ab}/\ell_{\rm Aa}$ are degenerate because KIC\,4851217 exhibits partial eclipses \citep[e.g.,][]{Jennings_KIC985}. Thus, since our solutions are in agreement with the spectroscopic light ratio derived in Section \ref{sec: spectral disentangling}, and indeed our spectroscopic light ratio was used to guide us in obtaining the correct light curve solution (see Section \ref{sec:lc:wd}), this is evidence that the scenario found here is an improvement on that of \citet{Liakos_2020_KIC4851217}. 

We have reported the detection of a tertiary M-dwarf companion (star B) in the KIC\,4851217 system from the analysis of the primary and secondary mid-eclipse times of the inner EB, which were measured from \emph{Kepler} and TESS light curves and show ETVs due to an outer orbit and apsidal motion of the EB orbit. The relatively low amplitude of the ETV signatures mean that the outer orbit is undetected in the time-span of our spectroscopic observations. In addition, the M-dwarf contributes negligible light to the system; both the current work and that of \citet{Liakos_2020_KIC4851217} measure a negligible value for third light. Hence, the tertiary M-dwarf was not detected by previous authors using \emph{Kepler} data alone. This is another example of the advantages associated with not only the high precision, but also the long time-base monitoring of stars provided by the combination of both the \emph{Kepler} and TESS observations. We analysed the ETVs jointly with the light curves and RVs measured from high (HERMES; R $\sim$ 85\,000) and moderate (ISIS; R $\sim$ 20\,000) resolution spectra in Section \ref{sec: combined anlysis}, where we report estimates for the mass, radius and \Teff\ of star B to precisions of 15, 16 and 5 per cent precision on average, respectively.

\begin{figure}
    \centering
    \includegraphics[width = \columnwidth]{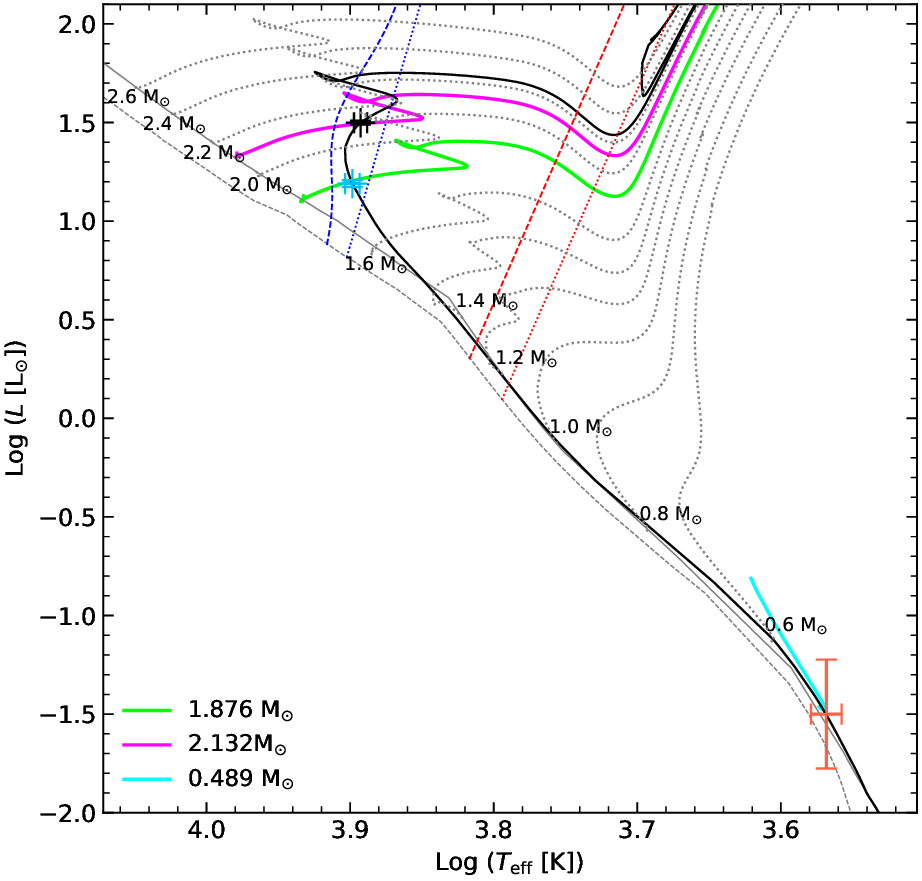}
    \caption{Theoretical Hertzsprung-Russell diagram showing the evolutionary tracks corresponding to the results for the masses of the components of KIC\,4851217 in Table \ref{tab: syntheticfit_KIC4851217}. The evolutionary tracks are shown in green, purple and blue for star Aa, star Ab and star B respectively, and their observed locations are indicated by the blue, black and red markers. The blue and red edges of the instability domains for low-order p- and g-modes calculated by \citet{Xiong_2016} for \dsct\ (dashed lines) and \gdor\ (dotted lines) stars are indicated by blue and red lines, respectively. The thin, black line is the solar ZAMS and the grey dashed line is the ZAMS for a metallicity of [Fe/H] $= -0.25$ dex. The thick, black line represents the MIST isochrone for the estimated age (0.82 Gyr) of the KIC\,4851217 system. Transparent grey dotted lines show solar-metallicity evolutionary tracks for stars with other labelled masses.}
    \label{fig:KIC485 Teff_l}
\end{figure}

Regarding our model-dependent measurement methods, i.e., the SED fitting in Section \ref{sec: SED fitting} and method 2 of the combined analysis, we find they agree with an age estimate of 0.82\,Gyr. We plotted the corresponding MIST isochrone \citep{dotter16, choi16, paxton19} in the HR plane in Fig.\,\ref{fig:KIC485 Teff_l}, along with the average locations of all three components based on the two solutions for each component in Table \ref{tab: syntheticfit_KIC4851217}. Note that each solution for the three components in Table \ref{tab: syntheticfit_KIC4851217} should be treated as an independent set of results, in general, and the average values plotted in Fig.\,\ref{fig:KIC485 Teff_l} simply serve to summarize an inferred evolutionary status of the objects based on this work. Also plotted are MIST evolutionary tracks for the corresponding mass estimates of each component as well as the \dsct\ and \gdor\ instability domains calculated by \citet{Xiong_2016}. The MIST models are calculated assuming single stars, so the fine alignment between the locations of each component with the evolutionary tracks and isochrone is evidence that each component has evolved as such, i.e., without prior mass transfer. The figure shows that star Ab is larger and more massive than star Aa, but cooler as it is approaching the end of the MS. Furthermore, we notice that the components of the EB are just within the blue edge of the \dsct\ instability strip but slightly outside that of the \gdor\ instability strip; this is in line with our observation of pulsations only at higher frequencies. 

We analysed the \emph{Kepler} LC light curves and extracted a list of pulsation frequencies which we presented in Table \ref{tab:freq}. Many of these frequencies are spaced by multiples of the orbital frequency and form vertical ridges in the \'echelle diagram of Fig.\,\ref{fig:echelle} plotted with respect to the orbital frequency. These multiplets are evidence that KIC\,4851217 has tidally tilted pulsations. 

We finally note that, in contrast to our methods, \citet{Liakos_2020_KIC4851217} derived the distance to the system using the pulsations; he used the detected dominant pulsation modes to calculate the absolute magnitude $M_{\rm V}$ using the pulsation period\,--\,luminosity relation for \dsct\ stars by \citet{Ziaali_2019}, and then the distance modulus with the apparent magnitude $m_{\rm V}$ from the Tycho-2 catalogue \citep{Hog_2000}. The resulting value for the distance is $579^{+38}_{-35}$\,pc, which is roughly half of every estimate made in this work, i.e., $\sim$ $1126 \pm 17 $\,pc, $1074 \pm 6$\,pc and $1115 \pm 17$\,pc in Sections \ref{sec: SED fitting}, \ref{sec: combined anlysis} and \ref{sec: physical properties individual}, respectively. The current estimates are in agreement with each other as well as the distance estimate from \emph{Gaia} of $1127 \pm 20$\,pc.

\section{Conclusion}\label{sec:conclusion}
We have presented the most comprehensive characterization of the KIC\,4851217 system to date. Our HERMES spectroscopic observations used to extract most of the RVs and precisely characterize the components' atmospheres are of the highest quality reported for the object, while the inclusion of the TESS light curves in addition to \emph{Kepler} data yields the longest time-base of photometric data studied for this system; this allowed us to discover a previously undetected tertiary M-dwarf companion, for which we present its fundamental characterization. 

We highlighted the results obtainable from the modelling of the individual subsets of spectroscopic and photometric data available for such a system, and compared those with the results obtained from their combined analysis; in general, we find good agreement. For partially eclipsing systems, the degeneracy between the radius ratio and light ratio means a spectroscopic light ratio is crucial for confirming the correct light curve solution is reached. 

The near-equal masses of the components of KIC\,4851217 combined with their differing evolutionary status makes the object excellent for constraining stellar theory \citep{Torres_2010}. While detailed evolutionary modelling is beyond the scope of the current work, the large list of model-independent results presented in this work make KIC\,4851217 well-suited for such a study. 

KIC\,4851217 is a precisely-characterized \dsct\ star so its contribution to the literature aligns with the broader objective to derive constraints on the internal structures of intermediate-mass stars, while also contributing an ideal candidate for developing our view of hierarchichal triples and TTPs. We note that the topic of TTPs is still in its infancy, and potentially why \citet{fedurco_parimucha_gajdos_2019} and \citet{Liakos_2020_KIC4851217} did not interpret the modes in KIC\,4851217 as such. 

Until the detection of TTPs in the subdwarf B star HD 265435 by \citet{Jayaraman_2022}, there were only three conclusively identified TTP stars reported in the literature \citep{Handler_2020, Kurtz_2020, Rappaport_2021}, with each of them being \dsct\ stars. Thus, the former precludes the possibility that tidal tilting of the pulsation axis is a phenomenon unique to \dsct\ stars, and this is in line with theory \citep{Fuller_2020}; it is noteworthy that while some stars exhibit TTPs, others exhibit tidally \emph{perturbed} pulsations \citep[e.g.,][]{Bowman_2019, Steindl_Zwintz_Bowman_2021, Johnston_2023}. Detecting and modelling more TTPs is in order to advance our understanding of this phenomenon.

\section*{Data Availability}
This paper includes data collected by the \emph{Kepler} and TESS missions which is publicly available at the Mikilski Archive for Space Telescopes (MAST) at the Space Telescope Science institute (STScl) (\href{https://mast.stsci.edu/portal/Mashup/Clients/Mast/Portal.html}{https://mast.stsci.edu}).

\section*{Acknowledgements}
We gratefully acknowledge financial support from the Science and Technology Facilities Council. This research has made use of the SIMBAD and CDS databases operated by the Centre de Donn\'ees astronomiques de Strasbourg, France. This paper also made use of data from the \emph{Kepler} and TESS missions obtained from the MAST data archive at the Space Telescope Science Institue (STScI). Funding for the \emph{Kepler} and TESS missions is provided by the NASA Science Mission Directorate and NASA Explorer Program, respectively. STScI is operated by the Association of Universities for Research in Astronomy, Inc., under NASA contract NAS 5–26555. GH thanks the Polish National Center for Science (NCN) for support through grant 2021/43/B/ST9/02972. TB acknowledges the financial supports of the HUN-REN Hungarian Research Network and the Hungarian National Research, Development and Innovation Office -- NKFIH Grant K-147131.

\bibliographystyle{mnras}
\bibliography{refs.bib}

\begin{thebibliography}{}
\makeatletter
\relax
\def\mn@urlcharsother{\let\do\@makeother \do\$\do\&\do\#\do\^\do\_\do\%\do\~}
\def\mn@doi{\begingroup\mn@urlcharsother \@ifnextchar [ {\mn@doi@}
  {\mn@doi@[]}}
\def\mn@doi@[#1]#2{\def\@tempa{#1}\ifx\@tempa\@empty \href
  {http://dx.doi.org/#2} {doi:#2}\else \href {http://dx.doi.org/#2} {#1}\fi
  \endgroup}
\def\mn@eprint#1#2{\mn@eprint@#1:#2::\@nil}
\def\mn@eprint@arXiv#1{\href {http://arxiv.org/abs/#1} {{\tt arXiv:#1}}}
\def\mn@eprint@dblp#1{\href {http://dblp.uni-trier.de/rec/bibtex/#1.xml}
  {dblp:#1}}
\def\mn@eprint@#1:#2:#3:#4\@nil{\def\@tempa {#1}\def\@tempb {#2}\def\@tempc
  {#3}\ifx \@tempc \@empty \let \@tempc \@tempb \let \@tempb \@tempa \fi \ifx
  \@tempb \@empty \def\@tempb {arXiv}\fi \@ifundefined
  {mn@eprint@\@tempb}{\@tempb:\@tempc}{\expandafter \expandafter \csname
  mn@eprint@\@tempb\endcsname \expandafter{\@tempc}}}

\bibitem[\protect\citeauthoryear{{Aerts}, {Christensen-Dalsgaard}  \&
  {Kurtz}}{{Aerts} et~al.}{2010}]{Aerts_book}
{Aerts} C.,  {Christensen-Dalsgaard} J.,   {Kurtz} D.~W.,  2010,
  {Asteroseismology}

\bibitem[\protect\citeauthoryear{{Antoci} et~al.,}{{Antoci}
  et~al.}{2014}]{Antoci_2014}
{Antoci} V.,  et~al., 2014, \mn@doi [\apj] {10.1088/0004-637X/796/2/118}, \href
  {https://ui.adsabs.harvard.edu/abs/2014ApJ...796..118A} {796, 118}

\bibitem[\protect\citeauthoryear{{Balona}}{{Balona}}{2014}]{2014MNRAS.439.3453B}
{Balona} L.~A.,  2014, \mn@doi [\mnras] {10.1093/mnras/stu193}, \href
  {https://ui.adsabs.harvard.edu/abs/2014MNRAS.439.3453B} {439, 3453}

\bibitem[\protect\citeauthoryear{{Blanco-Cuaresma}, {Soubiran}, {Heiter}  \&
  {Jofr{\'e}}}{{Blanco-Cuaresma} et~al.}{2014}]{ispec}
{Blanco-Cuaresma} S.,  {Soubiran} C.,  {Heiter} U.,   {Jofr{\'e}} P.,  2014,
  \mn@doi [\aap] {10.1051/0004-6361/201423945}, \href
  {https://ui.adsabs.harvard.edu/abs/2014A&A...569A.111B} {569, A111}

\bibitem[\protect\citeauthoryear{{Borkovits}}{{Borkovits}}{2022}]{Borkovits_2022}
{Borkovits} T.,  2022, \mn@doi [Galaxies] {10.3390/galaxies10010009}, \href
  {https://ui.adsabs.harvard.edu/abs/2022Galax..10....9B} {10, 9}

\bibitem[\protect\citeauthoryear{{Borkovits} et~al.,}{{Borkovits}
  et~al.}{2013}]{Borkovits_2013}
{Borkovits} T.,  et~al., 2013, \mn@doi [\mnras] {10.1093/mnras/sts146}, \href
  {https://ui.adsabs.harvard.edu/abs/2013MNRAS.428.1656B} {428, 1656}

\bibitem[\protect\citeauthoryear{{Borkovits} et~al.,}{{Borkovits}
  et~al.}{2014}]{Borkovits_2014}
{Borkovits} T.,  et~al., 2014, \mn@doi [\mnras] {10.1093/mnras/stu1379}, \href
  {https://ui.adsabs.harvard.edu/abs/2014MNRAS.443.3068B} {443, 3068}

\bibitem[\protect\citeauthoryear{{Borkovits}, {Rappaport}, {Hajdu}  \&
  {Sztakovics}}{{Borkovits} et~al.}{2015}]{Borkovits_2015}
{Borkovits} T.,  {Rappaport} S.,  {Hajdu} T.,   {Sztakovics} J.,  2015, \mn@doi
  [\mnras] {10.1093/mnras/stv015}, \href
  {https://ui.adsabs.harvard.edu/abs/2015MNRAS.448..946B} {448, 946}

\bibitem[\protect\citeauthoryear{{Borkovits}, {Hajdu}, {Sztakovics},
  {Rappaport}, {Levine}, {B{\'\i}r{\'o}}  \& {Klagyivik}}{{Borkovits}
  et~al.}{2016}]{Borkovits_2016}
{Borkovits} T.,  {Hajdu} T.,  {Sztakovics} J.,  {Rappaport} S.,  {Levine} A.,
  {B{\'\i}r{\'o}} I.~B.,   {Klagyivik} P.,  2016, \mn@doi [\mnras]
  {10.1093/mnras/stv2530}, \href
  {https://ui.adsabs.harvard.edu/abs/2016MNRAS.455.4136B} {455, 4136}

\bibitem[\protect\citeauthoryear{{Borkovits} et~al.,}{{Borkovits}
  et~al.}{2019}]{Borkovits_2019}
{Borkovits} T.,  et~al., 2019, \mn@doi [\mnras] {10.1093/mnras/sty3157}, \href
  {https://ui.adsabs.harvard.edu/abs/2019MNRAS.483.1934B} {483, 1934}

\bibitem[\protect\citeauthoryear{{Borkovits}, {Rappaport}, {Hajdu}, {Maxted},
  {P{\'a}l}, {Forg{\'a}cs-Dajka}, {Klagyivik}  \& {Mitnyan}}{{Borkovits}
  et~al.}{2020}]{Borkovits_2020}
{Borkovits} T.,  {Rappaport} S.~A.,  {Hajdu} T.,  {Maxted} P.~F.~L.,  {P{\'a}l}
  A.,  {Forg{\'a}cs-Dajka} E.,  {Klagyivik} P.,   {Mitnyan} T.,  2020, \mn@doi
  [\mnras] {10.1093/mnras/staa495}, \href
  {https://ui.adsabs.harvard.edu/abs/2020MNRAS.493.5005B} {493, 5005}

\bibitem[\protect\citeauthoryear{{Borucki} et~al.,}{{Borucki}
  et~al.}{2010}]{Borucki_2010}
{Borucki} W.~J.,  et~al., 2010, \mn@doi [Science] {10.1126/science.1185402},
  \href {https://ui.adsabs.harvard.edu/abs/2010Sci...327..977B} {327, 977}

\bibitem[\protect\citeauthoryear{{Bowman}}{{Bowman}}{2017}]{Bowman_2017}
{Bowman} D.~M.,  2017, {Amplitude Modulation of Pulsation Modes in Delta Scuti
  Stars}, \mn@doi{10.1007/978-3-319-66649-5.
}

\bibitem[\protect\citeauthoryear{{Bowman}, {Kurtz}, {Breger}, {Murphy}  \&
  {Holdsworth}}{{Bowman} et~al.}{2016}]{2016MNRAS.460.1970B}
{Bowman} D.~M.,  {Kurtz} D.~W.,  {Breger} M.,  {Murphy} S.~J.,   {Holdsworth}
  D.~L.,  2016, \mn@doi [\mnras] {10.1093/mnras/stw1153}, \href
  {https://ui.adsabs.harvard.edu/abs/2016MNRAS.460.1970B} {460, 1970}

\bibitem[\protect\citeauthoryear{{Bowman}, {Johnston}, {Tkachenko},
  {Mkrtichian}, {Gunsriwiwat}  \& {Aerts}}{{Bowman} et~al.}{2019}]{Bowman_2019}
{Bowman} D.~M.,  {Johnston} C.,  {Tkachenko} A.,  {Mkrtichian} D.~E.,
  {Gunsriwiwat} K.,   {Aerts} C.,  2019, \mn@doi [\apjl]
  {10.3847/2041-8213/ab3fb2}, \href
  {https://ui.adsabs.harvard.edu/abs/2019ApJ...883L..26B} {883, L26}

\bibitem[\protect\citeauthoryear{{Bowman}, {Hermans},
  {Daszy{\'n}ska-Daszkiewicz}, {Holdsworth}, {Tkachenko}, {Murphy}, {Smalley}
  \& {Kurtz}}{{Bowman} et~al.}{2021}]{Bowman_2021}
{Bowman} D.~M.,  {Hermans} J.,  {Daszy{\'n}ska-Daszkiewicz} J.,  {Holdsworth}
  D.~L.,  {Tkachenko} A.,  {Murphy} S.~J.,  {Smalley} B.,   {Kurtz} D.~W.,
  2021, \mn@doi [\mnras] {10.1093/mnras/stab1124}, \href
  {https://ui.adsabs.harvard.edu/abs/2021MNRAS.504.4039B} {504, 4039}

\bibitem[\protect\citeauthoryear{{Breger} et~al.,}{{Breger}
  et~al.}{1993}]{1993A&A...271..482B}
{Breger} M.,  et~al., 1993, \aap, \href
  {https://ui.adsabs.harvard.edu/abs/1993A&A...271..482B} {271, 482}

\bibitem[\protect\citeauthoryear{{Bressan}, {Marigo}, {Girardi}, {Salasnich},
  {Dal Cero}, {Rubele}  \& {Nanni}}{{Bressan} et~al.}{2012}]{PARSEC}
{Bressan} A.,  {Marigo} P.,  {Girardi} L.,  {Salasnich} B.,  {Dal Cero} C.,
  {Rubele} S.,   {Nanni} A.,  2012, \mn@doi [\mnras]
  {10.1111/j.1365-2966.2012.21948.x}, \href
  {https://ui.adsabs.harvard.edu/abs/2012MNRAS.427..127B} {427, 127}

\bibitem[\protect\citeauthoryear{{Carpenter}}{{Carpenter}}{2001}]{Carpenter01aj}
{Carpenter} J.~M.,  2001, AJ, \href {2001AJ....121.2851C} {121, 2851}

\bibitem[\protect\citeauthoryear{Castelli \& Kurucz}{Castelli \&
  Kurucz}{2004}]{castelli03}
Castelli F.,  Kurucz R.~L.,  2004, New Grids of ATLAS9 Model Atmospheres
  (\mn@eprint {arXiv} {astro-ph/0405087})

\bibitem[\protect\citeauthoryear{{Chen} et~al.,}{{Chen}
  et~al.}{2022}]{Chen_2022}
{Chen} X.,  et~al., 2022, \mn@doi [\apjs] {10.3847/1538-4365/aca284}, \href
  {https://ui.adsabs.harvard.edu/abs/2022ApJS..263...34C} {263, 34}

\bibitem[\protect\citeauthoryear{{Chen} et~al.,}{{Chen}
  et~al.}{2023}]{Chen_2023}
{Chen} X.,  et~al., 2023, VizieR Online Data Catalog, \href
  {https://ui.adsabs.harvard.edu/abs/2023yCat..22630034C} {p. J/ApJS/263/34}

\bibitem[\protect\citeauthoryear{{Choi}, {Dotter}, {Conroy}, {Cantiello},
  {Paxton}  \& {Johnson}}{{Choi} et~al.}{2016}]{choi16}
{Choi} J.,  {Dotter} A.,  {Conroy} C.,  {Cantiello} M.,  {Paxton} B.,
  {Johnson} B.~D.,  2016, \mn@doi [\apj] {10.3847/0004-637X/823/2/102}, \href
  {https://ui.adsabs.harvard.edu/abs/2016ApJ...823..102C} {823, 102}

\bibitem[\protect\citeauthoryear{{Claret}, {Gim{\'e}nez}, {Baroch}, {Ribas},
  {Morales}  \& {Anglada-Escud{\'e}}}{{Claret} et~al.}{2021}]{Claret_2021}
{Claret} A.,  {Gim{\'e}nez} A.,  {Baroch} D.,  {Ribas} I.,  {Morales} J.~C.,
  {Anglada-Escud{\'e}} G.,  2021, \mn@doi [\aap] {10.1051/0004-6361/202141484},
  \href {https://ui.adsabs.harvard.edu/abs/2021A&A...654A..17C} {654, A17}

\bibitem[\protect\citeauthoryear{{Conroy}, {Pr{\v{s}}a}, {Stassun}, {Orosz},
  {Fabrycky}  \& {Welsh}}{{Conroy} et~al.}{2014}]{Conroy_2014}
{Conroy} K.~E.,  {Pr{\v{s}}a} A.,  {Stassun} K.~G.,  {Orosz} J.~A.,  {Fabrycky}
  D.~C.,   {Welsh} W.~F.,  2014, \mn@doi [\aj] {10.1088/0004-6256/147/2/45},
  \href {https://ui.adsabs.harvard.edu/abs/2014AJ....147...45C} {147, 45}

\bibitem[\protect\citeauthoryear{{Cutri} et~al.,}{{Cutri}
  et~al.}{2003}]{Cutri+03book}
{Cutri} R.~M.,  et~al., 2003, {2MASS All Sky Catalogue of Point Sources}.
The IRSA 2MASS All-Sky Point Source Catalogue, NASA/IPAC Infrared Science
  Archive, Caltech, US

\bibitem[\protect\citeauthoryear{{Czesla}, {Schr{\"o}ter}, {Schneider},
  {Huber}, {Pfeifer}, {Andreasen}  \& {Zechmeister}}{{Czesla}
  et~al.}{2019}]{pyastronomy}
{Czesla} S.,  {Schr{\"o}ter} S.,  {Schneider} C.~P.,  {Huber} K.~F.,  {Pfeifer}
  F.,  {Andreasen} D.~T.,   {Zechmeister} M.,  2019, {PyA: Python
  astronomy-related packages} (\mn@eprint {ascl} {1906.010})

\bibitem[\protect\citeauthoryear{{Dotter}}{{Dotter}}{2016}]{dotter16}
{Dotter} A.,  2016, \mn@doi [\apjs] {10.3847/0067-0049/222/1/8}, \href
  {https://ui.adsabs.harvard.edu/abs/2016ApJS..222....8D} {222, 8}

\bibitem[\protect\citeauthoryear{{Fedurco}, {Parimucha}  \&
  {Gajdo{\v{s}}}}{{Fedurco} et~al.}{2019}]{fedurco_parimucha_gajdos_2019}
{Fedurco} M.,  {Parimucha} {\v{S}}.,   {Gajdo{\v{s}}} P.,  2019, \mn@doi
  [Proceedings of the International Astronomical Union]
  {10.1017/S1743921318002806}, 14, 295–298

\bibitem[\protect\citeauthoryear{{Feng} et~al.,}{{Feng}
  et~al.}{2021}]{Feng_2021}
{Feng} G.,  et~al., 2021, \mn@doi [\mnras] {10.1093/mnras/stab2063}, \href
  {https://ui.adsabs.harvard.edu/abs/2021MNRAS.508..529F} {508, 529}

\bibitem[\protect\citeauthoryear{{Ford}}{{Ford}}{2005}]{Ford_2005}
{Ford} E.~B.,  2005, \mn@doi [\aj] {10.1086/427962}, \href
  {https://ui.adsabs.harvard.edu/abs/2005AJ....129.1706F} {129, 1706}

\bibitem[\protect\citeauthoryear{{Fuller}}{{Fuller}}{2017}]{Fuller_2017}
{Fuller} J.,  2017, \mn@doi [\mnras] {10.1093/mnras/stx2135}, \href
  {https://ui.adsabs.harvard.edu/abs/2017MNRAS.472.1538F} {472, 1538}

\bibitem[\protect\citeauthoryear{{Fuller}, {Kurtz}, {Handler}  \&
  {Rappaport}}{{Fuller} et~al.}{2020}]{Fuller_2020}
{Fuller} J.,  {Kurtz} D.~W.,  {Handler} G.,   {Rappaport} S.,  2020, \mn@doi
  [\mnras] {10.1093/mnras/staa2376}, \href
  {https://ui.adsabs.harvard.edu/abs/2020MNRAS.498.5730F} {498, 5730}

\bibitem[\protect\citeauthoryear{{Gaia Collaboration}}{{Gaia
  Collaboration}}{2016}]{Gaia16aa}
{Gaia Collaboration} 2016, A\&A, \href {2016A+A...595A...1G} {595, A1}

\bibitem[\protect\citeauthoryear{{Gaia Collaboration}}{{Gaia
  Collaboration}}{2021}]{Gaia21aa}
{Gaia Collaboration} 2021, A\&A, \href {2021A+A...649A...1G} {649, A1}

\bibitem[\protect\citeauthoryear{{Gaulme} \& {Guzik}}{{Gaulme} \&
  {Guzik}}{2019a}]{Gaulme_Guzik_2019b}
{Gaulme} P.,  {Guzik} J.~A.,  2019a, VizieR Online Data Catalog, \href
  {https://ui.adsabs.harvard.edu/abs/2019yCat..36300106G} {pp J/A+A/630/A106}

\bibitem[\protect\citeauthoryear{{Gaulme} \& {Guzik}}{{Gaulme} \&
  {Guzik}}{2019b}]{Gaulme_Guzik_2019}
{Gaulme} P.,  {Guzik} J.~A.,  2019b, \mn@doi [\aap]
  {10.1051/0004-6361/201935821}, \href
  {https://ui.adsabs.harvard.edu/abs/2019A&A...630A.106G} {630, A106}

\bibitem[\protect\citeauthoryear{Gies, Williams, Matson, Guo, Thomas, Orosz  \&
  Peters}{Gies et~al.}{2012}]{Gies_2012}
Gies D.~R.,  Williams S.~J.,  Matson R.~A.,  Guo Z.,  Thomas S.~M.,  Orosz
  J.~A.,   Peters G.~J.,  2012, \mn@doi [The Astronomical Journal]
  {10.1088/0004-6256/143/6/137}, 143, 137

\bibitem[\protect\citeauthoryear{Gies, Matson, Guo, Lester, Orosz  \&
  Peters}{Gies et~al.}{2015}]{Gies_2015}
Gies D.~R.,  Matson R.~A.,  Guo Z.,  Lester K.~V.,  Orosz J.~A.,   Peters
  G.~J.,  2015, \mn@doi [\aj] {10.1088/0004-6256/150/6/178}, 150, 178

\bibitem[\protect\citeauthoryear{Gray}{Gray}{2005}]{gray_2005}
Gray D.~F.,  2005, The Observation and Analysis of Stellar Photospheres, 3 edn.
Cambridge University Press, \mn@doi{10.1017/CBO9781316036570}

\bibitem[\protect\citeauthoryear{{Grevesse}, {Asplund}  \& {Sauval}}{{Grevesse}
  et~al.}{2007}]{Grevesse_2007}
{Grevesse} N.,  {Asplund} M.,   {Sauval} A.~J.,  2007, \mn@doi [\ssr]
  {10.1007/s11214-007-9173-7}, \href
  {https://ui.adsabs.harvard.edu/abs/2007SSRv..130..105G} {130, 105}

\bibitem[\protect\citeauthoryear{{Guo}, {Fuller}, {Shporer}, {Li}, {Hambleton},
  {Manuel}, {Murphy}  \& {Isaacson}}{{Guo} et~al.}{2019}]{Guo_2019}
{Guo} Z.,  {Fuller} J.,  {Shporer} A.,  {Li} G.,  {Hambleton} K.,  {Manuel} J.,
   {Murphy} S.,   {Isaacson} H.,  2019, \mn@doi [\apj]
  {10.3847/1538-4357/ab41f6}, \href
  {https://ui.adsabs.harvard.edu/abs/2019ApJ...885...46G} {885, 46}

\bibitem[\protect\citeauthoryear{{Gustafsson}, {Edvardsson}, {Eriksson},
  {J{\o}rgensen}, {Nordlund}  \& {Plez}}{{Gustafsson}
  et~al.}{2008}]{Gustaffson_2008}
{Gustafsson} B.,  {Edvardsson} B.,  {Eriksson} K.,  {J{\o}rgensen} U.~G.,
  {Nordlund} {\r{A}}.,   {Plez} B.,  2008, \mn@doi [\aap]
  {10.1051/0004-6361:200809724}, \href
  {https://ui.adsabs.harvard.edu/abs/2008A&A...486..951G} {486, 951}

\bibitem[\protect\citeauthoryear{{Hambleton} et~al.,}{{Hambleton}
  et~al.}{2013}]{Hambleton_2013}
{Hambleton} K.~M.,  et~al., 2013, \mn@doi [\mnras] {10.1093/mnras/stt886},
  \href {https://ui.adsabs.harvard.edu/abs/2013MNRAS.434..925H} {434, 925}

\bibitem[\protect\citeauthoryear{{Handler} et~al.,}{{Handler}
  et~al.}{2020}]{Handler_2020}
{Handler} G.,  et~al., 2020, \mn@doi [Nature Astronomy]
  {10.1038/s41550-020-1035-1}, \href
  {https://ui.adsabs.harvard.edu/abs/2020NatAs...4..684H} {4, 684}

\bibitem[\protect\citeauthoryear{{Hareter} et~al.,}{{Hareter}
  et~al.}{2008}]{Hareter_2008}
{Hareter} M.,  et~al., 2008, \mn@doi [\aap] {10.1051/0004-6361:200809996},
  \href {https://ui.adsabs.harvard.edu/abs/2008A&A...492..185H} {492, 185}

\bibitem[\protect\citeauthoryear{{He{\l}miniak}, {Konacki}, {Maehara}, {Kambe},
  {Ukita}, {Ratajczak}, {Pigulski}  \& {Koz{\l}owski}}{{He{\l}miniak}
  et~al.}{2019}]{Helminiak_2019}
{He{\l}miniak} K.~G.,  {Konacki} M.,  {Maehara} H.,  {Kambe} E.,  {Ukita} N.,
  {Ratajczak} M.,  {Pigulski} A.,   {Koz{\l}owski} S.~K.,  2019, \mn@doi
  [\mnras] {10.1093/mnras/sty3528}, \href
  {https://ui.adsabs.harvard.edu/abs/2019MNRAS.484..451H} {484, 451}

\bibitem[\protect\citeauthoryear{{Henden}, {Levine}, {Terrell}, {Smith}  \&
  {Welch}}{{Henden} et~al.}{2012}]{Henden+12javso}
{Henden} A.~A.,  {Levine} S.~E.,  {Terrell} D.,  {Smith} T.~C.,   {Welch} D.,
  2012, Journal of the American Association of Variable Star Observers, \href
  {2012JAVSO..40..430H} {40, 430}

\bibitem[\protect\citeauthoryear{{H{\o}g} et~al.,}{{H{\o}g}
  et~al.}{2000}]{Hog_2000}
{H{\o}g} E.,  et~al., 2000, \aap, \href
  {https://ui.adsabs.harvard.edu/abs/2000A&A...355L..27H} {355, L27}

\bibitem[\protect\citeauthoryear{{Ilijic}}{{Ilijic}}{2017}]{FD3}
{Ilijic} S.,  2017, {fd3: Spectral disentangling of double-lined spectroscopic
  binary stars} (\mn@eprint {ascl} {1705.012})

\bibitem[\protect\citeauthoryear{{Ilijic}, {Hensberge}, {Pavlovski}  \&
  {Freyhammer}}{{Ilijic} et~al.}{2004}]{fd32}
{Ilijic} S.,  {Hensberge} H.,  {Pavlovski} K.,   {Freyhammer} L.~M.,  2004, in
  {Hilditch} R.~W.,  {Hensberge} H.,   {Pavlovski} K.,  eds,  Astronomical
  Society of the Pacific Conference Series Vol. 318, Spectroscopically and
  Spatially Resolving the Components of the Close Binary Stars. pp 111--113

\bibitem[\protect\citeauthoryear{{Jayaraman}, {Handler}, {Rappaport}, {Fuller},
  {Kurtz}, {Charpinet}  \& {Ricker}}{{Jayaraman} et~al.}{2022}]{Jayaraman_2022}
{Jayaraman} R.,  {Handler} G.,  {Rappaport} S.~A.,  {Fuller} J.,  {Kurtz}
  D.~W.,  {Charpinet} S.,   {Ricker} G.~R.,  2022, \mn@doi [\apjl]
  {10.3847/2041-8213/ac5c59}, \href
  {https://ui.adsabs.harvard.edu/abs/2022ApJ...928L..14J} {928, L14}

\bibitem[\protect\citeauthoryear{Jennings, Southworth, Pavlovski  \&
  Van Reeth}{Jennings et~al.}{2023}]{Jennings_KIC985}
Jennings Z.,  Southworth J.,  Pavlovski K.,   Van Reeth T.,  2023, \mn@doi
  [Monthly Notices of the Royal Astronomical Society] {10.1093/mnras/stad3427},
  527, 4052

\bibitem[\protect\citeauthoryear{{Jofr{\'e}}, {Heiter}, {Blanco-Cuaresma}  \&
  {Soubiran}}{{Jofr{\'e}} et~al.}{2014}]{Jofre_2014}
{Jofr{\'e}} P.,  {Heiter} U.,  {Blanco-Cuaresma} S.,   {Soubiran} C.,  2014, in
  Astronomical Society of India Conference Series. pp 159--166 (\mn@eprint
  {arXiv} {1312.2943}), \mn@doi{10.48550/arXiv.1312.2943}

\bibitem[\protect\citeauthoryear{{Johnston}, {Tkachenko}, {Van Reeth},
  {Bowman}, {Pavlovski}, {Sana}  \& {Sekaran}}{{Johnston}
  et~al.}{2023}]{Johnston_2023}
{Johnston} C.,  {Tkachenko} A.,  {Van Reeth} T.,  {Bowman} D.~M.,  {Pavlovski}
  K.,  {Sana} H.,   {Sekaran} S.,  2023, \mn@doi [\aap]
  {10.1051/0004-6361/202244808}, \href
  {https://ui.adsabs.harvard.edu/abs/2023A&A...670A.167J} {670, A167}

\bibitem[\protect\citeauthoryear{{Kahraman Ali{\c{c}}avu{\c{s}}}, {Soydugan},
  {Smalley}  \& {Kub{\'a}t}}{{Kahraman Ali{\c{c}}avu{\c{s}}}
  et~al.}{2017}]{Kahraman_2017}
{Kahraman Ali{\c{c}}avu{\c{s}}} F.,  {Soydugan} E.,  {Smalley} B.,
  {Kub{\'a}t} J.,  2017, \mn@doi [\mnras] {10.1093/mnras/stx1241}, \href
  {https://ui.adsabs.harvard.edu/abs/2017MNRAS.470..915K} {470, 915}

\bibitem[\protect\citeauthoryear{{Kahraman Ali{\c{c}}avu{\c{s}}},
  {G{\"u}m{\"u}{\c{s}}}, {K{\i}rm{\i}z{\i}ta{\c{s}}}, {Ekinci},
  {{\c{C}}avu{\c{s}}}, {Kaya}  \& {Ali{\c{c}}avu{\c{s}}}}{{Kahraman
  Ali{\c{c}}avu{\c{s}}} et~al.}{2022}]{Kahraman_2022}
{Kahraman Ali{\c{c}}avu{\c{s}}} F.,  {G{\"u}m{\"u}{\c{s}}} D.,
  {K{\i}rm{\i}z{\i}ta{\c{s}}} {\"O}.,  {Ekinci} {\"O}.,  {{\c{C}}avu{\c{s}}}
  S.,  {Kaya} Y.~T.,   {Ali{\c{c}}avu{\c{s}}} F.,  2022, \mn@doi [Research in
  Astronomy and Astrophysics] {10.1088/1674-4527/ac71a4}, \href
  {https://ui.adsabs.harvard.edu/abs/2022RAA....22h5003K} {22, 085003}

\bibitem[\protect\citeauthoryear{{Kervella}, {Th{\'e}venin}, {Di Folco}  \&
  {S{\'e}gransan}}{{Kervella} et~al.}{2004}]{Kervella+04aa}
{Kervella} P.,  {Th{\'e}venin} F.,  {Di Folco} E.,   {S{\'e}gransan} D.,  2004,
  \aap, \href {2004A+A...426..297K} {426, 297}

\bibitem[\protect\citeauthoryear{{Kirby}}{{Kirby}}{2011}]{Kirby_2011}
{Kirby} E.~N.,  2011, \mn@doi [\pasp] {10.1086/660019}, \href
  {https://ui.adsabs.harvard.edu/abs/2011PASP..123..531K} {123, 531}

\bibitem[\protect\citeauthoryear{{Kirk} et~al.,}{{Kirk}
  et~al.}{2016}]{Kirk_2016}
{Kirk} B.,  et~al., 2016, \mn@doi [\aj] {10.3847/0004-6256/151/3/68}, \href
  {https://ui.adsabs.harvard.edu/abs/2016AJ....151...68K} {151, 68}

\bibitem[\protect\citeauthoryear{{Koch} et~al.,}{{Koch}
  et~al.}{2010}]{Koch_2010}
{Koch} D.~G.,  et~al., 2010, \mn@doi [\apjl] {10.1088/2041-8205/713/2/L79},
  \href {https://ui.adsabs.harvard.edu/abs/2010ApJ...713L..79K} {713, L79}

\bibitem[\protect\citeauthoryear{{Kostov} et~al.,}{{Kostov}
  et~al.}{2021}]{Kostov_2021}
{Kostov} V.~B.,  et~al., 2021, \mn@doi [\apj] {10.3847/1538-4357/ac04ad}, \href
  {https://ui.adsabs.harvard.edu/abs/2021ApJ...917...93K} {917, 93}

\bibitem[\protect\citeauthoryear{{Kurtz} et~al.,}{{Kurtz}
  et~al.}{2020}]{Kurtz_2020}
{Kurtz} D.~W.,  et~al., 2020, \mn@doi [\mnras] {10.1093/mnras/staa989}, \href
  {https://ui.adsabs.harvard.edu/abs/2020MNRAS.494.5118K} {494, 5118}

\bibitem[\protect\citeauthoryear{{Kurucz}}{{Kurucz}}{2005}]{Kurucz_2005}
{Kurucz} R.~L.,  2005, Memorie della Societa Astronomica Italiana Supplementi,
  \href {https://ui.adsabs.harvard.edu/abs/2005MSAIS...8...14K} {8, 14}

\bibitem[\protect\citeauthoryear{{Lampens}, {Vermeylen}, {De Cat}  \& {Van
  Cauteren}}{{Lampens} et~al.}{2019}]{Lampens_2019}
{Lampens} P.,  {Vermeylen} L.,  {De Cat} P.,   {Van Cauteren} P.,  2019,
  Bulletin de la Societe Royale des Sciences de Liege, \href
  {https://ui.adsabs.harvard.edu/abs/2019BSRSL..88...89L} {88, 89}

\bibitem[\protect\citeauthoryear{{Landstreet}, {Kupka}, {Ford}, {Officer},
  {Sigut}, {Silaj}, {Strasser}  \& {Townshend}}{{Landstreet}
  et~al.}{2009}]{Landstreet_2009}
{Landstreet} J.~D.,  {Kupka} F.,  {Ford} H.~A.,  {Officer} T.,  {Sigut}
  T.~A.~A.,  {Silaj} J.,  {Strasser} S.,   {Townshend} A.,  2009, \mn@doi
  [\aap] {10.1051/0004-6361/200912083}, \href
  {https://ui.adsabs.harvard.edu/abs/2009A&A...503..973L} {503, 973}

\bibitem[\protect\citeauthoryear{{Latham}, {Nordstroem}, {Andersen}, {Torres},
  {Stefanik}, {Thaller}  \& {Bester}}{{Latham} et~al.}{1996}]{Latham_1996}
{Latham} D.~W.,  {Nordstroem} B.,  {Andersen} J.,  {Torres} G.,  {Stefanik}
  R.~P.,  {Thaller} M.,   {Bester} M.~J.,  1996, \aap, \href
  {https://ui.adsabs.harvard.edu/abs/1996A&A...314..864L} {314, 864}

\bibitem[\protect\citeauthoryear{{Lee}, {Hong}, {Koo}  \& {Park}}{{Lee}
  et~al.}{2018}]{Lee_2018}
{Lee} J.~W.,  {Hong} K.,  {Koo} J.-R.,   {Park} J.-H.,  2018, \mn@doi [\aj]
  {10.3847/1538-3881/aa947e}, \href
  {https://ui.adsabs.harvard.edu/abs/2018AJ....155....5L} {155, 5}

\bibitem[\protect\citeauthoryear{{Lenz} \& {Breger}}{{Lenz} \&
  {Breger}}{2005}]{2005CoAst.146...53L}
{Lenz} P.,  {Breger} M.,  2005, \mn@doi [Communications in Asteroseismology]
  {10.1553/cia146s53}, \href
  {https://ui.adsabs.harvard.edu/abs/2005CoAst.146...53L} {146, 53}

\bibitem[\protect\citeauthoryear{Liakos}{Liakos}{2020}]{Liakos_2020_KIC4851217}
Liakos A.,  2020, \mn@doi [\aap] {10.1051/0004-6361/202038065}, 642, A91

\bibitem[\protect\citeauthoryear{{Liakos} \& {Niarchos}}{{Liakos} \&
  {Niarchos}}{2017}]{Liakos_Niarchos_2017}
{Liakos} A.,  {Niarchos} P.,  2017, \mn@doi [\mnras] {10.1093/mnras/stw2756},
  \href {https://ui.adsabs.harvard.edu/abs/2017MNRAS.465.1181L} {465, 1181}

\bibitem[\protect\citeauthoryear{{Liakos} \& {Niarchos}}{{Liakos} \&
  {Niarchos}}{2020}]{Liakos_2020}
{Liakos} A.,  {Niarchos} P.,  2020, \mn@doi [Galaxies]
  {10.3390/galaxies8040075}, \href
  {https://ui.adsabs.harvard.edu/abs/2020Galax...8...75L} {8, 75}

\bibitem[\protect\citeauthoryear{{Liakos}, {Niarchos}, {Soydugan}  \&
  {Zasche}}{{Liakos} et~al.}{2012}]{Liakos_2012}
{Liakos} A.,  {Niarchos} P.,  {Soydugan} E.,   {Zasche} P.,  2012, \mn@doi
  [\mnras] {10.1111/j.1365-2966.2012.20704.x}, \href
  {https://ui.adsabs.harvard.edu/abs/2012MNRAS.422.1250L} {422, 1250}

\bibitem[\protect\citeauthoryear{{Maceroni}, {Montalb{\'a}n}, {Gandolfi},
  {Pavlovski}  \& {Rainer}}{{Maceroni} et~al.}{2013}]{Maceroni_2013}
{Maceroni} C.,  {Montalb{\'a}n} J.,  {Gandolfi} D.,  {Pavlovski} K.,   {Rainer}
  M.,  2013, \mn@doi [\aap] {10.1051/0004-6361/201220755}, \href
  {https://ui.adsabs.harvard.edu/abs/2013A&A...552A..60M} {552, A60}

\bibitem[\protect\citeauthoryear{{Marsh}}{{Marsh}}{2014}]{Marsh_2014}
{Marsh} T.~R.,  2014, {\textsc{pamela}: Optimal extraction code for long-slit
  CCD spectroscopy}, Astrophysics Source Code Library, record ascl:1406.002
  (\mn@eprint {ascl} {1406.002})

\bibitem[\protect\citeauthoryear{{Marsh}}{{Marsh}}{2019}]{Marsh_2019}
{Marsh} T.~R.,  2019, {\textsc{molly}: 1D astronomical spectra analyzer},
  Astrophysics Source Code Library, record ascl:1907.012 (\mn@eprint {ascl}
  {1907.012})

\bibitem[\protect\citeauthoryear{{M{\'e}sz{\'a}ros} et~al.,}{{M{\'e}sz{\'a}ros}
  et~al.}{2012}]{Meszaros_2012}
{M{\'e}sz{\'a}ros} S.,  et~al., 2012, \mn@doi [\aj]
  {10.1088/0004-6256/144/4/120}, \href
  {https://ui.adsabs.harvard.edu/abs/2012AJ....144..120M} {144, 120}

\bibitem[\protect\citeauthoryear{{Montgomery} \& {O'Donoghue}}{{Montgomery} \&
  {O'Donoghue}}{1999}]{1999DSSN...13...28M}
{Montgomery} M.~H.,  {O'Donoghue} D.,  1999, Delta Scuti Star Newsletter, \href
  {https://ui.adsabs.harvard.edu/abs/1999DSSN...13...28M} {13, 28}

\bibitem[\protect\citeauthoryear{{Murphy}, {Saio}, {Takada-Hidai}, {Kurtz},
  {Shibahashi}, {Takata}  \& {Hey}}{{Murphy} et~al.}{2020}]{Murphy_2020}
{Murphy} S.~J.,  {Saio} H.,  {Takada-Hidai} M.,  {Kurtz} D.~W.,  {Shibahashi}
  H.,  {Takata} M.,   {Hey} D.~R.,  2020, \mn@doi [\mnras]
  {10.1093/mnras/staa2667}, \href
  {https://ui.adsabs.harvard.edu/abs/2020MNRAS.498.4272M} {498, 4272}

\bibitem[\protect\citeauthoryear{{Ochsenbein}, {Bauer}  \&
  {Marcout}}{{Ochsenbein} et~al.}{2000}]{ochsenbein00}
{Ochsenbein} F.,  {Bauer} P.,   {Marcout} J.,  2000, \mn@doi [\aaps]
  {10.1051/aas:2000169}, \href
  {https://ui.adsabs.harvard.edu/abs/2000A&AS..143...23O} {143, 23}

\bibitem[\protect\citeauthoryear{{Pamyatnykh}}{{Pamyatnykh}}{1999}]{Pamyatnykh_1999}
{Pamyatnykh} A.~A.,  1999, \actaa, \href
  {https://ui.adsabs.harvard.edu/abs/1999AcA....49..119P} {49, 119}

\bibitem[\protect\citeauthoryear{{Park}, {Lee}, {Hong}, {Koo}  \& {Kim}}{{Park}
  et~al.}{2020}]{ParkJang_2020}
{Park} J.-H.,  {Lee} J.~W.,  {Hong} K.,  {Koo} J.-R.,   {Kim} C.-H.,  2020,
  \mn@doi [\aj] {10.3847/1538-3881/abbef4}, \href
  {https://ui.adsabs.harvard.edu/abs/2020AJ....160..247P} {160, 247}

\bibitem[\protect\citeauthoryear{{Paxton}, {Bildsten}, {Dotter}, {Herwig},
  {Lesaffre}  \& {Timmes}}{{Paxton} et~al.}{2011}]{paxton11}
{Paxton} B.,  {Bildsten} L.,  {Dotter} A.,  {Herwig} F.,  {Lesaffre} P.,
  {Timmes} F.,  2011, \mn@doi [\apjs] {10.1088/0067-0049/192/1/3}, \href
  {https://ui.adsabs.harvard.edu/abs/2011ApJS..192....3P} {192, 3}

\bibitem[\protect\citeauthoryear{{Paxton} et~al.,}{{Paxton}
  et~al.}{2015}]{paxton15}
{Paxton} B.,  et~al., 2015, \mn@doi [\apjs] {10.1088/0067-0049/220/1/15}, \href
  {https://ui.adsabs.harvard.edu/abs/2015ApJS..220...15P} {220, 15}

\bibitem[\protect\citeauthoryear{{Paxton} et~al.,}{{Paxton}
  et~al.}{2019}]{paxton19}
{Paxton} B.,  et~al., 2019, \mn@doi [\apjs] {10.3847/1538-4365/ab2241}, \href
  {https://ui.adsabs.harvard.edu/abs/2019ApJS..243...10P} {243, 10}

\bibitem[\protect\citeauthoryear{{Polfliet} \& {Smeyers}}{{Polfliet} \&
  {Smeyers}}{1990}]{Polfliet_Smeyers_1990}
{Polfliet} R.,  {Smeyers} P.,  1990, \aap, \href
  {https://ui.adsabs.harvard.edu/abs/1990A&A...237..110P} {237, 110}

\bibitem[\protect\citeauthoryear{{Pr{\v s}a} et~al.,}{{Pr{\v s}a}
  et~al.}{2016}]{Prsa+16aj}
{Pr{\v s}a} A.,  et~al., 2016, AJ, \href {2016AJ....152...41P} {152, 41}

\bibitem[\protect\citeauthoryear{{Pr{\v{s}}a} \& {Zwitter}}{{Pr{\v{s}}a} \&
  {Zwitter}}{2005}]{Prsa_2005}
{Pr{\v{s}}a} A.,  {Zwitter} T.,  2005, \mn@doi [\apj] {10.1086/430591}, \href
  {https://ui.adsabs.harvard.edu/abs/2005ApJ...628..426P} {628, 426}

\bibitem[\protect\citeauthoryear{{Pr{\v{s}}a} et~al.,}{{Pr{\v{s}}a}
  et~al.}{2011}]{Prsa_2011}
{Pr{\v{s}}a} A.,  et~al., 2011, \mn@doi [\aj] {10.1088/0004-6256/141/3/83},
  \href {https://ui.adsabs.harvard.edu/abs/2011AJ....141...83P} {141, 83}

\bibitem[\protect\citeauthoryear{{Pr{\v{s}}a} et~al.,}{{Pr{\v{s}}a}
  et~al.}{2022}]{Prsa_2022}
{Pr{\v{s}}a} A.,  et~al., 2022, \mn@doi [\apjs] {10.3847/1538-4365/ac324a},
  \href {https://ui.adsabs.harvard.edu/abs/2022ApJS..258...16P} {258, 16}

\bibitem[\protect\citeauthoryear{{Rappaport}, {Deck}, {Levine}, {Borkovits},
  {Carter}, {El Mellah}, {Sanchis-Ojeda}  \& {Kalomeni}}{{Rappaport}
  et~al.}{2013}]{Rappaport_2013}
{Rappaport} S.,  {Deck} K.,  {Levine} A.,  {Borkovits} T.,  {Carter} J.,  {El
  Mellah} I.,  {Sanchis-Ojeda} R.,   {Kalomeni} B.,  2013, \mn@doi [\apj]
  {10.1088/0004-637X/768/1/33}, \href
  {https://ui.adsabs.harvard.edu/abs/2013ApJ...768...33R} {768, 33}

\bibitem[\protect\citeauthoryear{{Rappaport} et~al.,}{{Rappaport}
  et~al.}{2021}]{Rappaport_2021}
{Rappaport} S.~A.,  et~al., 2021, \mn@doi [\mnras] {10.1093/mnras/stab336},
  \href {https://ui.adsabs.harvard.edu/abs/2021MNRAS.503..254R} {503, 254}

\bibitem[\protect\citeauthoryear{Rappaport et~al.,}{Rappaport
  et~al.}{2022}]{Rappaport_2022}
Rappaport S.~A.,  et~al., 2022, \mn@doi [\mnras] {10.1093/mnras/stac957}, 513,
  4341

\bibitem[\protect\citeauthoryear{{Raskin} et~al.,}{{Raskin}
  et~al.}{2011}]{Raskin_2011}
{Raskin} G.,  et~al., 2011, \mn@doi [\aap] {10.1051/0004-6361/201015435}, \href
  {https://ui.adsabs.harvard.edu/abs/2011A&A...526A..69R} {526, A69}

\bibitem[\protect\citeauthoryear{{Reyniers} \& {Smeyers}}{{Reyniers} \&
  {Smeyers}}{2003a}]{Reyniers_Smeyers_2003b}
{Reyniers} K.,  {Smeyers} P.,  2003a, \mn@doi [\aap]
  {10.1051/0004-6361:20030501}, \href
  {https://ui.adsabs.harvard.edu/abs/2003A&A...404.1051R} {404, 1051}

\bibitem[\protect\citeauthoryear{{Reyniers} \& {Smeyers}}{{Reyniers} \&
  {Smeyers}}{2003b}]{Reyniers_Smeyers_2003a}
{Reyniers} K.,  {Smeyers} P.,  2003b, \mn@doi [\aap]
  {10.1051/0004-6361:20031098}, \href
  {https://ui.adsabs.harvard.edu/abs/2003A&A...409..677R} {409, 677}

\bibitem[\protect\citeauthoryear{{Ricker} et~al.,}{{Ricker}
  et~al.}{2015}]{Ricker_2015}
{Ricker} G.~R.,  et~al., 2015, \mn@doi [Journal of Astronomical Telescopes,
  Instruments, and Systems] {10.1117/1.JATIS.1.1.014003}, \href
  {https://ui.adsabs.harvard.edu/abs/2015JATIS...1a4003R} {1, 014003}

\bibitem[\protect\citeauthoryear{{Sekaran}, {Tkachenko}, {Johnston}  \&
  {Aerts}}{{Sekaran} et~al.}{2021}]{Sekaran_2021}
{Sekaran} S.,  {Tkachenko} A.,  {Johnston} C.,   {Aerts} C.,  2021, \mn@doi
  [\aap] {10.1051/0004-6361/202040154}, \href
  {https://ui.adsabs.harvard.edu/abs/2021A&A...648A..91S} {648, A91}

\bibitem[\protect\citeauthoryear{{Shi}, {Qian}  \& {Li}}{{Shi}
  et~al.}{2022}]{Shi_2022}
{Shi} X.-d.,  {Qian} S.-b.,   {Li} L.-J.,  2022, \mn@doi [\apjs]
  {10.3847/1538-4365/ac59b9}, \href
  {https://ui.adsabs.harvard.edu/abs/2022ApJS..259...50S} {259, 50}

\bibitem[\protect\citeauthoryear{{Silva Aguirre}}{{Silva
  Aguirre}}{2018}]{SilvaAguirre_2018}
{Silva Aguirre} V.,  2018, in {Campante} T.~L.,  {Santos} N.~C.,   {Monteiro}
  M. J.~P.~F.~G.,  eds,  Astrophysics and Space Science Proceedings Vol. 49,
  Asteroseismology and Exoplanets: Listening to the Stars and Searching for New
  Worlds. p.~3 (\mn@eprint {arXiv} {1711.04461}),
  \mn@doi{10.1007/978-3-319-59315-9_1}

\bibitem[\protect\citeauthoryear{{Smalley}}{{Smalley}}{2005}]{Smalley_2005}
{Smalley} B.,  2005, \mn@doi [Memorie della Societa Astronomica Italiana
  Supplementi] {10.48550/arXiv.astro-ph/0509535}, \href
  {https://ui.adsabs.harvard.edu/abs/2005MSAIS...8..130S} {8, 130}

\bibitem[\protect\citeauthoryear{{Southworth}}{{Southworth}}{2013}]{Me13aa}
{Southworth} J.,  2013, A\&A, \href {2013A+A...557A.119S} {557, A119}

\bibitem[\protect\citeauthoryear{{Southworth}}{{Southworth}}{2015}]{Southworth_2015_DEBCat}
{Southworth} J.,  2015, in {Rucinski} S.~M.,  {Torres} G.,   {Zejda} M.,  eds,
  Astronomical Society of the Pacific Conference Series Vol. 496, Living
  Together: Planets, Host Stars and Binaries. p.~164 (\mn@eprint {arXiv}
  {1411.1219})

\bibitem[\protect\citeauthoryear{{Southworth}}{{Southworth}}{2020}]{Me20obs}
{Southworth} J.,  2020, The Observatory, 140, 247

\bibitem[\protect\citeauthoryear{{Southworth} \& {Bowman}}{{Southworth} \&
  {Bowman}}{2022}]{MeBowman22mn}
{Southworth} J.,  {Bowman} D.~M.,  2022, MNRAS, \href {2022MNRAS.513.3191S}
  {513, 3191}

\bibitem[\protect\citeauthoryear{{Southworth} \& {Clausen}}{{Southworth} \&
  {Clausen}}{2007}]{Southworth_Clausen_2007}
{Southworth} J.,  {Clausen} J.~V.,  2007, \mn@doi [\aap]
  {10.1051/0004-6361:20065614}, \href
  {https://ui.adsabs.harvard.edu/abs/2007A&A...461.1077S} {461, 1077}

\bibitem[\protect\citeauthoryear{{Southworth} \& {Van Reeth}}{{Southworth} \&
  {Van Reeth}}{2022}]{MeVanreeth22mn}
{Southworth} J.,  {Van Reeth} T.,  2022, MNRAS, \href {2022MNRAS.515.2755S}
  {515, 2755}

\bibitem[\protect\citeauthoryear{{Southworth}, {Maxted}  \&
  {Smalley}}{{Southworth} et~al.}{2005}]{Me++05aa}
{Southworth} J.,  {Maxted} P.~F.~L.,   {Smalley} B.,  2005, A\&A, \href
  {2005A+A...429..645S} {429, 645}

\bibitem[\protect\citeauthoryear{{Southworth} et~al.,}{{Southworth}
  et~al.}{2011}]{Southworth_2011}
{Southworth} J.,  et~al., 2011, \mn@doi [\mnras]
  {10.1111/j.1365-2966.2011.18559.x}, \href
  {https://ui.adsabs.harvard.edu/abs/2011MNRAS.414.2413S} {414, 2413}

\bibitem[\protect\citeauthoryear{{Soydugan}, {Erdem}, {Do{\u{g}}ru},
  {Ali{\c{c}}avu{\c{s}}}, {Soydugan}, {{\c{C}}i{\c{c}}ek}  \&
  {Demircan}}{{Soydugan} et~al.}{2011}]{Soydugan_2011}
{Soydugan} F.,  {Erdem} A.,  {Do{\u{g}}ru} S.~S.,  {Ali{\c{c}}avu{\c{s}}} F.,
  {Soydugan} E.,  {{\c{C}}i{\c{c}}ek} C.,   {Demircan} O.,  2011, \mn@doi [\na]
  {10.1016/j.newast.2010.11.006}, \href
  {https://ui.adsabs.harvard.edu/abs/2011NewA...16..253S} {16, 253}

\bibitem[\protect\citeauthoryear{{Stancliffe}, {Fossati}, {Passy}  \&
  {Schneider}}{{Stancliffe} et~al.}{2015}]{Stancliffe_2015}
{Stancliffe} R.~J.,  {Fossati} L.,  {Passy} J.~C.,   {Schneider} F.~R.~N.,
  2015, \mn@doi [\aap] {10.1051/0004-6361/201425126}, \href
  {https://ui.adsabs.harvard.edu/abs/2015A&A...575A.117S} {575, A117}

\bibitem[\protect\citeauthoryear{{Steindl}, {Zwintz}  \& {Bowman}}{{Steindl}
  et~al.}{2021}]{Steindl_Zwintz_Bowman_2021}
{Steindl} T.,  {Zwintz} K.,   {Bowman} D.~M.,  2021, \mn@doi [\aap]
  {10.1051/0004-6361/202039093}, \href
  {https://ui.adsabs.harvard.edu/abs/2021A&A...645A.119S} {645, A119}

\bibitem[\protect\citeauthoryear{{Torres} \& {Ribas}}{{Torres} \&
  {Ribas}}{2002}]{Torres&Ribas_2002}
{Torres} G.,  {Ribas} I.,  2002, \mn@doi [\apj] {10.1086/338587}, \href
  {https://ui.adsabs.harvard.edu/abs/2002ApJ...567.1140T} {567, 1140}

\bibitem[\protect\citeauthoryear{{Torres}, {Stefanik}, {Andersen}, {Nordstrom},
  {Latham}  \& {Clausen}}{{Torres} et~al.}{1997}]{Torres_1997}
{Torres} G.,  {Stefanik} R.~P.,  {Andersen} J.,  {Nordstrom} B.,  {Latham}
  D.~W.,   {Clausen} J.~V.,  1997, \mn@doi [\aj] {10.1086/118685}, \href
  {https://ui.adsabs.harvard.edu/abs/1997AJ....114.2764T} {114, 2764}

\bibitem[\protect\citeauthoryear{Torres, Andersen, Nordström  \&
  Latham}{Torres et~al.}{2000}]{Torres_2000}
Torres G.,  Andersen J.,  Nordström B.,   Latham D.~W.,  2000, \mn@doi [\aj]
  {10.1086/301305}, 119, 1942

\bibitem[\protect\citeauthoryear{{Torres}, {Andersen}  \&
  {Gim{\'e}nez}}{{Torres} et~al.}{2010}]{Torres_2010}
{Torres} G.,  {Andersen} J.,   {Gim{\'e}nez} A.,  2010, \mn@doi [\aapr]
  {10.1007/s00159-009-0025-1}, \href
  {https://ui.adsabs.harvard.edu/abs/2010A&ARv..18...67T} {18, 67}

\bibitem[\protect\citeauthoryear{{Tout}, {Pols}, {Eggleton}  \& {Han}}{{Tout}
  et~al.}{1996}]{Tout_1996}
{Tout} C.~A.,  {Pols} O.~R.,  {Eggleton} P.~P.,   {Han} Z.,  1996, \mn@doi
  [\mnras] {10.1093/mnras/281.1.257}, \href
  {https://ui.adsabs.harvard.edu/abs/1996MNRAS.281..257T} {281, 257}

\bibitem[\protect\citeauthoryear{{Uytterhoeven} et~al.,}{{Uytterhoeven}
  et~al.}{2011}]{Uttyerhoeven_2011}
{Uytterhoeven} K.,  et~al., 2011, \mn@doi [\aap] {10.1051/0004-6361/201117368},
  \href {https://ui.adsabs.harvard.edu/abs/2011A&A...534A.125U} {534, A125}

\bibitem[\protect\citeauthoryear{Virtanen et~al.,}{Virtanen
  et~al.}{2020}]{2020SciPy-NMeth}
Virtanen P.,  et~al., 2020, \mn@doi [Nature Methods]
  {10.1038/s41592-019-0686-2}, \href {https://rdcu.be/b08Wh} {17, 261}

\bibitem[\protect\citeauthoryear{{Welsh} et~al.,}{{Welsh}
  et~al.}{2011}]{Welsh_2011}
{Welsh} W.~F.,  et~al., 2011, \mn@doi [\apjs] {10.1088/0067-0049/197/1/4},
  \href {https://ui.adsabs.harvard.edu/abs/2011ApJS..197....4W} {197, 4}

\bibitem[\protect\citeauthoryear{{Wilson}}{{Wilson}}{1979}]{Wilson79apj}
{Wilson} R.~E.,  1979, ApJ, \href {1979ApJ...234.1054W} {234, 1054}

\bibitem[\protect\citeauthoryear{{Wilson} \& {Devinney}}{{Wilson} \&
  {Devinney}}{1971}]{WilsonDevinney71apj}
{Wilson} R.~E.,  {Devinney} E.~J.,  1971, ApJ, \href {1971ApJ...166..605W}
  {166, 605}

\bibitem[\protect\citeauthoryear{{Wilson} \& {Van Hamme}}{{Wilson} \& {Van
  Hamme}}{2004}]{WilsonVanhamme04}
{Wilson} R.~E.,  {Van Hamme} W.,  2004, Computing Binary Star Observables
  (Wilson-Devinney program user guide), available at
  ftp://ftp.astro.ufl.edu/pub/wilson

\bibitem[\protect\citeauthoryear{{Xiong}, {Deng}, {Zhang}  \& {Wang}}{{Xiong}
  et~al.}{2016}]{Xiong_2016}
{Xiong} D.~R.,  {Deng} L.,  {Zhang} C.,   {Wang} K.,  2016, \mn@doi [\mnras]
  {10.1093/mnras/stw047}, \href
  {https://ui.adsabs.harvard.edu/abs/2016MNRAS.457.3163X} {457, 3163}

\bibitem[\protect\citeauthoryear{{Xu}, {Cisewski-Kehe}, {Davis}, {Fischer}  \&
  {Brewer}}{{Xu} et~al.}{2019}]{Xu_2019}
{Xu} X.,  {Cisewski-Kehe} J.,  {Davis} A.~B.,  {Fischer} D.~A.,   {Brewer}
  J.~M.,  2019, \mn@doi [\aj] {10.3847/1538-3881/ab1b47}, \href
  {https://ui.adsabs.harvard.edu/abs/2019AJ....157..243X} {157, 243}

\bibitem[\protect\citeauthoryear{{Yang}, {Zuo}, {Li}, {Bedding}, {Murphy}  \&
  {Joyce}}{{Yang} et~al.}{2021}]{Yang_2021}
{Yang} T.-Z.,  {Zuo} Z.-Y.,  {Li} G.,  {Bedding} T.~R.,  {Murphy} S.~J.,
  {Joyce} M.,  2021, \mn@doi [\aap] {10.1051/0004-6361/202142198}, \href
  {https://ui.adsabs.harvard.edu/abs/2021A&A...655A..63Y} {655, A63}

\bibitem[\protect\citeauthoryear{{Zhang}, {Rappaport}, {Jayaraman}, {Kurtz},
  {Handler}, {Fuller}  \& {Borkovits}}{{Zhang} et~al.}{2023}]{Zhang_2023}
{Zhang} V.,  {Rappaport} S.,  {Jayaraman} R.,  {Kurtz} D.~W.,  {Handler} G.,
  {Fuller} J.,   {Borkovits} T.,  2023, \mn@doi [arXiv e-prints]
  {10.48550/arXiv.2311.16248}, \href
  {https://ui.adsabs.harvard.edu/abs/2023arXiv231116248Z} {p. arXiv:2311.16248}

\bibitem[\protect\citeauthoryear{{Zhou}}{{Zhou}}{2010}]{Zhoe_2010}
{Zhou} A.~Y.,  2010, arXiv e-prints, \href
  {https://ui.adsabs.harvard.edu/abs/2010arXiv1002.2729Z} {p. arXiv:1002.2729}

\bibitem[\protect\citeauthoryear{{Ziaali}, {Bedding}, {Murphy}, {Van Reeth}  \&
  {Hey}}{{Ziaali} et~al.}{2019}]{Ziaali_2019}
{Ziaali} E.,  {Bedding} T.~R.,  {Murphy} S.~J.,  {Van Reeth} T.,   {Hey} D.~R.,
   2019, \mn@doi [\mnras] {10.1093/mnras/stz1110}, \href
  {https://ui.adsabs.harvard.edu/abs/2019MNRAS.486.4348Z} {486, 4348}

\bibitem[\protect\citeauthoryear{{Zucker} \& {Mazeh}}{{Zucker} \&
  {Mazeh}}{1994}]{Zucker&Mazeh_1994}
{Zucker} S.,  {Mazeh} T.,  1994, \mn@doi [\apj] {10.1086/173605}, \href
  {https://ui.adsabs.harvard.edu/abs/1994ApJ...420..806Z} {420, 806}

\bibitem[\protect\citeauthoryear{{Zucker}, {Mazeh}  \& {Alexander}}{{Zucker}
  et~al.}{2007}]{Zucker++07apj}
{Zucker} S.,  {Mazeh} T.,   {Alexander} T.,  2007, ApJ, 670, 1326

\bibitem[\protect\citeauthoryear{{del Burgo} \& {Allende Prieto}}{{del Burgo}
  \& {Allende Prieto}}{2018}]{delBurgo_2018}
{del Burgo} C.,  {Allende Prieto} C.,  2018, \mn@doi [\mnras]
  {10.1093/mnras/sty1371}, \href
  {https://ui.adsabs.harvard.edu/abs/2018MNRAS.479.1953D} {479, 1953}

\makeatother
\end{thebibliography}


\clearpage
\appendix
\section{Measured Pulsation Frequencies}
This appendix presents a table of the pulsation frequencies measured as described in Section \ref{sec: pulsation analysis}.
\begin{table*}
\caption{Multiple pulsation frequency solution for the {\it Kepler} 30-min photometry of KIC 4851217. The phases are calculated with respect to a time of primary minimum of BJD 2454953.900333. The formal errors on the frequencies and phases \citep{1999DSSN...13...28M} are given in braces in units of the last significant digit; the formal errors on the amplitudes are $\pm$0.008 mmag.} 
\begin{center}
\begin{tabular}{lccclccc}
\hline
ID & Freq. & Ampl. & Phase & ID & Freq. & Ampl. & Phase \\
 & (d$^{-1}$) & (mmag) & (rad) & & (d$^{-1}$) & (mmag) & (rad) \\
\hline
$\nu_1-$9$\nu_{orb}$ & 15.449121 & 0.033 & 1.8(3) & $\nu_6-$10$\nu_{orb}$ & 12.18782 & 0.019 & 1.4(5) \\
$\nu_1-$8$\nu_{orb}$ & 15.853932 & 0.021 & 1.8(5) & $\nu_6-$9$\nu_{orb}$ & 12.59263 & 0.021 & $-$0.2(4) \\
$\nu_1-$7$\nu_{orb}$ & 16.258744 & 0.033 & 1.8(3) & $\nu_6-$8$\nu_{orb}$ & 12.99744 & 0.027 & 1.5(3) \\
$\nu_1-$6$\nu_{orb}$ & 16.663556 & 0.030 & 1.7(3) & $\nu_6-$7$\nu_{orb}$ & 13.40225 & 0.026 & 0.0(3) \\
$\nu_1-$5$\nu_{orb}$ & 17.068368 & 0.044 & 1.8(2) & $\nu_6-$6$\nu_{orb}$ & 13.80706 & 0.035 & 1.6(2) \\
$\nu_1-$4$\nu_{orb}$ & 17.47318 & 0.043 & 2.0(2) & $\nu_6-$5$\nu_{orb}$ & 14.21188 & 0.033 & 0.1(2) \\
$\nu_1-$3$\nu_{orb}$ & 17.877991 & 0.076 & 1.5(1) & $\nu_6-$4$\nu_{orb}$ & 14.61669 & 0.038 & 1.7(2) \\
$\nu_1-$2$\nu_{orb}$ & 18.282803 & 0.135 & $-$0.03(6) & $\nu_6-$3$\nu_{orb}$ & 15.0215 & 0.036 & 0.2(2) \\
$\nu_1-\nu_{orb}$ & 18.687615 & 0.266 & 1.75(3) & $\nu_6-$2$\nu_{orb}$ & 15.426311 & 0.063 & 1.7(1) \\
$\nu_1$ & 19.092427(1) & 3.410 & 1.703(3) & $\nu_6-\nu_{orb}$ & 15.83112 & 0.022 & 0.5(4) \\
$\nu_1$+2$\nu_{orb}$ & 19.90205 & 0.055 & 0.5(2) & $\nu_6$ & 16.23593(1) & 0.303 & 1.93(3) \\
$\nu_1$+3$\nu_{orb}$ & 20.306862 & 0.248 & 1.86(3) & $\nu_6$+2$\nu_{orb}$ & 17.04556 & 0.082 & 1.8(1) \\
$\nu_1$+4$\nu_{orb}$ & 20.711674 & 0.062 & 1.8(2) & $\nu_6$+3$\nu_{orb}$ & 17.45037 & 0.034 & 0.6(3) \\
$\nu_1$+5$\nu_{orb}$ & 21.116486 & 0.032 & 1.9(4) & $\nu_6$+4$\nu_{orb}$ & 17.85518 & 0.043 & 1.7(2) \\
$\nu_1$+6$\nu_{orb}$ & 21.521298 & 0.039 & 1.8(3) & $\nu_6$+5$\nu_{orb}$ & 18.25999 & 0.026 & 0.8(4) \\
$\nu_1$+7$\nu_{orb}$ & 21.926109 & 0.030 & 1.7(4) & $\nu_6$+6$\nu_{orb}$ & 18.66481 & 0.028 & 2.2(4) \\
$\nu_1$+8$\nu_{orb}$ & 22.330921 & 0.028 & 1.9(4) & $\nu_6$+8$\nu_{orb}$ & 19.47443 & 0.019 & $-$0.7(5) \\
$\nu_2-$10$\nu_{orb}$ & 11.96951 & 0.024 & 1.6(5) & $\nu_7-$2$\nu_{orb}$ & 15.03273 & 0.062 & 1.4(1) \\
$\nu_2-$9$\nu_{orb}$ & 12.374322 & 0.026 & 2.2(4) & $\nu_7-\nu_{orb}$ & 15.43754 & 0.042 & $-$0.4(2) \\
$\nu_2-$8$\nu_{orb}$ & 12.779134 & 0.038 & 1.7(3) & $\nu_7$ & 15.842351(1) & 0.251 & 1.35(3) \\
$\nu_2-$7$\nu_{orb}$ & 13.183946 & 0.031 & 2.2(3) & $\nu_7$+$\nu_{orb}$ & 16.24716 & 0.062 & $-$0.1(1) \\
$\nu_2-$6$\nu_{orb}$ & 13.588757 & 0.060 & 1.7(2) & $\nu_7$+2$\nu_{orb}$ & 16.65197 & 0.093 & 1.4(1) \\
$\nu_2-$5$\nu_{orb}$ & 13.993569 & 0.033 & 2.2(3) & $\nu_8-$4$\nu_{orb}$ & 16.70908 & 0.026 & $-$0.5(4) \\
$\nu_2-$4$\nu_{orb}$ & 14.398381 & 0.084 & 1.8(1) & $\nu_8-$2$\nu_{orb}$ & 17.5187 & 0.535 & $-$0.11(2) \\
$\nu_2-$3$\nu_{orb}$ & 14.803193 & 0.077 & 1.9(1) & $\nu_8-\nu_{orb}$ & 17.92351 & 0.062 & 2.1(1) \\
$\nu_2-$2$\nu_{orb}$ & 15.208005 & 0.059 & $-$0.8(2) & $\nu_8$ & 18.32832(1) & 0.346 & $-$0.03(3) \\
$\nu_2-\nu_{orb}$ & 15.612816 & 1.176 & 1.876(7) & $\nu_8$+$\nu_{orb}$ & 18.73314 & 0.051 & 0.7(2) \\
$\nu_2$ & 16.017628(2) & 0.784 & 1.80(1) & $\nu_8$+2$\nu_{orb}$ & 19.13795 & 0.356 & 1.59(3) \\
$\nu_2+\nu_{orb}$ & 16.422440 & 1.829 & 1.946(5) & $\nu_8$+5$\nu_{orb}$ & 20.35238 & 0.022 & 0.0(5) \\
$\nu_2$+2$\nu_{orb}$ & 16.827252 & 0.630 & 2.06(1) & $\nu_9-$6$\nu_{orb}$ & 16.02389 & 0.023 & 1.0(5) \\
$\nu_2$+3$\nu_{orb}$ & 17.232064 & 0.139 & 2.04(6) & $\nu_9-$4$\nu_{orb}$ & 16.83352 & 0.030 & 1.3(4) \\
$\nu_2$+4$\nu_{orb}$ & 17.636875 & 0.069 & 2.1(1) & $\nu_9-$3$\nu_{orb}$ & 17.23832 & 0.061 & $-$0.4(1) \\
$\nu_2$+5$\nu_{orb}$ & 18.041687 & 0.020 & 1.7(6) & $\nu_9-$2$\nu_{orb}$ & 17.64314 & 0.055 & 1.3(2) \\
$\nu_2$+6$\nu_{orb}$ & 18.446499 & 0.066 & 2.1(1) & $\nu_9-\nu_{orb}$ & 18.04795 & 0.112 & $-$0.3(1) \\
$\nu_2$+7$\nu_{orb}$ & 18.851311 & 0.033 & 1.6(3) & $\nu_9$ & 18.45276(3) & 0.048 & 1.5(2) \\
$\nu_2$+8$\nu_{orb}$ & 19.256123 & 0.049 & 2.1(2) & $\nu_9$+$\nu_{orb}$ & 18.85757 & 0.123 & $-$0.2(1) \\
$\nu_2$+9$\nu_{orb}$ & 19.660934 & 0.020 & 1.6(6) & $\nu_9$+2$\nu_{orb}$ & 19.26239 & 0.033 & 1.7(3) \\
$\nu_2$+10$\nu_{orb}$ & 20.065746 & 0.035 & 2.1(3) & $\nu_9$+3$\nu_{orb}$ & 19.6672 & 0.022 & 0.3(4) \\
$\nu_2$+12$\nu_{orb}$ & 20.87537 & 0.024 & 2.1(5) & $\nu_9$+4$\nu_{orb}$ & 20.07201 & 0.029 & 1.7(3) \\
$\nu_3-$7$\nu_{orb}$ & 16.256245 & 0.023 & 0.7(5) & $\nu_{10}-$2$\nu_{orb}$ & 15.23858 & 0.032 & 0.0(3) \\
$\nu_3-$5$\nu_{orb}$ & 17.065869 & 0.024 & 0.4(5) & $\nu_{10}$ & 16.04820(1) & 0.246 & $-$0.35(4) \\
$\nu_3-$3$\nu_{orb}$ & 17.875493 & 0.040 & 0.5(3) & $\nu_{10}$+$\nu_{orb}$ & 16.45301 & 0.193 & $-$0.56(5) \\
$\nu_3-$2$\nu_{orb}$ & 18.685116 & 0.034 & 0.7(3) & $\nu_{10}$+2$\nu_{orb}$ & 16.85783 & 0.136 & 1.01(7) \\
$\nu_3-\nu_{orb}$ & 19.494740(2) & 1.761 & 0.693(5) & $\nu_{10}$+3$\nu_{orb}$ & 17.26264 & 0.099 & $-$0.72(9) \\
$\nu_3$ & 19.899552 & 0.065 & 0.2(1) & $\nu_{10}$+4$\nu_{orb}$ & 17.66745 & 0.081 & 0.6(1) \\
$\nu_3$+$\nu_{orb}$ & 20.304363 & 1.606 & 2.323(5) & $\nu_{10}$+5$\nu_{orb}$ & 18.07226 & 0.119 & $-$0.10(8) \\
$\nu_3$+2$\nu_{orb}$ & 20.709175 & 0.053 & 1.6(1) & $\nu_{10}$+6$\nu_{orb}$ & 18.47707 & 0.106 & 2.24(8) \\
$\nu_4-$4$\nu_{orb}$ & 14.618782 & 0.042 & 0.8(3) & $\nu_{10}$+7$\nu_{orb}$ & 18.88188 & 0.056 & $-$0.1(1) \\
$\nu_4-$3$\nu_{orb}$ & 15.023594 & 0.074 & 2.1(1) & $\nu_{10}$+10$\nu_{orb}$ & 20.09632 & 0.031 & 0.8(3) \\
$\nu_4-$2$\nu_{orb}$ & 15.428406 & 0.727 & $-$0.76(1) & $\nu_{11}-$2$\nu_{orb}$ & 15.24229 & 0.307 & $-$0.20(3) \\
$\nu_4-\nu_{orb}$ & 15.833218 & 0.268 & $-$0.72(3) & $\nu_{11}-\nu_{orb}$ & 15.6471 & 0.066 & $-$0.7(1) \\
$\nu_4$ & 16.238029(2) & 1.676 & 0.816(5) & $\nu_{11}$ & 16.05191(1) & 0.342 & $-$0.01(3) \\
$\nu_4$+$\nu_{orb}$ & 16.642841 & 0.310 & 0.68(3) & $\nu_{11}$+$\nu_{orb}$ & 16.45672 & 0.282 & 0.36(3) \\
$\nu_4$+2$\nu_{orb}$ & 17.047653 & 0.333 & $-$0.64(3) & $\nu_{11}$+2$\nu_{orb}$ & 16.86154 & 0.248 & 1.49(4) \\
$\nu_5-$7$\nu_{orb}$ & 12.938057 & 0.014 & 1.9(6) & $\nu_{11}$+3$\nu_{orb}$ & 17.26635 & 0.035 & $-$0.4(3) \\
$\nu_5-$5$\nu_{orb}$ & 13.74768 & 0.016 & 1.9(6) & $\nu_{11}$+5$\nu_{orb}$ & 18.07597 & 0.036 & $-$0.3(3) \\
$\nu_5-$2$\nu_{orb}$ & 14.962116 & 0.084 & 1.2(1) & $\nu_{12}-$4$\nu_{orb}$ & 17.80443 & 0.024 & 2.0(4) \\
$\nu_5-\nu_{orb}$ & 15.366927 & 0.022 & 0.4(4) & $\nu_{12}-$3$\nu_{orb}$ & 18.20924 & 0.037 & $-$0.2(3) \\
$\nu_5$ & 15.771739(5) & 0.715 & 1.22(1) & $\nu_{12}-$2$\nu_{orb}$ & 18.61406 & 0.059 & 0.9(2) \\
$\nu_5$+$\nu_{orb}$ & 16.176551 & 0.244 & 1.11(3) & $\nu_{12}-\nu_{orb}$ & 19.01887 & 0.052 & 0.2(2) \\
$\nu_5$+2$\nu_{orb}$ & 16.581363 & 0.083 & $-$0.42(9) & $\nu_{12}$ & 19.42368(1) & 0.318 & $-$0.77(3) \\
$\nu_5$+3$\nu_{orb}$ & 16.986175 & 0.045 & $-$0.5(2) & $\nu_{12}$+2$\nu_{orb}$ & 20.2333 & 0.066 & 1.0(1) \\
$\nu_5$+4$\nu_{orb}$ & 17.390986 & 0.080 & $-$0.1(1) \\ 
\hline
\end{tabular}
\end{center}
\label{tab:freq}
\end{table*}

\setcounter{table}{0}

\begin{table*}
\caption{Multiple pulsation frequency solution for the {\it Kepler} 30-min photometry of KIC 4851217 (continued). The phases are calculated with respect to a time of primary minimum of BJD 2454953.900333. The formal errors on the frequencies and phases \citep{1999DSSN...13...28M} are given in braces in units of the last significant digit; the formal errors on the amplitudes are $\pm$0.008 mmag.} 
\begin{center}
\begin{tabular}{lccclccc}
\hline
ID & Freq. & Ampl. & Phase & ID & Freq. & Ampl. & Phase \\
 & (d$^{-1}$) & (mmag) & (rad) & & (d$^{-1}$) & (mmag) & (rad) \\
\hline
$\nu_{13}-$2$\nu_{orb}$  & 17.012433 & 0.122 & 0.41(7) &  $\nu_{26}-$3$\nu_{orb}$  & 17.79352 & 0.038 & 0.4(2) \\
$\nu_{13}-\nu_{orb}$  & 17.417245 & 0.035 & -0.6(3) &  $\nu_{26}-$2$\nu_{orb}$  & 18.19833 & 0.030 & 1.9(3) \\
$\nu_{13}$  & 17.822057(5) & 0.650 & 0.38(1) &  $\nu_{26}-\nu_{orb}$  & 18.60314 & 0.055 & -0.3(2) \\
$\nu_{14}-\nu_{orb}$  & 17.86777 & 0.035 & 0.3(3) &  $\nu_{26}$  & 19.00795(1) & 0.316 & 0.49(3) \\
$\nu_{14}$  & 18.27258(2) & 0.211 & 0.68(4) &  $\nu_{26}$+$\nu_{orb}$  & 19.41277 & 0.049 & 2.2(2) \\
$\nu_{14}$+2$\nu_{orb}$  & 19.08220 & 0.176 & -0.77(5) &  $\nu_{26}$+2$\nu_{orb}$  & 19.81758 & 0.243 & 2.06(4) \\
$\nu_{15}-$4$\nu_{orb}$  & 17.959205 & 0.059 & 1.2(1) &  $\nu_{26}$+6$\nu_{orb}$  & 21.43682 & 0.024 & 2.2(4) \\
$\nu_{15}-$3$\nu_{orb}$  & 18.364016 & 0.054 & 0.5(1) &  $\nu_{27}$  & 19.50695(7) & 0.045 & 1.6(2) \\
$\nu_{15}-$2$\nu_{orb}$  & 18.768828 & 0.272 & -0.16(3) &  $\nu_{27}$+2$\nu_{orb}$  & 20.31658 & 0.043 & 0.1(2) \\
$\nu_{15}$  & 19.578452(8) & 0.431 & -0.07(2) &  $\nu_{28}-$2$\nu_{orb}$  & 19.84917 & 0.032 & 0.7(3) \\
$\nu_{15}$+2$\nu_{orb}$  & 20.388075 & 0.039 & 1.6(2) &  $\nu_{28}$  & 20.65880(6) & 0.052 & 0.8(2) \\
$\nu_{16}-\nu_{orb}$  & 19.21393 & 0.099 & 1.17(9) &  $\nu_{28}$+2$\nu_{orb}$  & 21.46842 & 0.035 & 0.8(3) \\
$\nu_{16}$  & 19.61874(2) & 0.041 & -0.1(2) &  $\nu_{29}-$2$\nu_{orb}$  & 21.88076 & 0.034 & 1.7(3) \\
$\nu_{16}$+$\nu_{orb}$  & 20.02355 & 0.140 & 1.28(7) &  $\nu_{29}$  & 22.69038(9) & 0.035 & 1.9(3) \\
$\nu_{17}-$2$\nu_{orb}$  & 19.23716 & 0.119 & 0.91(8) &  $\nu_{30}-$4$\nu_{orb}$  & 15.14377 & 0.037 & 0.1(3) \\
$\nu_{17}$  & 20.04678(2) & 0.141 & 0.98(7) &  $\nu_{30}$  & 16.76302(8) & 0.042 & 0.2(2) \\
$\nu_{18}-$7$\nu_{orb}$  & 15.61714 & 0.022 & 0.8(4) &  $\nu_{30}+\nu_{orb}$  & 17.16783 & 0.032 & 1.9(3) \\
$\nu_{18}-$5$\nu_{orb}$  & 16.42676 & 0.024 & 1.0(4) &  $\nu_{30}$+2$\nu_{orb}$  & 17.57264 & 0.031 & 0.2(3) \\
$\nu_{18}-\nu_{orb}$  & 18.04601 & 0.142 & 1.26(6) &  $\nu_{30}$+3$\nu_{orb}$  & 17.97746 & 0.034 & -0.2(3) \\
$\nu_{18}$  & 18.45082(2) & 0.032 & -0.1(3) &  $\nu_{31}-$3$\nu_{orb}$  & 17.29054 & 0.027 & 1.1(3) \\
$\nu_{18}$+$\nu_{orb}$  & 18.85563 & 0.180 & 1.32(5) &  $\nu_{31}-\nu_{orb}$  & 18.10016 & 0.027 & -0.6(3) \\
$\nu_{18}$+3$\nu_{orb}$  & 19.66526 & 0.025 & 1.4(4) &  $\nu_{31}$  & 18.50497(6) & 0.060 & -0.4(2) \\
$\nu_{18}$+5$\nu_{orb}$  & 20.47488 & 0.027 & 1.5(3) &  $\nu_{31}$+2$\nu_{orb}$  & 19.31460 & 0.044 & 1.3(2) \\
$\nu_{19}-$10$\nu_{orb}$  & 15.68755 & 0.030 & -0.6(3) &  $\nu_{32}-$2$\nu_{orb}$  & 20.04343 & 0.059 & 1.9(2) \\
$\nu_{19}-$8$\nu_{orb}$  & 16.49718 & 0.039 & 1.5(2) &  $\nu_{32}$  & 20.85306(6) & 0.032 & 2.1(3) \\
$\nu_{19}-$7$\nu_{orb}$  & 16.90199 & 0.027 & 0.9(3) &  $\nu_{32}$+2$\nu_{orb}$  & 21.66268 & 0.046 & 0.4(2) \\
$\nu_{19}-$6$\nu_{orb}$  & 17.30680 & 0.036 & -0.3(3) &  $\nu_{33}-\nu_{orb}$  & 16.81905 & 0.033 & -0.6(3) \\
$\nu_{19}-$5$\nu_{orb}$  & 17.71161 & 0.059 & 1.2(2) &  $\nu_{33}$  & 17.22387(8) & 0.032 & 1.2(3) \\
$\nu_{19}-$2$\nu_{orb}$  & 18.92605 & 0.028 & 1.6(3) &  $\nu_{33}$+$\nu_{orb}$  & 17.62868 & 0.041 & -0.4(2) \\
$\nu_{19}$  & 19.73567(4) & 0.081 & 1.7(1) &  $\nu_{34}-$2$\nu_{orb}$  & 22.13968 & 0.039 & 0.7(2) \\
$\nu_{20}-$2$\nu_{orb}$  & 17.31839 & 0.083 & 2.0(1) &  $\nu_{34}$  & 22.94931(8) & 0.031 & -0.7(3) \\
$\nu_{20}$  & 18.12801(4) & 0.061 & 1.4(1) &  $\nu_{35}-$2$\nu_{orb}$  & 22.14875 & 0.045 & 1.7(2) \\
$\nu_{20}$+2$\nu_{orb}$  & 18.93764 & 0.038 & 1.4(2) &  $\nu_{35}$  & 22.95838(7) & 0.042 & 0.1(2) \\
$\nu_{21}$  & 16.10628(8) & 0.040 & -0.4(2) &  $\nu_{36}$  & 20.61582(7) & 0.050 & 2.0(2) \\
$\nu_{21}$+$\nu_{orb}$  & 16.51109 & 0.037 & 1.1(2) &  $\nu_{36}$+2$\nu_{orb}$  & 21.42545 & 0.033 & 2.0(3) \\
$\nu_{22}-$3$\nu_{orb}$  & 18.15684 & 0.027 & 1.9(3) &  $\nu_{37}$  & 17.39852(7) & 0.054 & -0.3(2) \\
$\nu_{22}$  & 19.37128(3) & 0.099 & 0.46(9) &  $\nu_{37}$+2$\nu_{orb}$  & 18.20814 & 0.024 & 1.3(4) \\
$\nu_{22}$+2$\nu_{orb}$  & 20.18090 & 0.094 & 1.96(9) &  $\nu_{38}-$5$\nu_{orb}$  & 15.69757 & 0.030 & 0.6(3) \\
$\nu_{23}$-3$\nu_{orb}$  & 18.56518 & 0.039 & -0.7(2) &  $\nu_{38}-$4$\nu_{orb}$  & 16.10238 & 0.028 & 2.3(3) \\
$\nu_{23}$  & 19.77961(8) & 0.027 & 0.9(3) &  $\nu_{38}-$2$\nu_{orb}$  & 16.91200 & 0.024 & -0.7(4) \\
$\nu_{23}$+3$\nu_{orb}$  & 20.99405 & 0.031 & -0.4(3) &  $\nu_{38}$  & 17.72163(2) & 0.207 & -0.48(4) \\
$\nu_{24}-$2$\nu_{orb}$  & 17.36238 & 0.038 & -0.1(2) &  $\nu_{38}$+2$\nu_{orb}$  & 18.53125 & 0.029 & -0.5(3) \\
$\nu_{24}-\nu_{orb}$  & 17.76720 & 0.029 & 1.9(3) &  $\nu_{39}-$2$\nu_{orb}$  & 15.71829 & 0.040 & 1.8(2) \\
$\nu_{24}$  & 18.17201(5) & 0.072 & -0.3(1) &  $\nu_{39}$  & 16.52792(6) & 0.054 & 2.1(2) \\
$\nu_{24}$+$\nu_{orb}$  & 18.57682 & 0.029 & 2.1(3) &  $\nu_{40}$  & 19.49752(3) & 0.100 & 0.89(9) \\
$\nu_{24}$+2$\nu_{orb}$  & 18.98163 & 0.071 & -0.1(1) &  $\nu_{41}$  & 21.19934(4) & 0.091 & 1.7(1) \\
$\nu_{25}-$2$\nu_{orb}$  & 19.81179 & 0.043 & 1.1(2) &  $\nu_{42}$  & 21.72483(6) & 0.056 & 1.4(2) \\
$\nu_{25}$  & 20.62141(4) & 0.085 & 1.2(1) &  $\nu_{43}$  & 21.72880(4) & 0.079 & -0.7(1) \\
$\nu_{25}$+2$\nu_{orb}$  & 21.43104 & 0.061 & 1.2(1) &  $\nu_{44}$  & 20.14523(2) & 0.152 & 1.56(6) \\
& & & & $\nu_{45}$  & 15.85114(3) & 0.096 & 0.1(1) \\
& & & & $\nu_{46}$  & 19.09031(1) & 0.275 & -0.09(3) \\
\hline
\end{tabular}
\end{center}
\end{table*}

\clearpage
\FloatBarrier
\section{Results from previous authors}
\begin{table}\caption{\label{tab: appendix- liakos comparisons} Comparisons of the model-independent results obtained in this work against those derived by \citet{Liakos_2020_KIC4851217}. The first column compares our results obtained from the individual analyses of the data subsets against those reported by \citet{Liakos_2020_KIC4851217} ($\Delta_1$), i.e., a negative value means our result is smaller than his. The second column gives the comparisons for our results using method 1 of the combined analysis ($\Delta_2$). Also given are these discrepancies in units of their mutual uncertainties $\sigma$.}
\centering
\begin{tabular}{>{$}l<{$} r@{\,$\pm$\,}l r@{\,\%\,\,(}l@{\,$\sigma$)} c r@{\,\%\,\,(}l@{\,$\sigma$)}}

\hline
       &   \multicolumn{2}{c}{previous result}&
           \multicolumn{2}{c}{$\Delta_1$ } & &
           \multicolumn{2}{c}{$\Delta_2$} \\
\hline
e              &0.036 & 0.001 &  -11.1     &  -2.0     &&   -13   &   -5.0 \\
q              &1.14 & 0.04&  -0.4  &  -0.1  &&    -0.2 &   -0.1 \\
T_{\rm eff,Aa} &8000 & 250&       -2.1   &  -0.6   &&    -2.0 &   -0.7 \\
T_{\rm eff,Ab} &7890 & 98&       -2.4   &  -1.5   &&    -1.9 &   -1.5 \\
k          &1.026 & 0.027&       36.6     &  10.0     &&    42.4 &   12,5 \\
M_{\rm Aa} &1.92 & 0.10&       -1.1   &  -0.2   &&    -2.3 &   -0.4 \\
M_{\rm Ab} &2.19 & 0.18&       -1.6   &  -0.2   &&    -2.6 &   -0.3 \\
R_{\rm Aa} &2.61 & 0.05&      -15.9     &  -7.1     &&   -19.5 &   -8.7 \\
R_{\rm Ab} &2.68 & 0.05&       14.8     &   6.2     &&    14.5 &    7.5 \\
\log(g)_{\rm Aa} &3.89 & 0.03&        3.7   &   4.5   &&     4.6 &    5.5 \\
\log(g)_{\rm Ab} &3.92 & 0.04&       -3.2   &  -3.0   &&    -3.2 &   -3.1 \\
\log(L)_{\rm Aa} &1.398 & 0.052&      -13.2     &  -3.3     &&   -16.0 &   -4.2 \\
\log(L)_{\rm Ab} &1.398 & 0.052&        5.7   &   1.4   &&     6.0 &    1.6 \\
\hline
\end{tabular}
\end{table}


\bsp	
\label{lastpage}
\end{document}